\documentclass[useAMS]{aa}

\usepackage{times}
\usepackage{graphicx,multicol}
\usepackage{caption}
\usepackage{subcaption}

\usepackage{amsmath,amsfonts,amsthm,bm}
\usepackage{txfonts}
\usepackage{relsize}
\usepackage[colorlinks=true,citecolor=cyan, urlcolor=cyan, linkcolor=cyan]{hyperref}
\usepackage{cleveref}
\usepackage[section]{placeins}

\newcommand\Tstrut{\rule{0pt}{2.9ex}}         
\newcommand\Bstrut{\rule[-1.2ex]{0pt}{0pt}}   
\newcommand\TBstrut{\Tstrut\Bstrut}

\begin{document}

\title{First Core Properties: From Low- to High-mass Star Formation}

\author{Asmita Bhandare$^{1,4}$
	\and Rolf Kuiper$^{2,1}$ 
	\and Thomas Henning$^1$
	\and Christian Fendt$^1$
	\and Gabriel-Dominique Marleau$^{2,3,1}$
	\and Anders K\"olligan$^2$ 	
}

\institute{$^1$ Max-Planck-Institut f\"ur Astronomie, K\"onigstuhl 17, 69117 Heidelberg, Germany \\
\email{bhandare@mpia.de} \\
$^2$ Institut f\"ur Astronomie und Astrophysik, Universit\"at T\"ubingen, Auf der Morgenstelle 10, 72076 T\"ubingen, Germany	\\
$^3$ Physikalisches Institut, Universit\"at Bern, Sidlerstr. 5, 3012 Bern, Switzerland \\
$^4$ Member of the International Max-Planck Research School for Astronomy and Cosmic Physics at the University of Heidelberg (IMPRS-HD), Germany   
}

\date{Submitted: January 13, 2018 / Accepted: July 16, 2018}

\abstract{}{In this study, the main goal is to understand the molecular cloud core collapse through the stages of first and second hydrostatic core formation. We investigate the properties of Larson’s first and second cores following the evolution of the molecular cloud core until formation of Larson's cores. We expand these collapse studies for the first time to span a wide range of initial cloud masses from 0.5 to 100 $\mathrm{M_{\odot}}$.}{Understanding the complexity of the numerous physical processes involved in the very early stages of star formation requires detailed thermodynamical modeling in terms of radiation transport and phase transitions. For this we use a realistic gas equation of state via a density and temperature-dependent adiabatic index and mean molecular weight to model the phase transitions. We use a gray treatment of radiative transfer coupled with hydrodynamics to simulate Larson’s collapse in spherical symmetry.} {We reveal a dependence of a variety of first core properties on the initial cloud mass. The first core radius and mass increase from the low-mass to the intermediate-mass regime and decrease from the intermediate-mass to the high-mass regime. The lifetime of first cores strongly decreases towards the intermediate- and high-mass regime.} {Our studies show the presence of a transition region in the intermediate-mass regime. Low-mass protostars tend to evolve through two distinct stages of formation which are related to the first and second hydrostatic cores. In contrast, in the high-mass star formation regime, the collapsing cloud cores rapidly evolve through the first collapse phase and essentially immediately form Larson's second cores. }

\keywords{Stars: formation - Methods: numerical - Hydrodynamics - Radiative transfer - Gravitation - Equation of state}

\authorrunning{A. Bhandare et al.}

\maketitle

\graphicspath{{}}

\section{Introduction}
\label{sec:intro}

Stars are formed by the gravitational collapse of dense, gaseous and dusty cores within magnetized molecular clouds. Details of the earliest epochs of star formation and protostellar evolution however remain far from fully understood due to the complexity of the physical processes involved, such as hydrodynamics, radiative transfer and magnetic fields. Understanding how stars form has thus been one of the most fundamental questions raised for several decades \citep[see detailed reviews by][]{Larson2003,Mckee2007,Inutsuka2012}. Owing to the optically thick regime, it is still very challenging to obtain reliable observational constraints during the earliest phases of star formation \citep[e.g.][]{Nielbock2012, Launhardt2013, Dunham2014}. Numerical studies by \citet{Larson1969} were among the first to indicate the presence of two quasi-hydrostatic cores that are formed during a non-homologous collapse of the molecular cloud. They investigated the faster collapse of the denser inner region in comparison to the less dense outer region with one-dimensional (1D) hydrodynamic simulations using the diffusion approximation for radiative transfer. Since then there have been detailed numerical investigations using both grid-based \citep[][and references therein]{Bodenheimer1968, Winkler1980a, Winkler1980b, Stahler1980a, Stahler1980b, Stahler1981, Masunaga1998, Masunaga2000, Tomida2010b, Commercon2011, Vaytet2012, Tomida2013, Vaytet2013, Vaytet2017} and smoothed particle hydrodynamics (SPH) simulations \citep{Whitehouse2006, Stamatellos2007, Bate2014} to better understand the isolated collapse scenario. The early phases of low- and intermediate-mass star formation can briefly be summarized as follows. 

Initially, the optically thin cloud collapses isothermally under its own gravity due to the efficient thermal emission from dust grains during this phase. The collapse may be initiated either by the ambipolar diffusion of magnetic fields that once supported the cloud against gravitational collapse \citep[e.g.,][]{Shu1987} or by the dissipation of turbulence which reduces the effective speed of sound in cloud cores \citep[e.g.,][]{Nakano1998}. It can start from a contracting, marginally stable Bonnor--Ebert sphere or the collapse can be triggered by an external shock wave running over the previously stable cloud \citep{Masunaga2000}. 

As the density increases, the optical depth becomes greater than unity and radiation cooling becomes inefficient. As the cloud compresses further, the temperature in this dense central region gradually increases. This almost halts the collapse and leads to the formation of the first hydrostatic core which subsequently contracts adiabatically with an adiabatic index $\gamma \approx$~5/3. With a rise in temperature, the rotational degrees of freedom of the diatomic gas get excited and the adiabatic index changes to $\approx$ 7/5. Once the temperature inside the first core reaches $\sim$ 2000~K, $\mathrm{H_2}$ molecules begin to dissociate. Gravity wins over pressure since $\mathrm{H_2}$ dissociation is a strongly endothermic process. The core thus becomes unstable which leads to the second collapse phase. Once most of the $\mathrm{H_2}$ has been dissociated, it is followed by the formation of the second hydrostatic core. The second core forms almost instantaneously and hence the second collapse phase lasts only for a few years, in comparison to the first collapse phase which lasts for about $10^4$~years for an initially 1 $\mathrm{M_{\odot}}$ cloud. The collapse is then stopped by an increase in thermal pressure. The surrounding envelope continues to fall onto the central core as the core grows in mass through the main accretion phase with a further increase in temperature. A star is born when the core reaches ignition temperatures for nuclear reactions. 

In the studies presented here we simulate the gravitational collapse for isolated gas spheres with a uniform temperature and initial Bonnor--Ebert density profile. Thermodynamical modeling in terms of radiation transport and phase transitions is crucial to better understand the complex physical mechanisms involved. Hence, we use the gray flux-limited diffusion (FLD) radiative transfer \citep{Levermore1981} coupled with hydrodynamics to simulate Larson's collapse. Using one-dimensional spherically symmetric collapse simulations, we investigate properties of Larson's first and second core. One-dimensional studies are proven to be of importance in understanding the role of different physical processes involved while three-dimensional studies can still be computationally very expensive. In this work, we focus on properties of the hydrostatic cores governed by gravity and thermal pressure and not of the environment. Since the thermal pressure is isotropic, a one-dimensional approach is quite a good approximation for these objects, even though the collapsing environment is not described accurately. 

The different chemical species affect the gas hydrodynamics via heat capacity, line cooling and chemical energy and the radiation via gas and dust opacities. In order to take into account effects such as dissociation, ionization, rotational and vibrational degrees of freedom for the molecules in our studies, we use a realistic gas equation of state with a density and temperature-dependent adiabatic index and mean molecular weight to model phase transitions. Using a non-constant adiabatic index is particularly important since it has a strong influence on the thermal evolution of the gas and in general also on the stability of the gas against gravitational collapse \citep{Stamatellos2009}. The specific heat and mean molecular weight are computed as a function of temperature by solving partition functions for rotational, vibrational and translational energy levels of $\mathrm{H_2}$ instead of using a constant value. 

The main goal of this paper is to understand the entire collapse phase through the stages of first and second core formation by incorporating a realistic gas equation of state and appropriate opacity tables for a wide range of initial cloud masses from 0.5~to~100.0 $\mathrm{M_{\odot}}$. In doing so, we quantify the dependence of the first core properties on the initial cloud mass. 

The paper is organized as follows. The microphysics used in our studies is detailed in \cref{sec:RHDeqns}. We describe our numerical scheme and initial setup in \cref{sec:Method}. The evolution of the cloud through various stages until the formation of the second hydrostatic core is presented in \cref{sec:results}. We first provide a detailed description of the collapse of an initial 1.0 $\mathrm{M_{\odot}}$ cloud in \cref{sec:solarmass} and then extend this explanation to all other cases for different initial cloud masses from 0.5 up to 100 $\mathrm{M_{\odot}}$ in \cref{sec:initcloudmass}. We further discuss the dependence of the first core properties on the initial cloud mass in \cref{sec:firstcore}. In \cref{sec:initialsetup} we extend our parameter space to study the influence of initial cloud properties on the first core properties. Our results are in good agreement with previous work and comparisons are provided in \cref{sec:comparisons}. We note the limitations of our method and discuss the outlook in \cref{sec:limitations}. Section \ref{sec:Summary} summarizes the results presented herein.

\section{Equations of radiation hydrodynamics}
\label{sec:RHDeqns}
Gas thermodynamics is considered under the approximation of local thermodynamic equilibrium (LTE) and a two-temperature approach (2T), for the gas and radiation. The basic hydrodynamics equations that account for the conservation of mass, momentum and energy i.e. the continuity, Euler's and energy equation, respectively, are given as 
\begin{align}
\mbox{$ \partial_{t} ~\rho + \bm \nabla \cdot (\rho ~ \bm u)$ = 0 },
\end{align}
\begin{align}
\mbox{$ \partial_{t} ~(\rho \bm u) + \bm\nabla \cdot (\rho ~\bm u \otimes ~\bm u + P)  = \rho \bm a $}, 
\end{align}
\begin{align}
\mbox{$ \partial_{t} ~E +  \bm\nabla \cdot ((E + P) ~\bm u) = \rho ~\bm u \cdot \bm a$}, 
\label{eq:energyconservation}
\end{align}
where $\rho$ is the density, $\bm u$ is the dynamical velocity, \textit{P} is the thermal pressure, and $\bm{a}$ denotes the acceleration source term due to self-gravity given by
\begin{align}
\mbox{$ \bm{a} = - \bm\nabla \mathrm{\bm\Phi_{sg}}$},
\end{align}
where $\mathrm{\bm\Phi_{sg}}$ is the gravitational potential determined using Poisson's equation, expressed as
\begin{align}
\mbox{$ \bm\nabla^2 \mathrm{\bm\Phi_{sg}} = 4 \pi G \rho$}.
\end{align}
The total energy \textit{E} = $\textit{E}_\mathrm{int}$ + $\textit{E}_\mathrm{kin}$ is the sum of internal and kinetic energy. The kinetic energy density $\textit{E}_\mathrm{kin} = \tfrac{1}{2} \rho \bm u^2$, whereas the internal energy density is calculated by taking into account the contributions from different hydrogen and helium species. This is described in \cref{sec:EOS}. 

The time-dependent radiation transport equation in case of locally isotropic radiation when neglecting small contributions due to scattering can be written as
\begin{align}
\mbox{$ \partial_t ~E_\mathrm{rad} + \bm\nabla \cdot \bm F_\mathrm{rad} = c ~\chi_\mathrm{abs} ~(B_\mathrm{rad} - E_\mathrm{rad}) $}, 
\label{eq:radiationeqn}
\end{align}
where $E_\mathrm{rad}$ is the radiation energy density, $\bm F_\mathrm{rad}$ is the radiation energy flux, \textit{c} is the speed of light, $\chi_\mathrm{abs}$ is the coefficient of absorption and $B_\mathrm{rad}$ is the integral of the black-body Planck spectrum. 
The flux of radiation energy density $\bm F_\mathrm{rad}$ in the FLD approximation is determined as
\begin{align}
\mbox{$ \bm F_\mathrm{rad} = - ~D_\mathrm{rad} ~\bm\nabla E_\mathrm{rad} = - ~\dfrac{\lambda c}{\kappa_\mathrm{R} \rho}  ~\bm\nabla E_\mathrm{rad}$},
\label{eq:flux}
\end{align}
where $\kappa_\mathrm{R}$ is the Rosseland mean opacity and the flux limiter $\lambda$ is chosen following \citet{Levermore1981}. The flux limiter recovers the limiting cases of diffusion and free streaming, respectively. 

Using Eq. \eqref{eq:flux} in the conservation Eq. \eqref{eq:radiationeqn} gives the time evolution of radiation energy density as
\begin{align}
\mbox{$ \partial_t ~E_\mathrm{rad} - \bm\nabla \cdot (D_\mathrm{rad} ~\bm\nabla  E_\mathrm{rad}) = c ~\chi_\mathrm{abs} ~(B_\mathrm{rad} - E_\mathrm{rad})$}.
\label{eq:diffusioneqn}  
\end{align}
The two unknowns in the above Eq. \eqref{eq:diffusioneqn} namely, the radiation energy density $E_\mathrm{rad}$ and the local temperature of the medium $\textit{B}_\mathrm{rad} = a T^4$, where \textit{a} is the radiation constant are coupled to each other via heating and cooling processes. The time evolution of the local internal energy is given by
\begin{align}
\mbox{$\partial_t E_\mathrm{int} = - ~c ~\chi_\mathrm{abs} ~(B_\mathrm{rad} - E_\mathrm{rad})$}. 
\label{eq:Eint}
\end{align}
For the two-temperature model, the coupled equations \eqref{eq:diffusioneqn} and \eqref{eq:Eint} can be reduced to a single equation using a linearization approach where the radiation and medium temperatures evolve as two different quantities \citep{CommerconTeyssier2011}. Additionally, the specific heat capacity is taken to be constant over the course of a single main iteration. 

A detailed description of the numerical code in use can be found in \citet{Mignone2007} and \citet{Mignone2012} for the hydrodynamics, \citet{Vaidya2015} and Marleau et al. (in prep.) for the gas equation of state (see also following section), \citet{KuiperRT2010} and Kuiper et al. (subm.) for the radiation transport, and \citet{Kuiper2010} and \citet{Kuiper2011} for the self-gravity. 

\subsection{Gas equation of state}
\label{sec:EOS}

We use the gas equation of state (EOS) of \citet{Dangelo2013} to account for effects such as ionization of atomic hydrogen and helium, dissociation of molecular hydrogen ($\mathrm{H_2}$) as well as the molecular vibrations and rotations. This is a realistic approach for modeling the second collapse phase where $\mathrm{H_2}$ begins to dissociate depending on the pressure, temperature and density. This EOS has been implemented in the PLUTO code by \citet{Vaidya2015} and we have now updated the radiation transport module to make use of this (see details in Marleau et al., in prep.).  

The adiabatic index or ratio of specific heats $\gamma$,
which takes into account the translational, rotational, and vibrational degrees of freedom, is defined as
\begin{align}
\mbox{$\gamma = \dfrac{C_P}{C_V}$}.
\end{align}  

\begin{figure}[!tp]
\centering
\begin{subfigure}{0.5\textwidth}
\includegraphics[width=\textwidth]{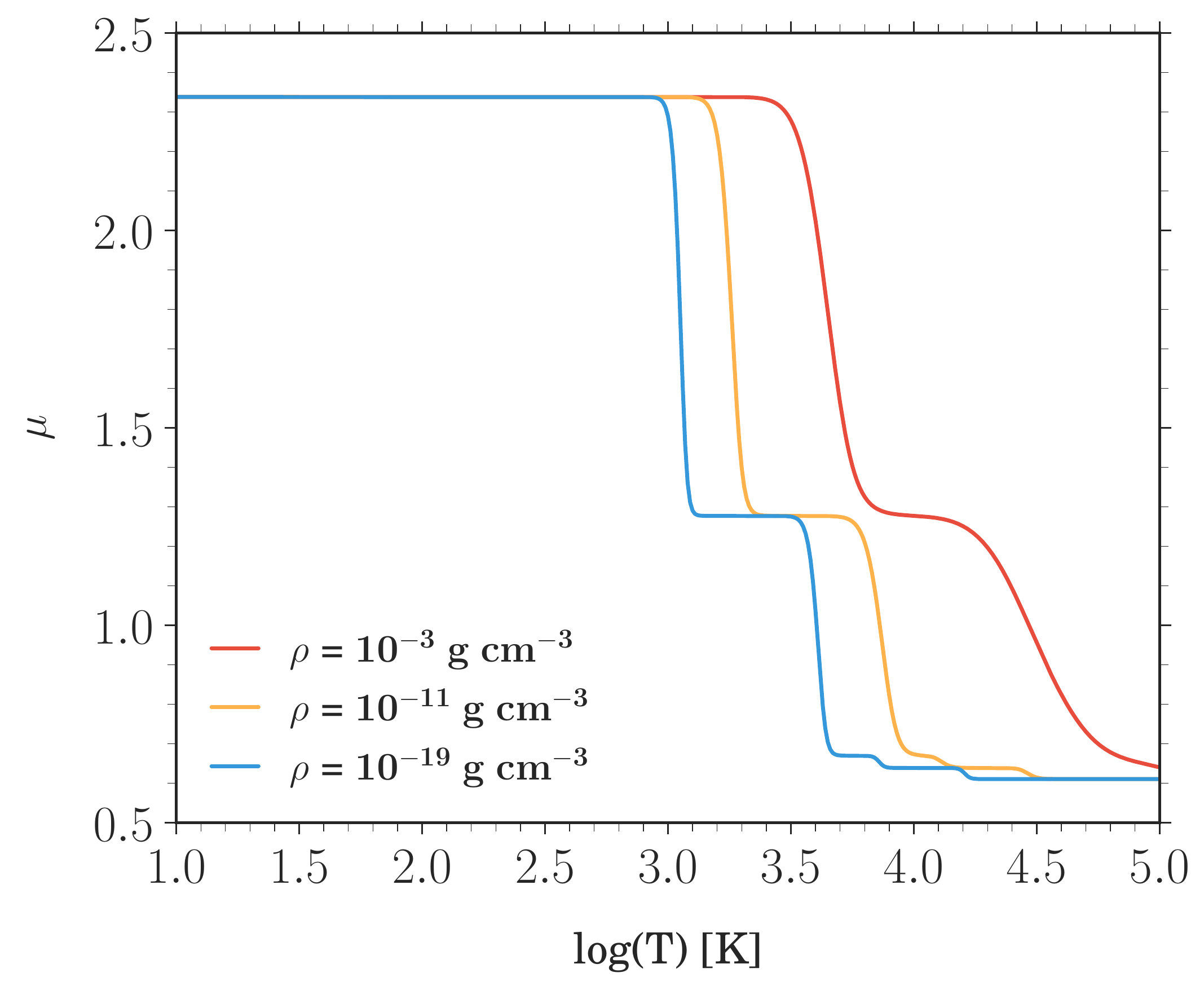}
\end{subfigure}
\begin{subfigure}{0.5\textwidth}
\includegraphics[width=\textwidth]{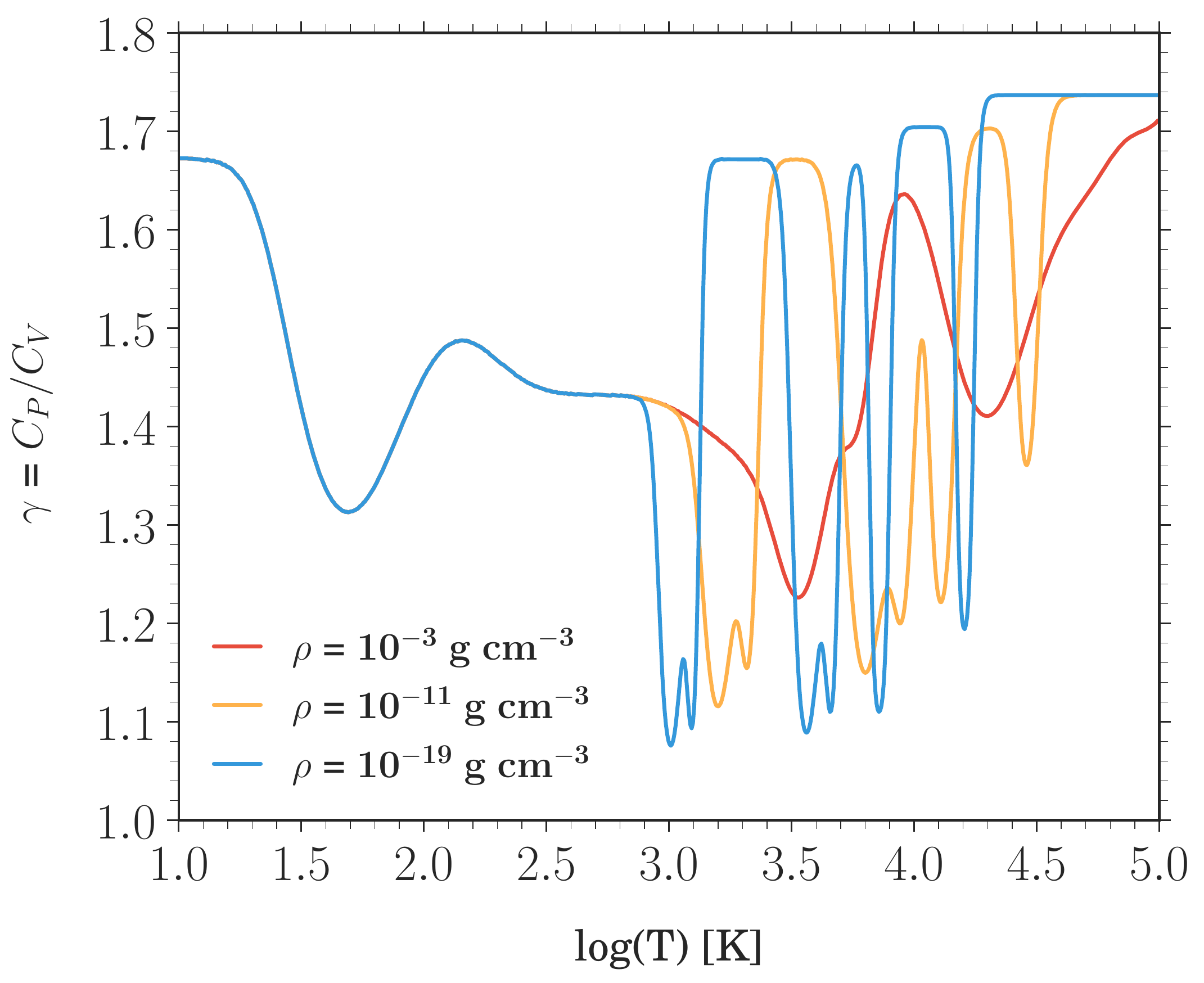}
\end{subfigure} 
\caption{Mean molecular weight $\mu$ and adiabatic index $\gamma$ as a function of temperature for three different gas densities (\mbox{$\rho = 10^{-3}$, $10^{-11}$ and $10^{-19}$ g cm$^{-3}$)}.}
\label{fig:eos}
\end{figure}

Figure~\ref{fig:eos} shows the mean molecular weight $\mu$ and $\gamma$ as a function of temperature and also indicates the dependence on gas density. The mean molecular weight $\mu$ has an upper limit of $\sim$~2.3 and lower limit of $\sim$ 0.6. The first transition (i.e. the plateau region) indicates the dissociation of $\mathrm{H_2}$ whereas the second transition shows the ionization phase. In the plot showing $\gamma$ as a function of temperature, the gas behaves as a monatomic ideal gas with $\gamma \approx$ 5/3 at lower temperatures. The transition from a monatomic gas $\gamma \approx$ 5/3 (where rotational degrees of freedom of $\mathrm{H_2}$ are frozen) to a diatomic gas $\gamma \approx$ 7/5 and further to the dissociation phase where $\gamma \approx 1.1$ is also seen as dips in $\gamma$. Following the curve to higher temperatures, the other dips occur at the ionization of hydrogen and at the first and second ionization of helium. Increasing density raises the temperature at which these processes occur. Since the range in $\log T$ over which they occur widens and the $\gamma$ dips become shallower, the dips gradually blend. This is clear when comparing the curves at $\rho=10^{-19}$~g~cm$^{-3}$ and at~$\rho=10^{-3}$~g~cm$^{-3}$.

In our studies, we assume the ortho:para ratio of molecular hydrogen to be in thermal equilibrium at all temperatures. \citet{Vaytet2014} show that the ortho:para ratio influences the thermal evolution of the first core but has negligible effects on the core properties. An appropriate ortho:para ratio as initial conditions for star formation still remains unclear. 

Considering LTE, the ionization-recombination and dissociation processes for hydrogen are given by
\begin{equation}
\begin{split}
&\mathrm{H + e^- \rightleftharpoons H^+ + 2e^-} \\
&\mathrm{H_2 \rightleftharpoons H + H}, 	
\end{split}
\end{equation}
respectively. The degree of ionization of atomic hydrogen $x$, the degree of dissociation of molecular hydrogen $y$ and the degrees of single $z_1$ and double $z_2$ ionization of helium are defined from \citet{Dangelo2013} as
\begin{align}
\mbox{$x = \dfrac{\rho_\mathrm{H^+}}{\rho_\mathrm{H^+} + \rho_\mathrm{H}}$}
\end{align}
\begin{align}
\mbox{$y = \dfrac{\rho_\mathrm{H}}{\rho_\mathrm{H} + \rho_\mathrm{H_2}}$}
\end{align}
\begin{align}
\mbox{$z_1 = \dfrac{\rho_\mathrm{He^+}}{\rho_\mathrm{He^+} + \rho_\mathrm{He}}$}
\end{align}
\begin{align}
\mbox{$z_2 = \dfrac{\rho_\mathrm{He^{2+}}}{\rho_\mathrm{He^{2+}} + \rho_\mathrm{He^+}}$}.
\end{align}
Following the Boltzmann law of the energy distribution, the ionization and dissociation degrees using Saha equations is given as \citep[see for e.g.,][]{Black1975}
\begin{align}
\mbox{$\dfrac{x^2}{1-x} = \dfrac{m_\mathrm{H}}{X \rho} ~\Bigg( {\dfrac{m_\mathrm{e} k_\mathrm{B} T}{2 \pi \hbar^2}}\Bigg)^{3/2} ~e^{-13.60 ~\mathrm{eV} / (k_\mathrm{B} T)}$}
\label{eq:H}
\end{align}
\begin{align}
\mbox{$\dfrac{y^2}{1-y} = \dfrac{m_\mathrm{H}}{2 X \rho} ~\Bigg( {\dfrac{m_\mathrm{H} k_\mathrm{B} T}{4 \pi \hbar^2}}\Bigg)^{3/2} ~e^{-4.48 ~\mathrm{eV} / (k_\mathrm{B} T)}$}
\label{eq:H2}
\end{align}
\begin{align}
\mbox{$\dfrac{z_1}{1-z_1} = \dfrac{4m_\mathrm{H}}{\rho} ~\Bigg( {\dfrac{m_\mathrm{e} k_\mathrm{B} T}{2 \pi \hbar^2}} \Bigg)^{3/2} \dfrac{e^{-24.59 ~\mathrm{eV} / (k_\mathrm{B} T)}}{X + z_1 Y/4}$}
\end{align}
\begin{align}
\mbox{$\dfrac{z_2}{1-z_2} = \dfrac{m_\mathrm{H}}{\rho} ~\Bigg( {\dfrac{m_\mathrm{e} k_\mathrm{B} T}{2 \pi \hbar^2}}\Bigg)^{3/2} \dfrac{e^{-54.42 ~\mathrm{eV} / (k_\mathrm{B} T)}}{X + (z_2 + 1) ~Y/4}$},
\end{align}
where $m_\mathrm{e}$ is the electron mass, $m_\mathrm{H}$ is the hydrogen mass, $k_\mathrm{B}$ is the Boltzmann constant, $\hbar$ is Planck's constant divided by 2$\pi$ and \mbox{$\rho$ = $n \mu m_\mathrm{u}$} is the total gas density. The hydrogen and helium mass fractions are taken as $X = 0.711$ and $Y = 0.289$, respectively. 

For a gas mixture mainly consisting of hydrogen (atoms, molecules $\&$ ions), helium and a negligible fraction of metals, the mean molecular weight $\mu$ is given as (e.g., \citealt{Black1975})
\begin{align}
\mbox{$\dfrac{\mu}{4} = [2X (1 + y + 2xy) + Y (1 + z_1 + z_1 z_2)]^{-1} $},
\end{align}
and the gas internal energy density $(\rho e)_\mathrm{gas}$ is given by
\begin{align}
\mbox{$(\rho e)_\mathrm{gas} = (\epsilon_\mathrm{H_2} + \epsilon_\mathrm{H} + \epsilon_\mathrm{He} + \epsilon_\mathrm{H + H} + \epsilon_\mathrm{H^+} + \epsilon_\mathrm{He^+} + \epsilon_\mathrm{He^{2+}}) ~\dfrac{\rho k_\mathrm{B} T}{m_\mathrm{u}}$}.
\end{align}
Here, the quantity $m_\mathrm{u}$ is the atomic mass unit and contributions from different species in the parenthesis are dimensionless and can be obtained using an appropriate partition function $\zeta$ by taking into account the translational, rotational and vibrational degrees of freedom as detailed in \citet{Dangelo2013}. 

The stability condition needed for numerical calculations requires an estimate of the sound speed $c_\mathrm{s}$ which relates pressure and density and is defined as
\begin{align}
\mbox{$ c_\mathrm{s}^2 = \dfrac{\Gamma_1 P}{\rho}$}.
\end{align} 
Here, $\Gamma_1$ is the first adiabatic index, which has a functional dependence on temperature and density, given as
\begin{align}
\mbox{$ \Gamma_1 = \dfrac{1}{C_\mathrm{V} (T)} ~\Bigg( \dfrac{P}{\rho T} \Bigg) ~\chi^2_\mathrm{T} + \chi_\mathrm{\rho} $},
\end{align}
where $C_\mathrm{V}(T)$ is obtained by taking the derivative of the specific gas internal energy $e(T)$ with respect to temperature at a constant volume and the temperature $\chi_\mathrm{T}$ and density $\chi_\mathrm{\rho}$ exponents (see \citealt{Dangelo2013}) are defined by
\begin{align}
\mbox{$\chi_\mathrm{T} = {\Bigg( \dfrac{\partial \mathrm{ln} ~P}{\partial \mathrm{ln} ~T} \Bigg)}_{\rho} = 1 - \dfrac{\partial \mathrm{ln} \mu} {\partial \mathrm{ln} T}$}
\end{align}
\begin{align}
\mbox{$\chi_\mathrm{\rho} = {\Bigg( \dfrac{\partial \mathrm{ln} ~P}{\partial \mathrm{ln} ~\rho} \Bigg)}_{T} = 1 - \dfrac{\partial \mathrm{ln} \mu} {\partial \mathrm{ln} \rho}$}.
\end{align}
Note that for an ideal gas where phase transitions are ignored, i.e., with constant $\mu$ and $\gamma$, $\Gamma_1$ is equal to the adiabatic index $\gamma$.  

With all of the above considerations, the thermal EOS (relating pressure, temperature and volume) and the caloric EOS (relating internal energy, volume and temperature) can be expressed as
\begin{equation}
\begin{split}
&P = \dfrac{\rho k_\mathrm{B} T}{m_\mathrm{u} \mu(X)} \\
&e = e(T, X),
\end{split}
\label{eq:pvteEOS}
\end{equation}
where the mean molecular weight $\mu (X)$ depends on the gas composition. Here, the chemical fractions are not solved independently and can be expressed as $X = X(T, \rho)$ under equilibrium assumptions. Thus the thermal and caloric EOS can also be expressed as a function of temperature and density, $P = P(\rho, T)$ and $e = e(T, \rho)$, respectively. Owing to the explicit temperature dependence, the conversion between pressure and internal energy density and vice-versa, is preceded by computing temperatures using the thermal EOS and pre-computed lookup tables of pressure and internal energy density. Further details on the implementation of lookup tables can be found in \citet{Vaidya2015}.

\subsection{Opacities}
\label{sec:opacity}
We make use of tabulated dust opacities from \citet{Ossenkopf1994} and tabulated gas opacities from \citet{Malygin2014}. At lower temperatures the contribution from dust dominates whereas at higher temperatures this is negligible since the dust is evaporated. 

Code-wise, we updated the evaporation and sublimation module to consider a time-dependent evolution of the dust. The dust is treated as being perfectly coupled to the gas, i.e. the dust is moving with the gas flow, but the dust content is allowed to change in time due to evaporation and sublimation of dust grains. Hence, we store -- in addition to the gas mass density -- the local dust-to-gas mass ratio $R(t) = M_\mathrm{dust} / M_\mathrm{gas}$. 

The evaporation temperature $T_\mathrm{evap}$ is computed based on \citet{Pollack1994}
utilizing the power-law formula by \citet{Isella2005}, their Eq.~(16):
\begin{align}
\mbox{$T_\mathrm{evap} = \beta_1 \times \Bigg({\dfrac{\rho_\mathrm{gas}}{1\mathrm{~g ~cm^{-3}}}}\Bigg)^{\beta_2}$},
\end{align}
with $\beta_1 = 2000$ K and $\beta_2 = 1.95 \times 10^{-2}$.
In the sublimation regime $T_\mathrm{dust} < T_\mathrm{evap}$, the temporal evolution of the dust-to-gas mass ratio R(t) is described by
\begin{align}
\mbox{$R(t+\Delta t) = R(t) + dR ~\times ~R_\mathrm{max} \times \left( 1 - \exp\left(- \dfrac{\Delta t }{t_\mathrm{subl}} \times dT \times dR \right) \right)$},
\end{align}
with
$dT = | T_\mathrm{evap} - T_\mathrm{dust}| / T_\mathrm{evap}$ and
$dR = (R_\mathrm{max} - R(t)) / R_\mathrm{max}$.
In the evaporation regime $T_\mathrm{dust} > T_\mathrm{evap}$, the temporal evolution of the dust-to-gas mass ratio is described by
\begin{align}
\mbox{$R(t+\Delta t) = R(t) \times \exp\left(- \dfrac{\Delta t}{t_\mathrm{evap}} \times dT \times \dfrac{1}{dR + \omega} \right)$}.
\end{align}
Here, $\omega$ serves as a lower limit to the $dR$ term, which prevents the $dR^{-1}$ term from diverging. 

In a nutshell, evaporation and sublimation becomes more efficient for higher temperature differences between dust and evaporation temperature. Furthermore, the evaporation efficiency decreases towards lower dust-to-gas mass ratios, and the sublimation efficiency decreases towards the maximum dust-to-gas mass ratio allowed. For all simulations performed, we used $R_\mathrm{max} =~0.01$, $t_\mathrm{subl} = 10~\mbox{yr}$, $t_\mathrm{evap} = 100~\mbox{yr}$, and $\omega = 0.01$.

\section{Numerics and initial setup}
\label{sec:Method}

In this study, one-dimensional spherically symmetric collapse simulations are performed using the (magneto) hydrodynamic code PLUTO \citep{Mignone2007} combined with the gray flux-limited diffusion (FLD) radiation transport module \mbox{MAKEMAKE}. The theory and numerics of the radiation transfer scheme are described and tested in \citet{KuiperRT2010} and \mbox{Kuiper et al. (subm.).} 

\cite{Vaytet2012, Vaytet2013} have indicated slight differences in the core properties between gray and multigroup method. However, they argue that the gray method proves sufficient for the 1D case and the multigroup radiative transfer may be more important in the later evolutionary stages of the protostar. The hydrodynamic equations are solved using a shock capturing Riemann solver within a conservative finite volume scheme whereas the FLD equation is solved in an implicit way using the generalized minimal residual solver (GMRES). We use the Harten-Lax-VanLeer approximate Riemann solver that restores with the middle contact discontinuity (hllc), a monotonized central difference (MC) flux limiter using piecewise linear interpolation and a Runge-Kutta 2 (RK2) time integration. 

As an initial density distribution we use a stable Bonnor--Ebert \citep{Bonnor1956, Ebert1955} sphere like density profile. Comparisons to an initially uniform density cloud are described in \cref{sec:density}. 

Given an initial cloud mass $M_0$ and outer radius $R_\mathrm{out}$, the initial sound speed $c_\mathrm{s0}$ is computed as
\begin{align}
\mbox{$c_\mathrm{s0}^2 = \dfrac{G M_0}{\mathrm{ln}(14.1) ~R_\mathrm{out}} $},
\end{align}
where $G$ is the gravitational constant. The initial cloud masses range from 0.5 to 100.0 $\mathrm{M_{\odot}}$. 

The temperature $T_\mathrm{BE}$ for the stable sphere is calculated as
\begin{align}
\mbox{$T_\mathrm{BE} = \dfrac{\mu ~c_\mathrm{s0}^2}{\gamma ~\Re}$},
\end{align}
where $\mu$ = 2.353, $\gamma$ = 5/3 and $\Re$ is the universal gas constant.  

The initial outer $\rho_\mathrm{o}$ and central $\rho_\mathrm{c}$ densities are determined by
\begin{equation}
\begin{split}
&\rho_\mathrm{o} = \Bigg (\dfrac{1.18 ~c_\mathrm{s0}^3}{M_0 ~G^{3/2}} \Bigg)^2 \\
&\rho_\mathrm{c} = 14.1 ~\rho_\mathrm{o}. 
\end{split}
\end{equation}
The density contrast between the center and edge of the sphere corresponds to a dimensionless radius of $\xi$ = 6.45 where $\xi$ is defined as 
\begin{align}
\mbox{$ \xi = \sqrt{\dfrac{4 \pi G \rho_\mathrm{o}}{c_\mathrm{s}^2}} R_\mathrm{cloud} $},
\end{align}
where $R_\mathrm{cloud}$ is the cloud radius. The integrated mass of the cloud is the same as that of a critical Bonnor--Ebert sphere. The thermal pressure is computed using Eq. \eqref{eq:pvteEOS} for a fixed lower temperature $T_\mathrm{0}$ in comparison to the stable Bonnor--Ebert sphere setup, which causes gravity to dominate and initiates the collapse. This temperature $T_\mathrm{0}$ varies from 5 -- 100~K. The radiation temperature is initially in equilibrium with the gas temperature and the dust and gas temperatures are closely coupled.

For the first set of numerical calculations (see Table~\ref{tab:initialparams}), the inner radius is fixed to $10^{-4}$ au and the outer radius is fixed to 3000 au. This setup implies that the central density of the different Bonnor-Ebert spheres scales as a function of initial cloud mass. 

\begin{table}[t]
\centering
\caption{Initial cloud properties}
\begin{tabular}[t]{ccccc}
\hline	
$M_{0} ~\mathrm{[M_{\odot}]}$ & $R_\mathrm{out}$ [au] & $T_{\mathrm{0}}$ [K] & $M_{\mathrm{BE}}/M_{\mathrm{0}}$ & $\rho_\mathrm{c} ~[\mathrm{g ~cm^{-3}}]$  
\TBstrut\\ \hline \hline 
0.5   & 3000   & 10.0       & 1.05e-00    & ~1.16e-17  \Tstrut \\
1.0   & 3000   & 10.0       & 5.27e-01    & 2.33e-17   \\
2.0   & 3000   & 10.0       & 2.64e-01    & 4.66e-17   \\
5.0   & 3000   & 10.0       & 1.05e-01	  & 1.17e-16   \\
8.0	  & 3000   & 10.0       & 6.58e-02    & 1.87e-16   \\
10.0  & 3000   & 10.0       & 5.27e-02    & 2.33e-16   \\
12.0  & 3000   & 10.0       & 4.39e-02	  & 2.80e-16   \\
14.0  & 3000   & 10.0       & 3.76e-02	  & 3.26e-16   \\
15.0  & 3000   & 10.0       & 3.51e-02    & 3.50e-16   \\
16.0  & 3000   & 10.0       & 3.29e-02    & 3.73e-16   \\
18.0  & 3000   & 10.0       & 2.93e-02    & 4.20e-16   \\
20.0  & 3000   & 10.0       & 2.63e-02    & 4.66e-16   \\
30.0  & 3000   & 10.0       & 1.76e-02    & 6.99e-16   \\
40.0  & 3000   & 10.0       & 1.32e-02    & 9.33e-16   \\
60.0  & 3000   & 10.0       & 8.78e-03    & 1.40e-15   \\
80.0  & 3000   & 10.0       & 6.58e-03    & 1.87e-15   \\
100.0 & 3000   & 10.0       & 5.27e-03    & ~2.33e-15  \Bstrut \\ \hline
\end{tabular}
\vspace{0.2cm}
\caption*{Note: Listed above are the cloud properties for runs with different initial cloud mass $M_{0} ~\mathrm{[M_{\odot}]}$, outer radius $R_\mathrm{out}$ [au], temperature $T_{\mathrm{0}}$ [K], stability parameter $M_{\mathrm{BE}}/M_{\mathrm{0}}$ and central density $\rho_\mathrm{c} ~[\mathrm{g ~cm^{-3}}]$. }
\label{tab:initialparams}
\end{table}

The computational grid for these simulations is comprised of 4416 cells. We use 320 uniformly spaced cells from $10^{-4}$ to $10^{-2}$~au and 4096 logarithmically spaced cells from $10^{-2}$ to 3000~au. We make sure that the last uniform cell and the first logarithmic cell are identical in size. The integration time step in the inner dense core regions becomes very small if a logarithmic binning is used throughout, which would become computationally very expensive. Therefore, we choose a linear grid in the very inner part of the computational domain. We performed convergence tests using different resolutions (see \cref{sec:resolution}) and different inner radii $R_\mathrm{in}$ (see \cref{sec:rin}) in order to test our approach. These tests show that the applied resolution is fully sufficient and hence there is no need to use higher resolution in the inner parts. 

\begin{figure*}[!htp]
\centering
\begin{subfigure}{0.25\textwidth}
\includegraphics[width=1.2\textwidth]{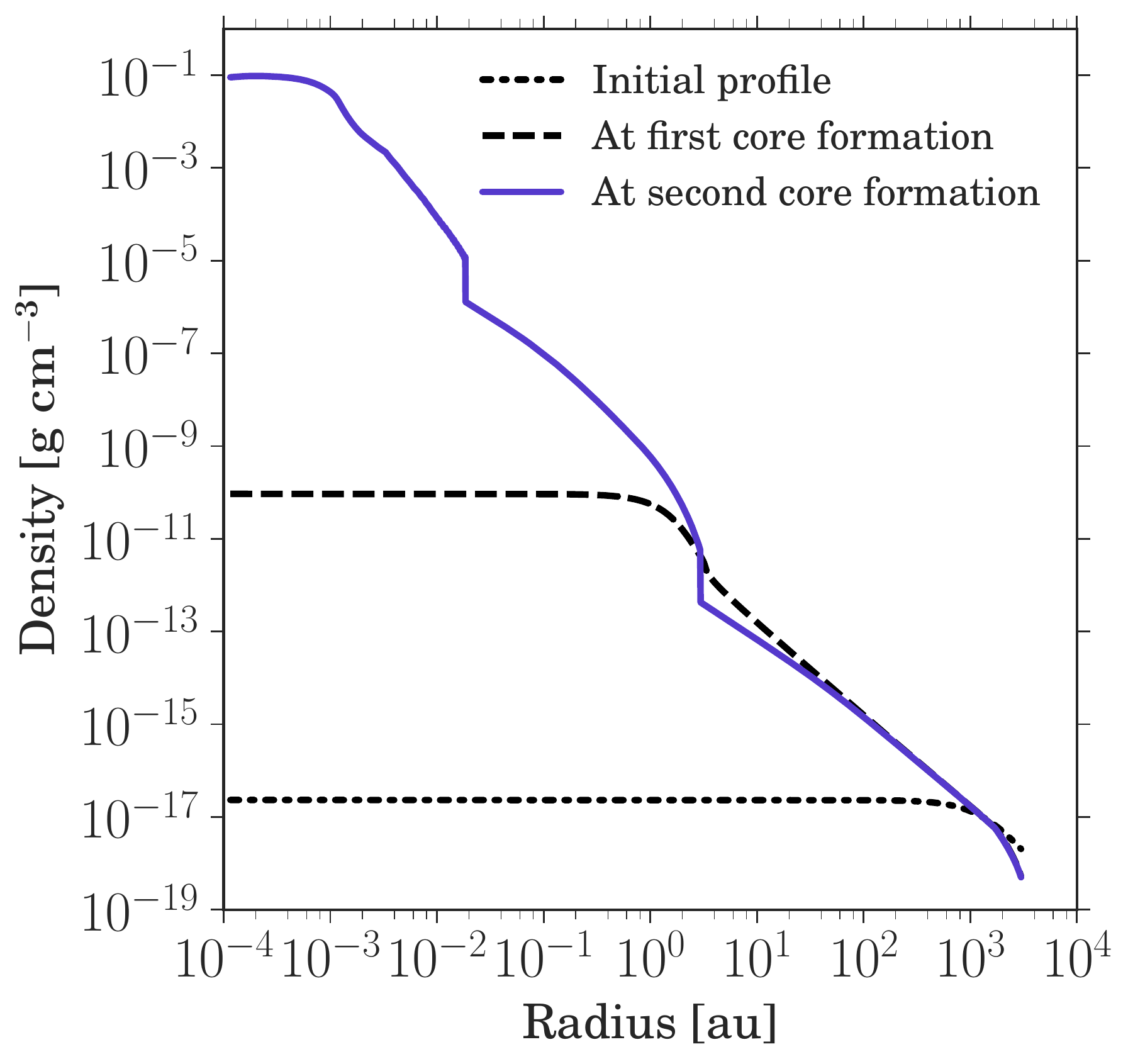}
\end{subfigure}
\hspace{0.5in}
\begin{subfigure}{0.25\textwidth}
\includegraphics[width= 1.2\textwidth]{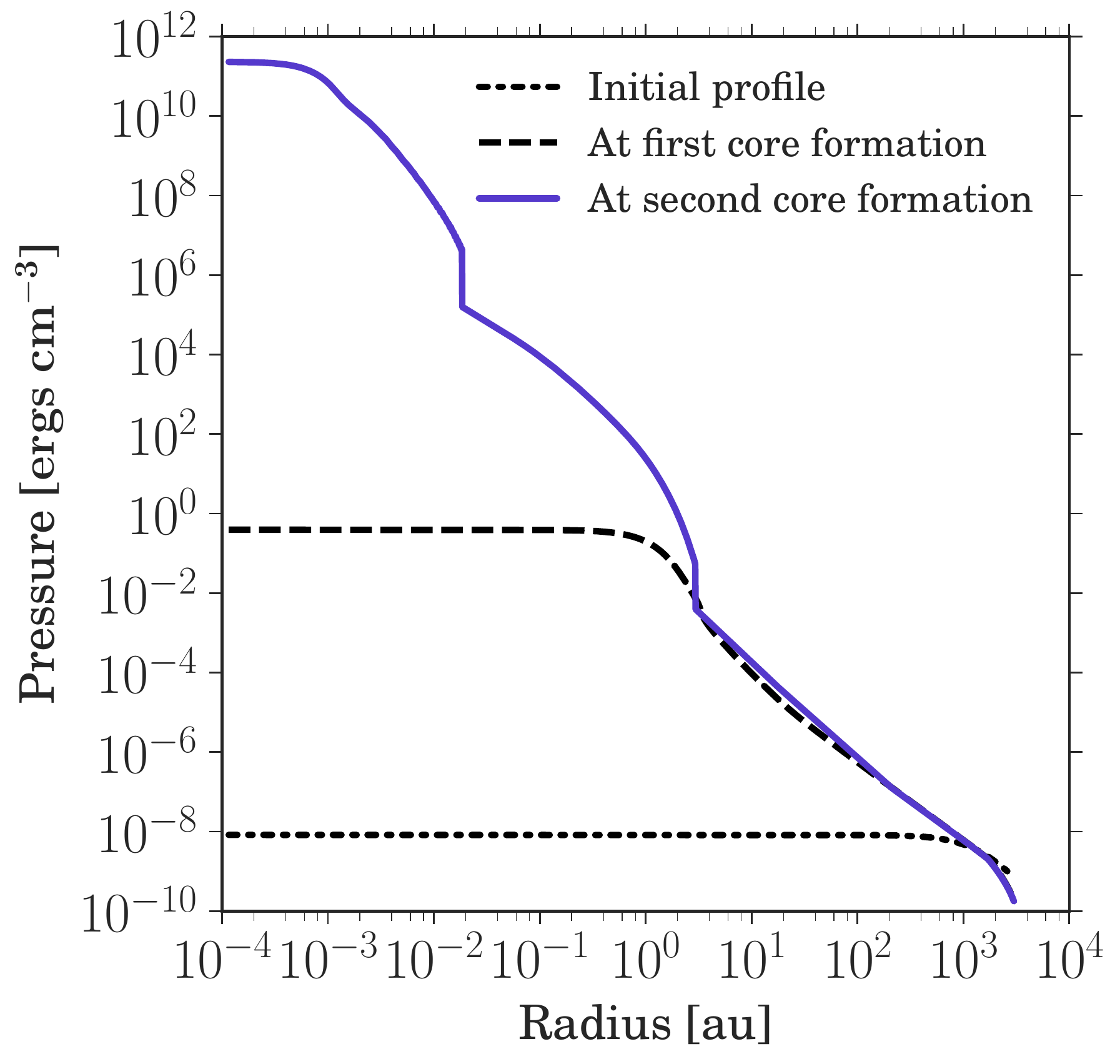}
\end{subfigure}
\hspace{0.5in}
\begin{subfigure}{0.25\textwidth}
\includegraphics[width= 1.2\textwidth]{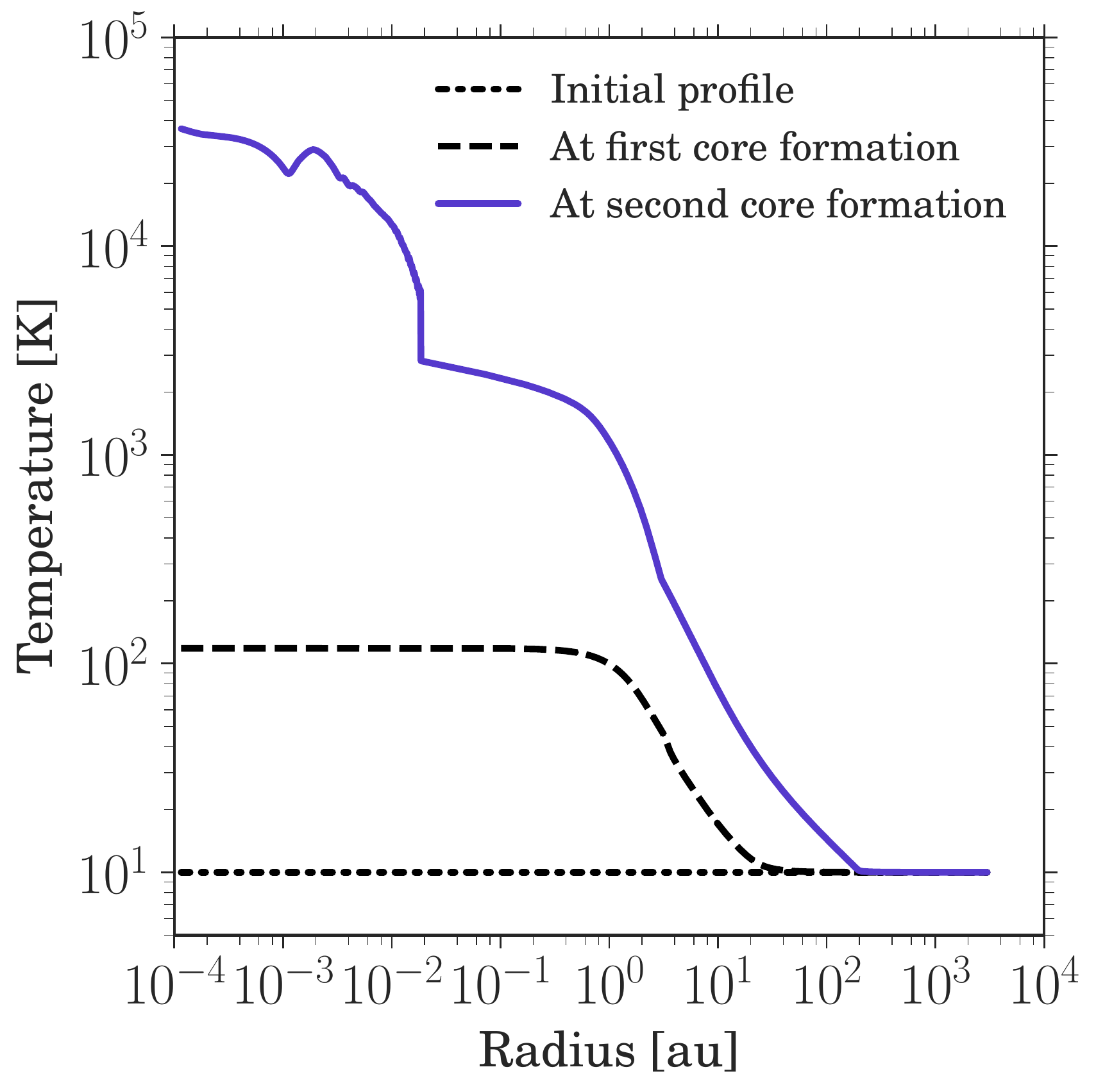}
\end{subfigure}
\begin{subfigure}{0.25\textwidth}
\includegraphics[width= 1.2\textwidth]{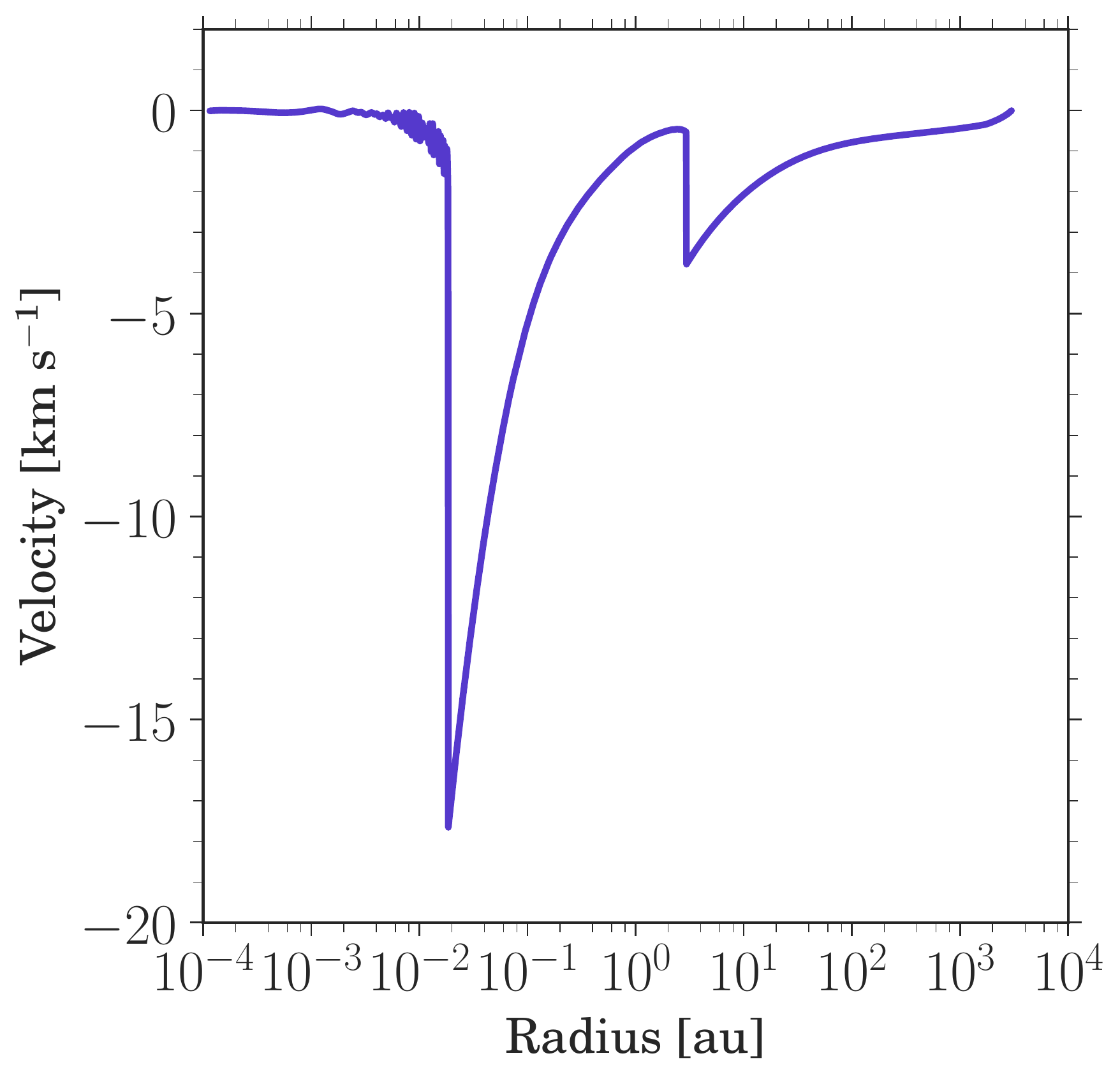}
\end{subfigure}
\hspace{0.5in}
\begin{subfigure}{0.25\textwidth}
\includegraphics[width= 1.2\textwidth]{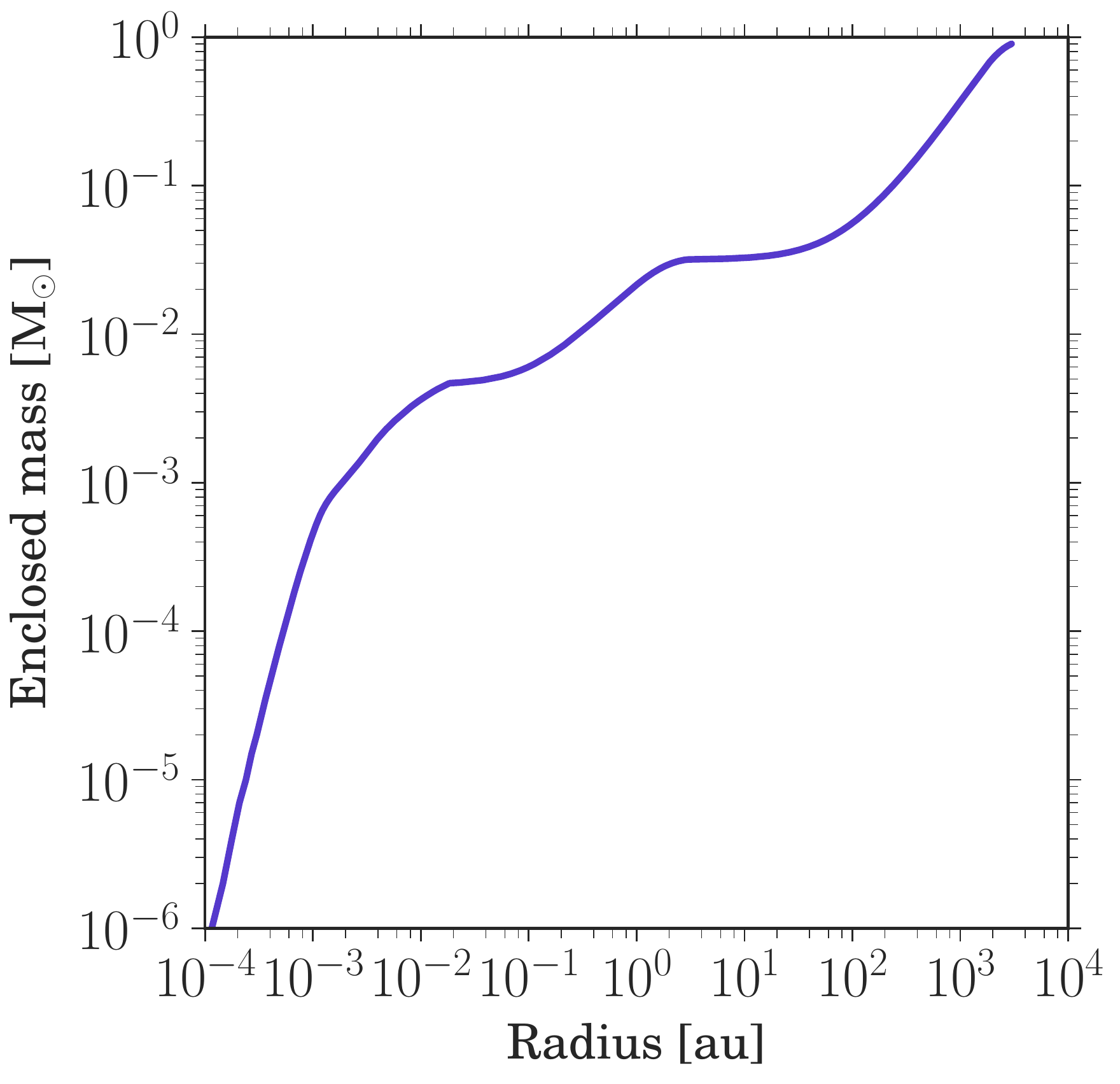}
\end{subfigure}
\hspace{0.5in}
\begin{subfigure}{0.25\textwidth}
\includegraphics[width= 1.2\textwidth]{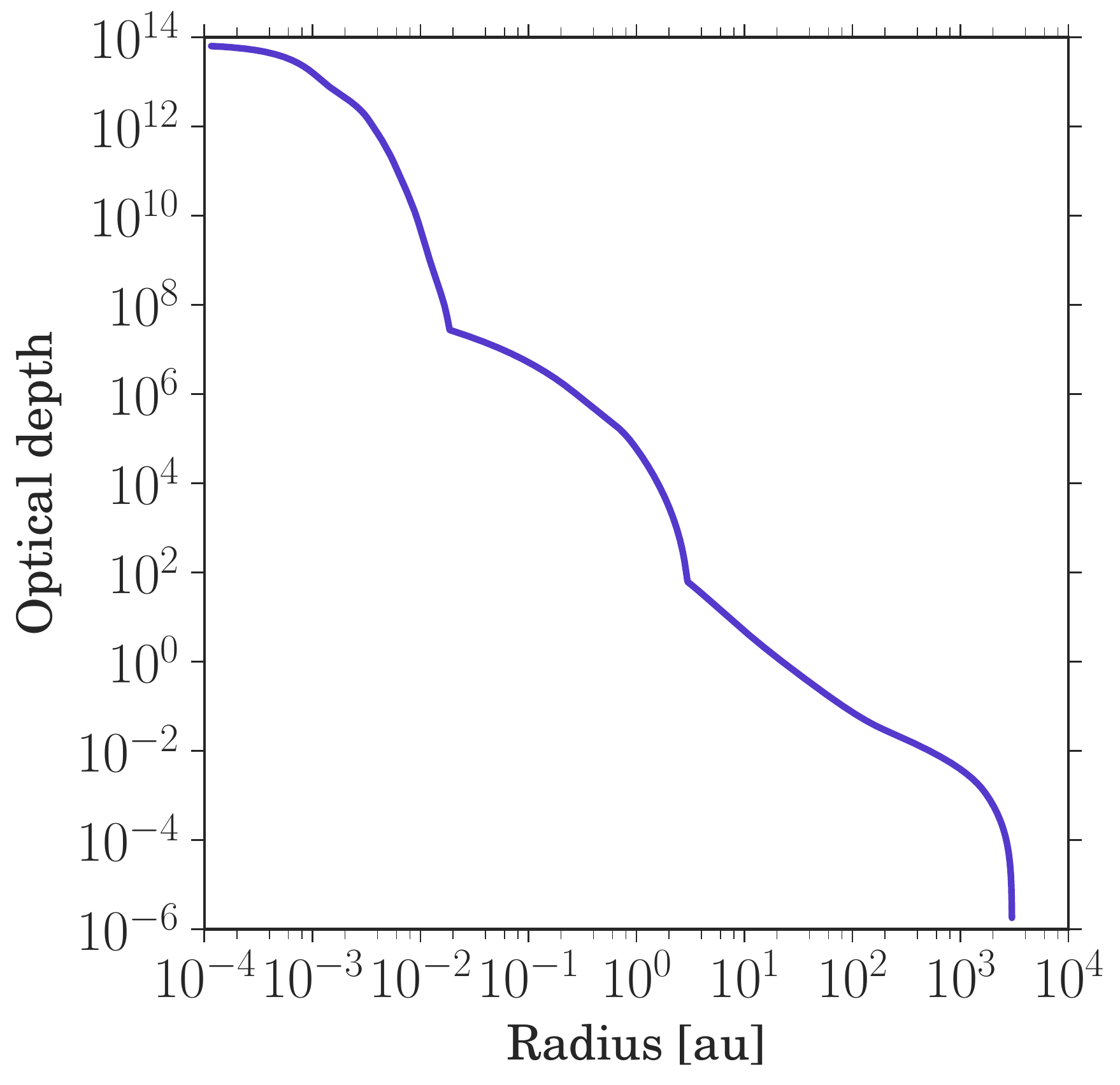}
\end{subfigure}
\begin{subfigure}{0.25\textwidth}
\includegraphics[width= 1.2\textwidth]{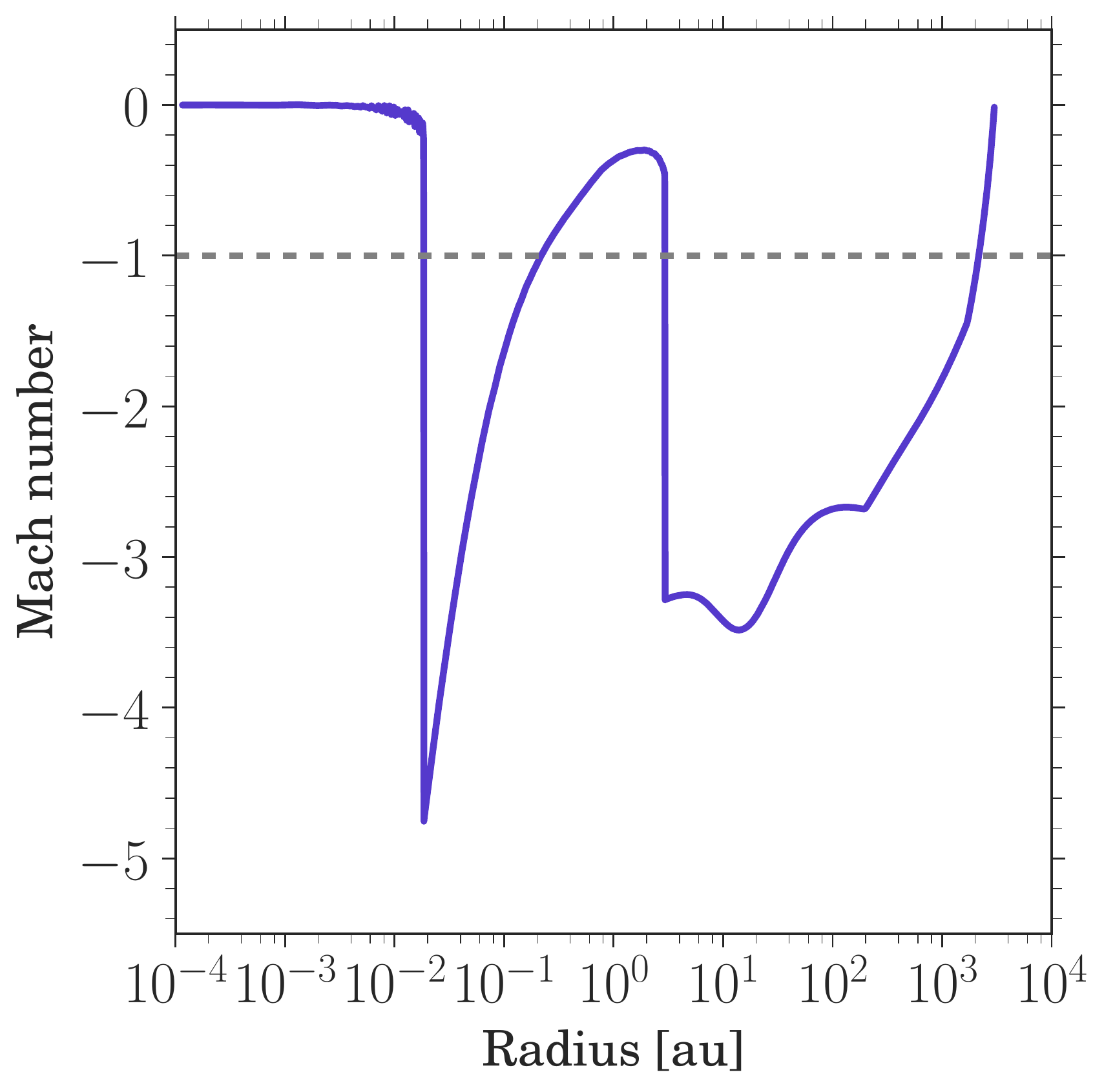}
\end{subfigure}
\hspace{0.5in}
\begin{subfigure}{0.25\textwidth}
\includegraphics[width= 1.2\textwidth]{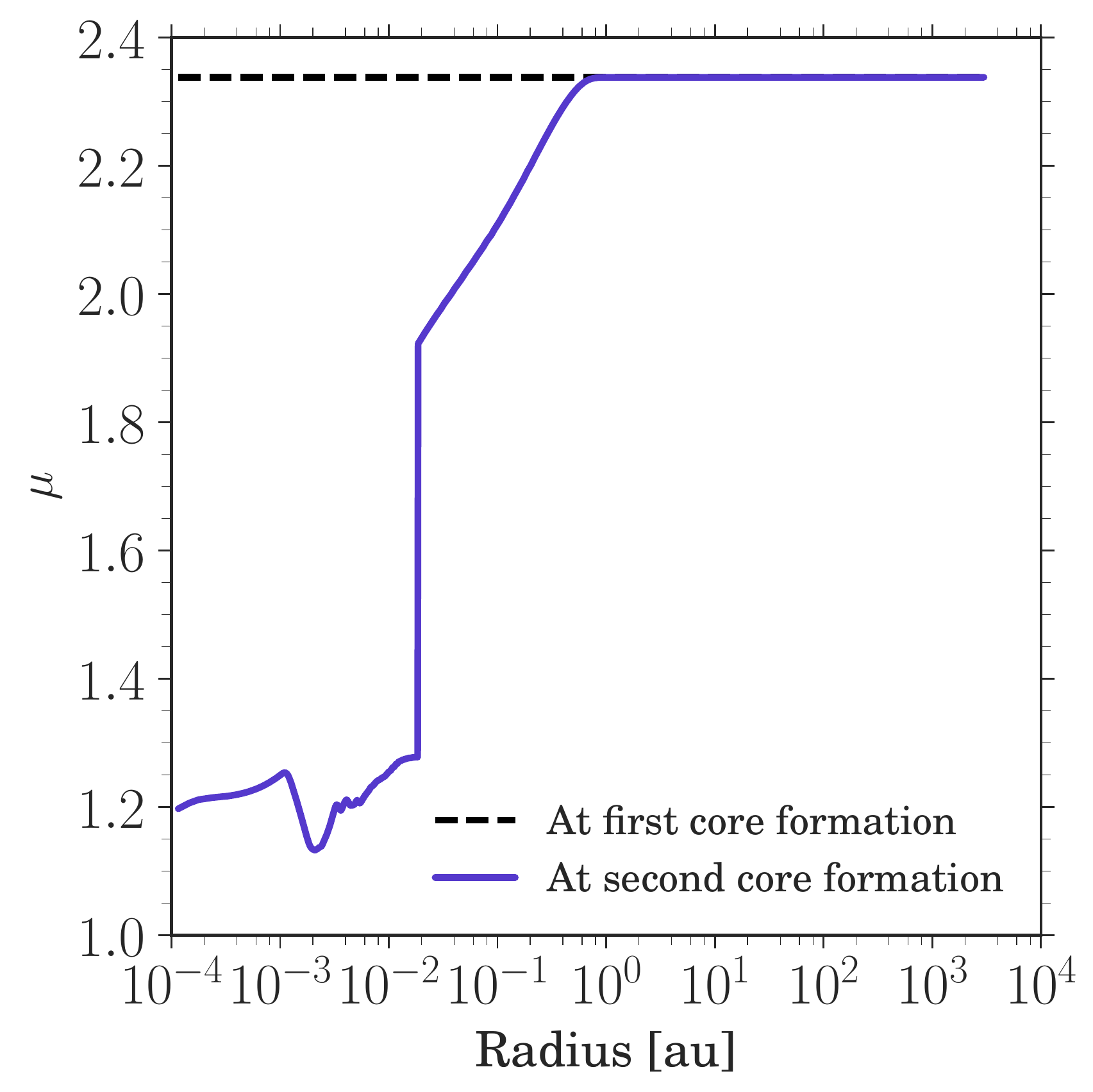}
\end{subfigure}
\hspace{0.5in}
\begin{subfigure}{0.25\textwidth}
\includegraphics[width= 1.2\textwidth]{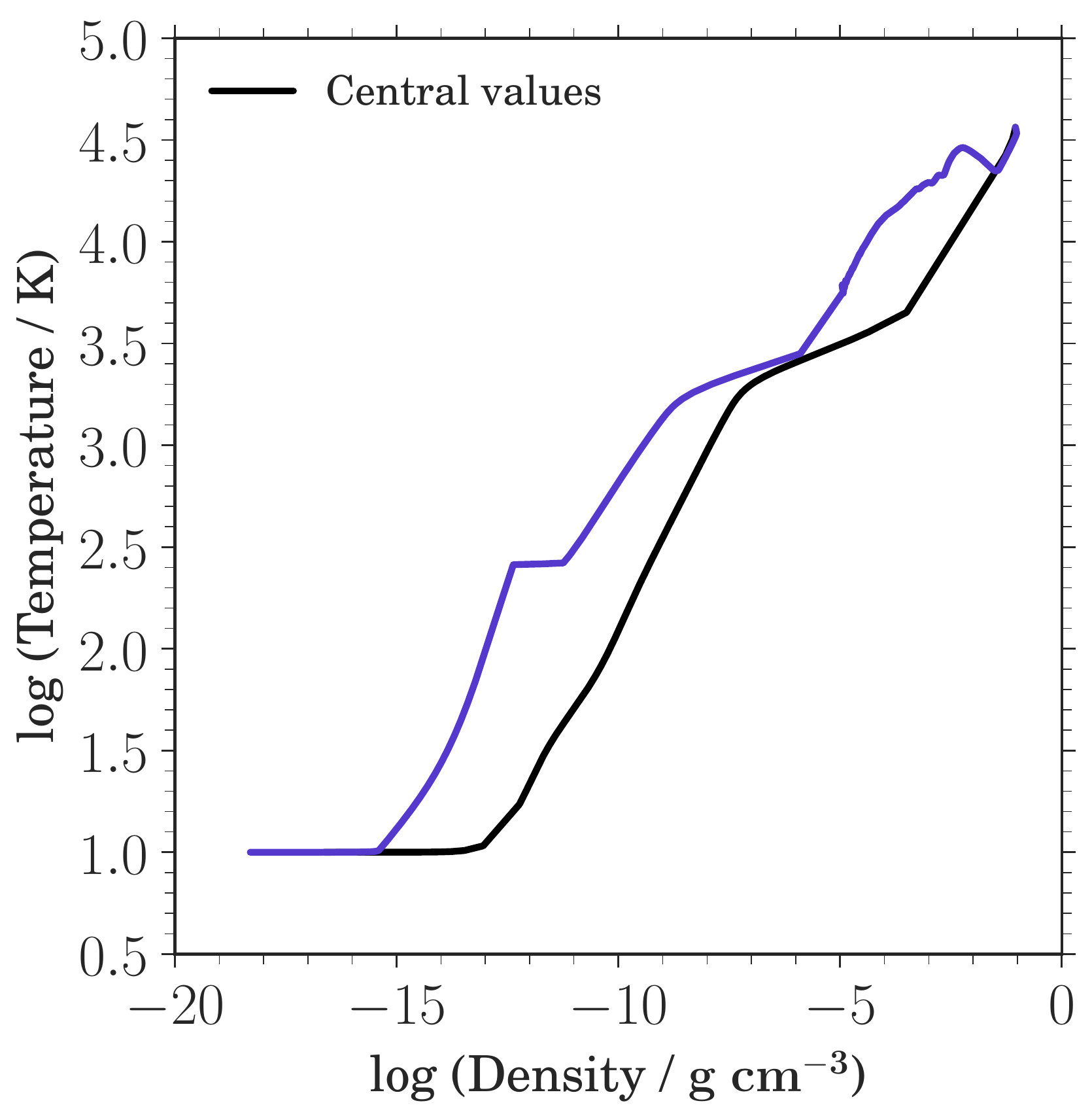}
\end{subfigure}
\caption{Collapse of a 1 $\mathrm{M_{\odot}}$ cloud. Radial profiles (across and down) of \mbox{\bf a)} density, \mbox{\bf b)} pressure, \mbox{\bf c)} gas temperature, \mbox{\bf d)} velocity, \mbox{\bf e)}~enclosed mass, \mbox{\bf f)} optical depth, \mbox{\bf g)} Mach number, \mbox{\bf h)} mean molecular weight, and \mbox{\bf i)} thermal structure are shown at the snapshot after second core formation. The black line in sub-figure \mbox{\bf (i)} shows the temporal evolution of the central temperature and density. The initial profile is shown by the black dot dashed line, the first collapse phase is indicated by the black dashed line and the bluish purple line describes the structure after formation of the second hydrostatic core.}
\label{fig:radialprofiles}
\end{figure*}

Additionally, in order to investigate the dependence on the initial cloud properties, we explore a range of initial conditions by performing three different set of simulations using a different constant stability parameter $M_{\mathrm{BE}}/M_{\mathrm{0}}$ for the low-mass (0.5~to~10~$\mathrm{M_{\odot}}$), intermediate-mass (8 to 20 $\mathrm{M_{\odot}}$)  and high-mass regime (30 to 100 $\mathrm{M_{\odot}}$), respectively. For these runs we fix the outer radius to 3000 au but vary the initial cloud temperature from 5 -- 100 K. The initial cloud properties for the selected parameter space are listed in Table~\ref{tab:newruns} and the implications on the first core properties are described in \cref{sec:initialsetup}. We also perform a set of simulations detailed in Table~\ref{tab:R5000} with an outer radius of 5000 au and a constant initial temperature of 10 K for initial cloud masses ranging from 1~to~100~$\mathrm{M_{\odot}}$.  

At the inner edge $R_\mathrm{in}$ we use a reflective boundary condition for the hydrodynamics as well as for radiation transport (no radiative flux should cross the inner boundary interface). At the outer edge $R_\mathrm{out}$, we use a Dirichlet boundary condition on the radiation temperature with a constant boundary value of $T_\mathrm{0}$ and an outflow-no-inflow condition for the hydrodynamics which includes a zero-gradient (i.e. no-force) boundary condition for the thermal pressure given as 
\begin{align}
\mbox{$\dfrac{dP}{dr} = 0 $}.
\end{align}

\section{Results: From clouds to cores}
\label{sec:results}

\subsection{Fiducial 1 $\mathrm{M_{\odot}}$ case}
\label{sec:solarmass}

\begin{figure*}[htp]
\centering
\includegraphics[width=0.8\linewidth]{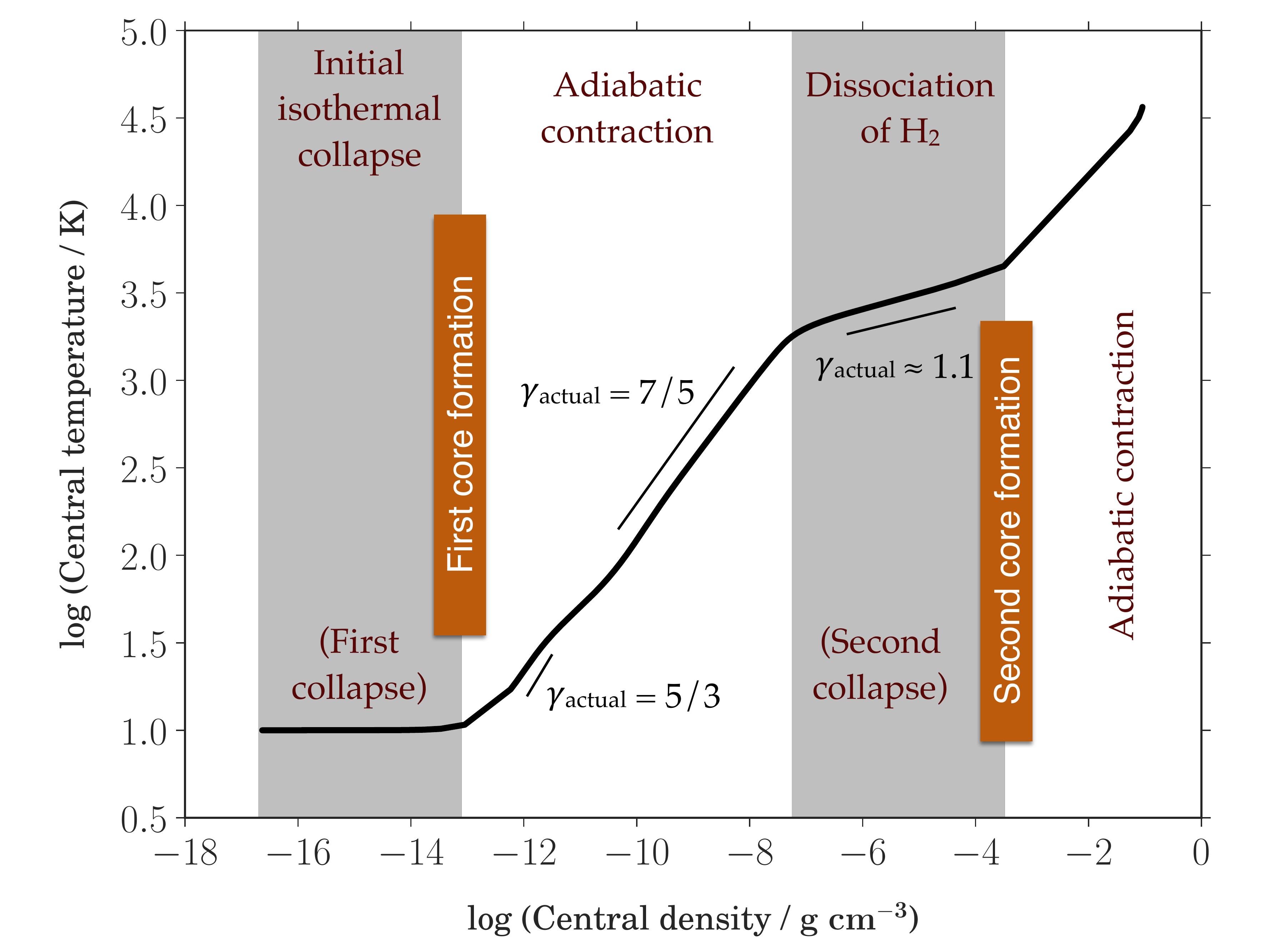}
\caption{Thermal evolution showing the first and second collapse phase for a 1 $\mathrm{M_{\odot}}$ cloud. The change in adiabatic index $\mathrm{\gamma_{actual}}$ indicates the importance of using a realistic gas equation of state.}
\label{fig:thermalevolution}	
\end{figure*}

In this section, we present a general overview of the collapse evolution and its effects on various properties for an initial 1 $\mathrm{M_{\odot}}$ cloud. Figure \ref{fig:radialprofiles} shows the radial profiles of the density, pressure, gas temperature, velocity, enclosed mass, optical depth, Mach number, mean molecular weight $\mu$, and the thermal structure at a time step right after second core formation. We consider an initial Bonnor--Ebert sphere like density profile (as described in \cref{sec:Method}) where the initial central density is $\rho_\mathrm{c} \approx 10^{-17} \mathrm{~g ~cm^{-3}}$. The evolution of the cloud through its first and second collapse phase can be understood as follows:

\begin{itemize}
\item Initially the optically thin cloud collapses isothermally with $\mathrm{\gamma_{actual}} \approx$ 1 under its own gravity, where $\mathrm{\gamma_{actual}}$ is the change in the slope of the temperature evolution with density (see Fig.~\ref{fig:thermalevolution}). 
\item During the first collapse phase, as the density and pressure increases, the optical depth becomes greater than unity \citep{Masunaga1999} and the cloud compresses adiabatically. The cloud starts absorbing the thermal radiation and heats up leading to an adiabatic collapse phase. 
\item These conditions lead to the formation of the first hydrostatic core after about $10^4$ years with initial values of $R_{\mathrm{fc}} \approx$ 2 au, $M_{\mathrm{fc}} \approx$ $10^{-3}$ $\mathrm{M_{\odot}}$ which subsequently contracts adiabatically with $\mathrm{\gamma_{actual}} \approx$ 5/3. 
\item The strong compression leads to the first shock at the border of the first core as seen in the velocity profile (Fig. \ref{fig:radialprofiles}d). Comparing this to the temperature profile (Fig. \ref{fig:radialprofiles}c), the first shock is supercritical, i.e. pre- and post-shock temperatures are similar as discussed in \citet{Commercon2011}.
\item The first core mainly consists of $\mathrm{H_2}$ molecules and neutral He, with a constant mean molecular weight $\mu$ of 2.353. 
\item With a rise in temperature, the adiabatic index $\gamma$ changes from its initial monatomic value $\gamma=5/3$ to the value for a diatomic gas, $\gamma=7/5$, once the gas is warm enough to excite the rotational degrees of freedom. As Fig.~\ref{fig:thermalevolution} indicates, $\mathrm{\gamma_{actual}}$ undergoes the same evolution.
\item Once the temperature inside the first core reaches $\sim$ 2000 K, $\mathrm{H_2}$ molecules begin to dissociate which leads to the second collapse phase. During this phase, $\mathrm{\gamma_{actual}}$ changes roughly to 1.1, which is well below the critical value of 4/3 for stability of a self-gravitational sphere.
\item As the molecular hydrogen and neutral helium concentration changes and the fraction of atomic hydrogen increases during the dissociation phase, $\mu$ gradually decreases in the inner regions, as seen in Fig. \ref{fig:radialprofiles}h.
\item Once most of the $\mathrm{H_2}$ has been dissociated, it is followed by the formation of the second hydrostatic core with initial values of $R_{\mathrm{sc}} \approx$ 1.8 $\times$ $10^{-2}$ au $\approx$ 3.9 $\mathrm{R_{\odot}}$, $M_{\mathrm{sc}} \approx$ 4.6 $\times$ $10^{-3}$ $\mathrm{M_{\odot}}$.
\item The second shock at the border of the second core is seen in the velocity profile (Fig. \ref{fig:radialprofiles}d). Comparing this to the temperature profile, the second shock is seen to be subcritical with the pre-shock temperature being higher than the post-shock temperature, suggesting that the accretion energy is transferred onto the second core and not radiated away. 
\item The central density rapidly rises up to $\rho_\mathrm{c} \approx 10^{-1} \mathrm{~g ~cm^{-3}}$ at the end of the second collapse phase which lasts only for a few years since the second hydrostatic core forms almost instantaneously.
\item At later times when $\rho_\mathrm{c} \approx 10^{-1} \mathrm{~g ~cm^{-3}}$, the outer layers tend to have higher temperatures due to the effects of shock heating and absorption of radiation from the hot central region. Differences in the thermal evolution of the central region (black line) and the thermal structure at a time when $\rho_\mathrm{c} \approx 10^{-1} \mathrm{~g ~cm^{-3}}$ (bluish purple line) can be seen in Fig.~\ref{fig:radialprofiles}i.
\item Finally (not simulated here), once the temperature inside the second core reaches ignition temperatures ($\geq 10^6$ K) for nuclear reactions, it eventually leads to the birth of a star. 
\end{itemize}
Figure~\ref{fig:thermalevolution} summarizes the different evolutionary stages that the molecular cloud undergoes to form the first and second Larson's cores and indicates the phase transition from monatomic to diatomic gas i.e. the change in the adiabatic index $\mathrm{\gamma_{actual}}$.

\begin{figure*}[!]
\centering
\begin{subfigure}{0.243\textwidth}
\includegraphics[width= 1.2\textwidth]{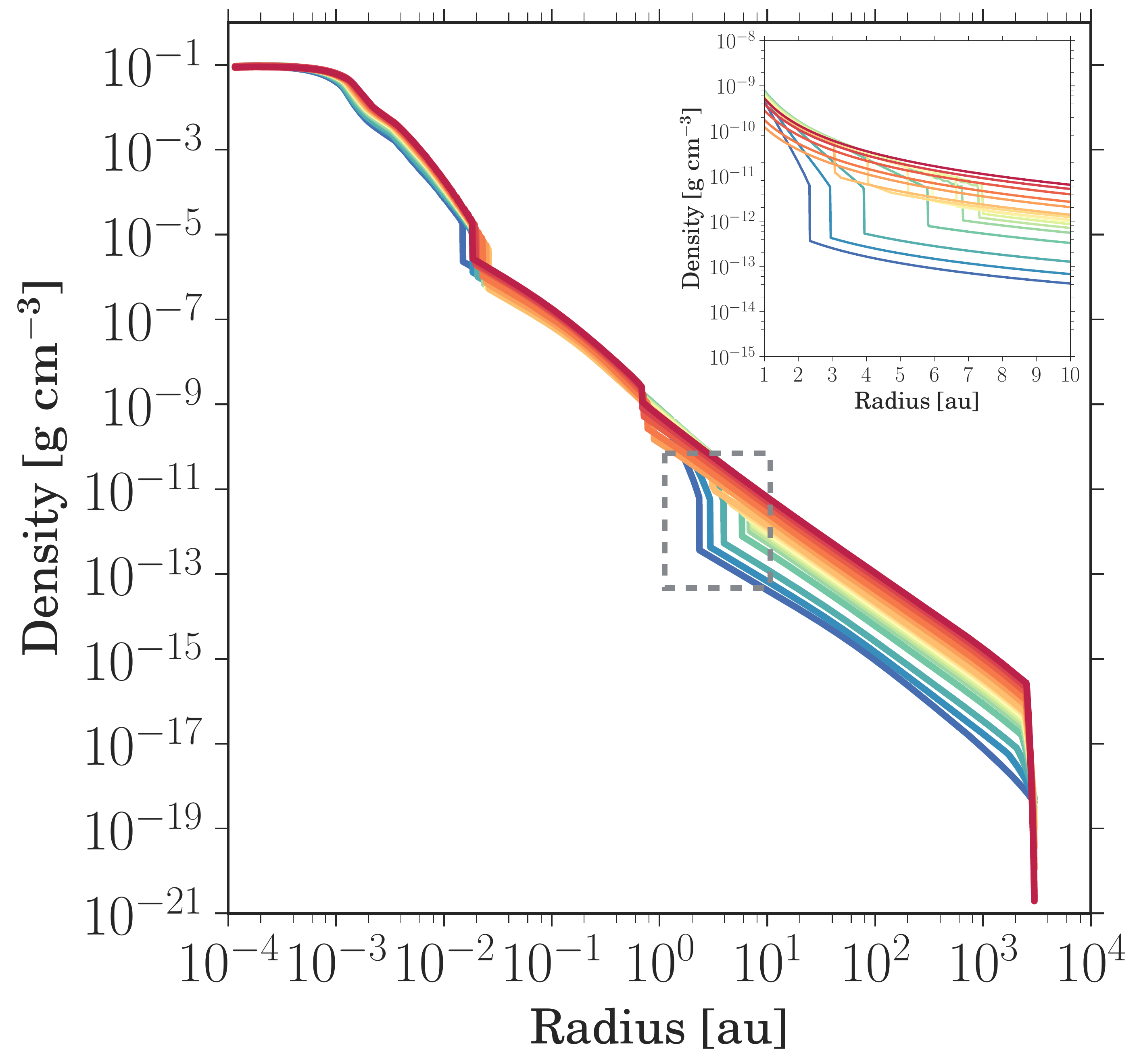}
\end{subfigure}
\hspace{0.5in}
\begin{subfigure}{0.243\textwidth}
\includegraphics[width= 1.2\textwidth]{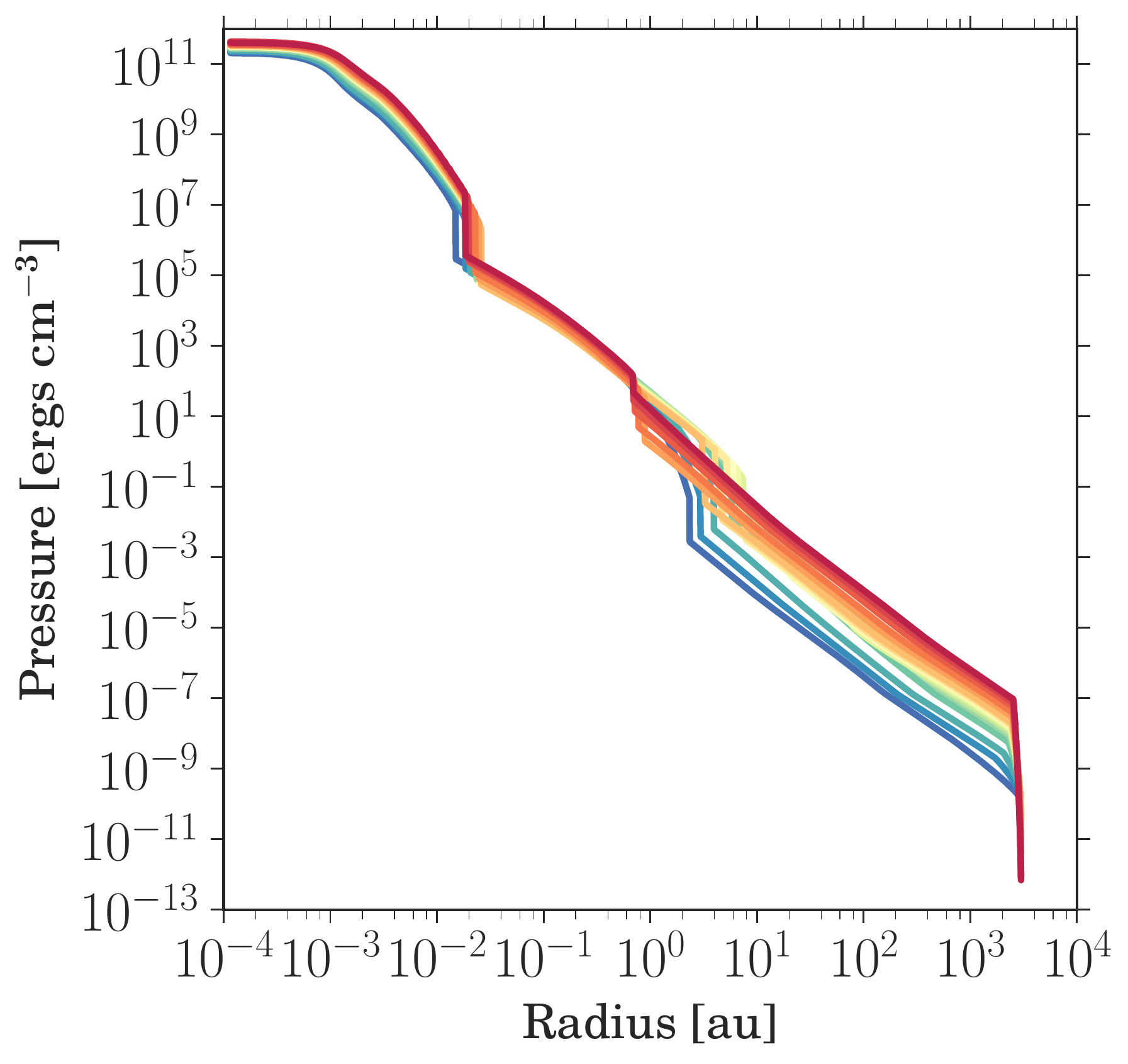}
\end{subfigure}
\hspace{0.5in}
\begin{subfigure}{0.243\textwidth}
\includegraphics[width= 1.2\textwidth]{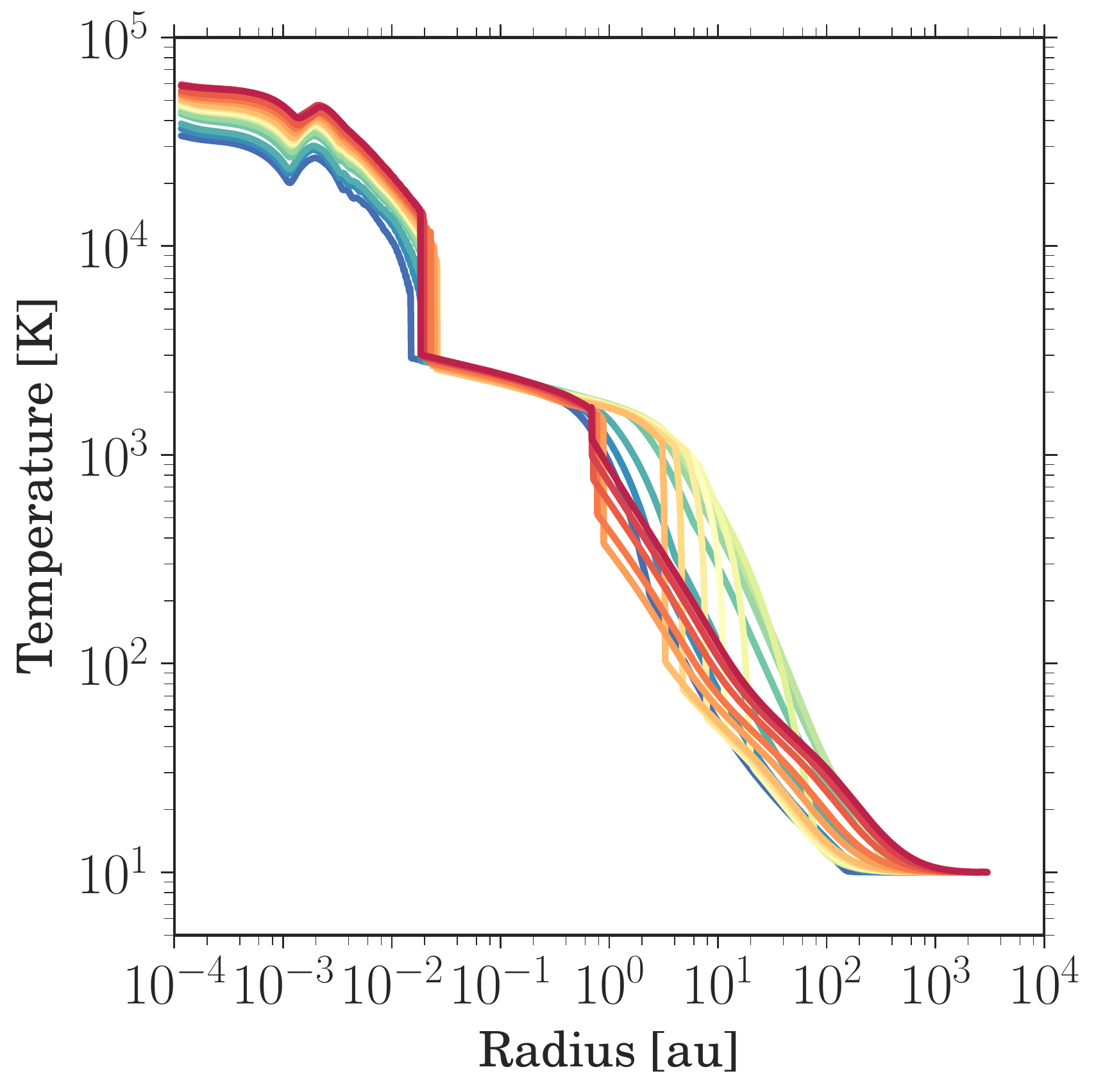}
\end{subfigure}
\begin{subfigure}{0.243\textwidth}
\includegraphics[width= 1.2\textwidth]{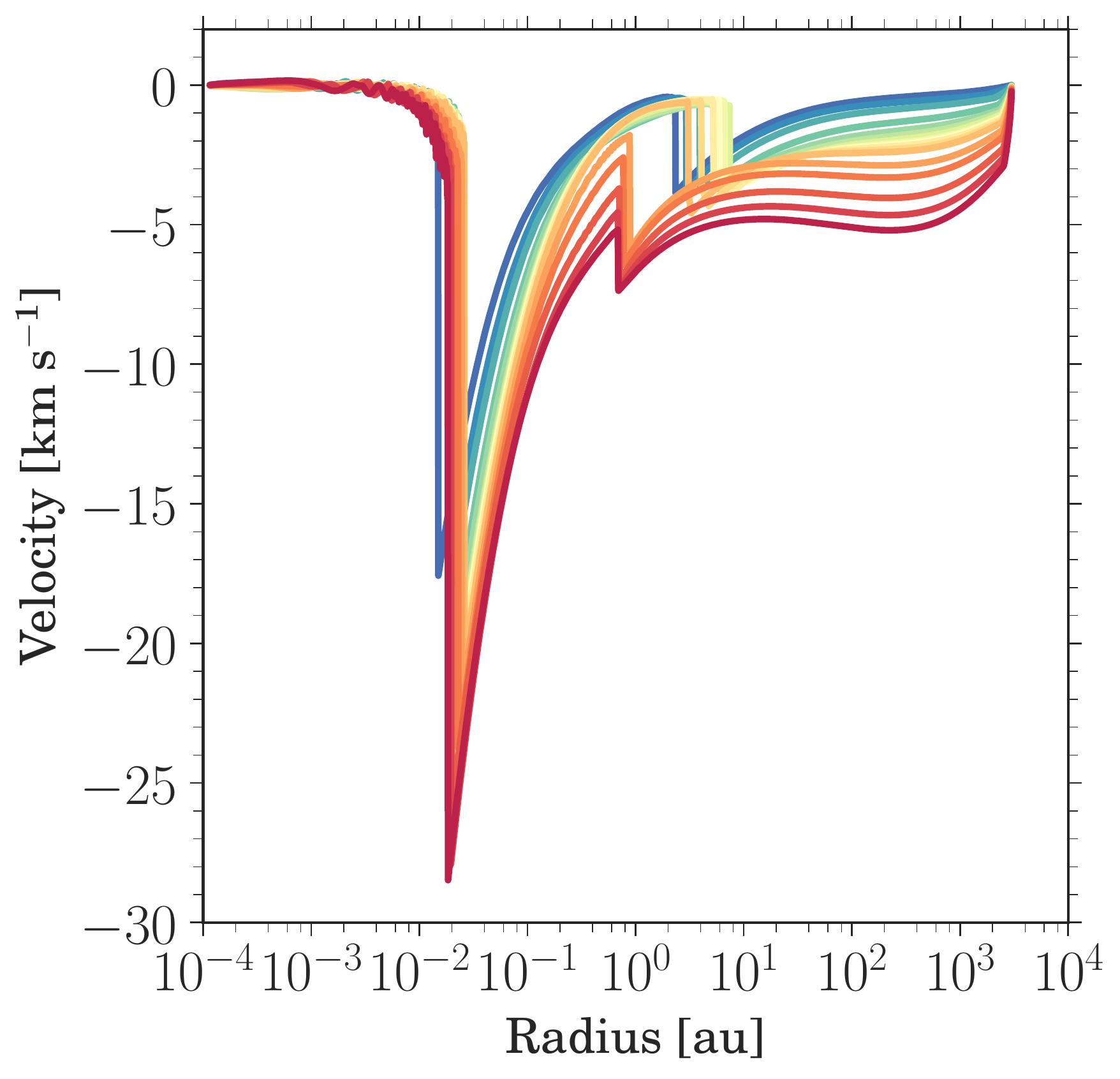}
\end{subfigure}
\hspace{0.5in}
\begin{subfigure}{0.243\textwidth}
\includegraphics[width= 1.2\textwidth]{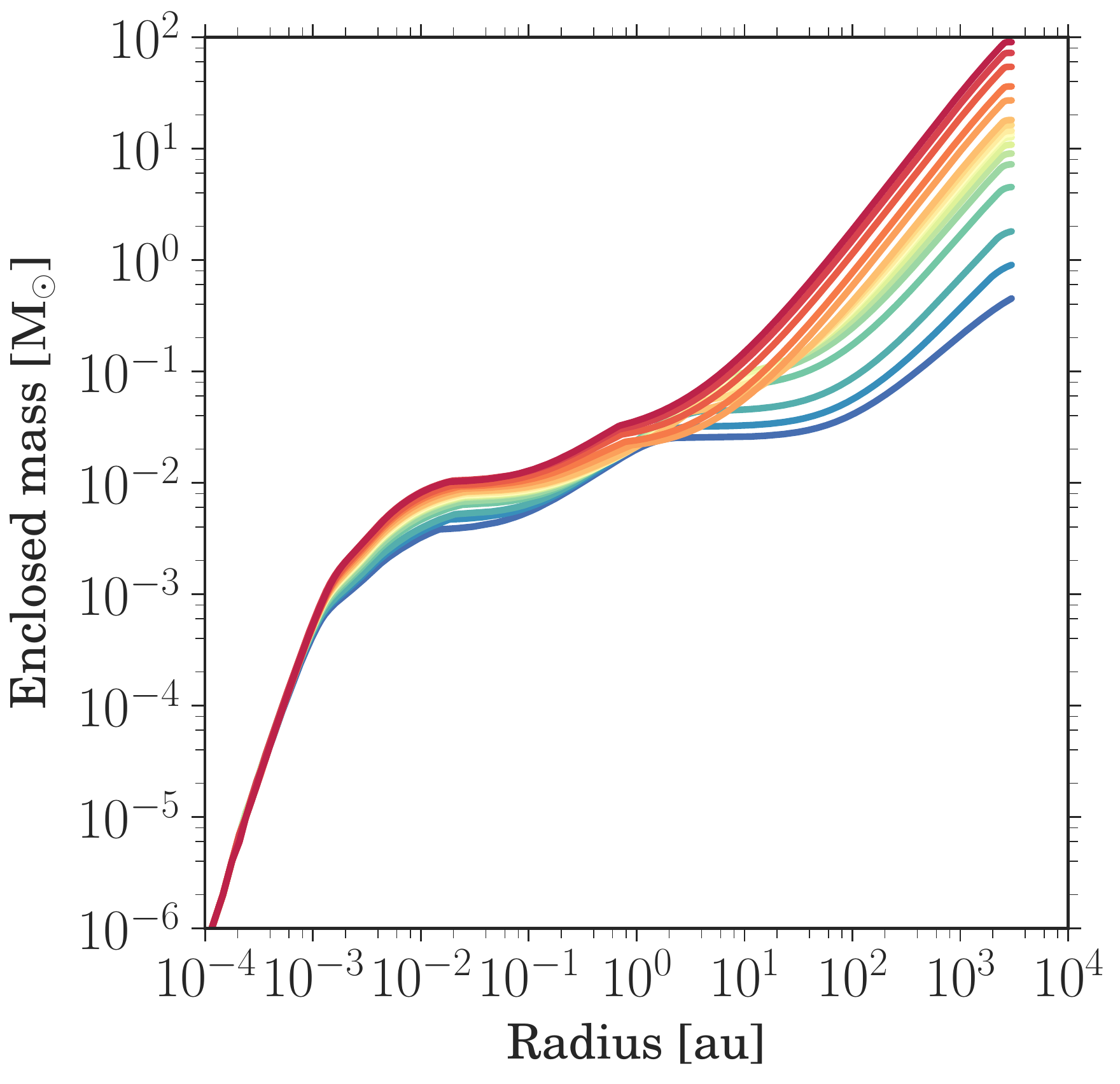}
\end{subfigure}
\hspace{0.5in}
\begin{subfigure}{0.243\textwidth}
\includegraphics[width= 1.2\textwidth]{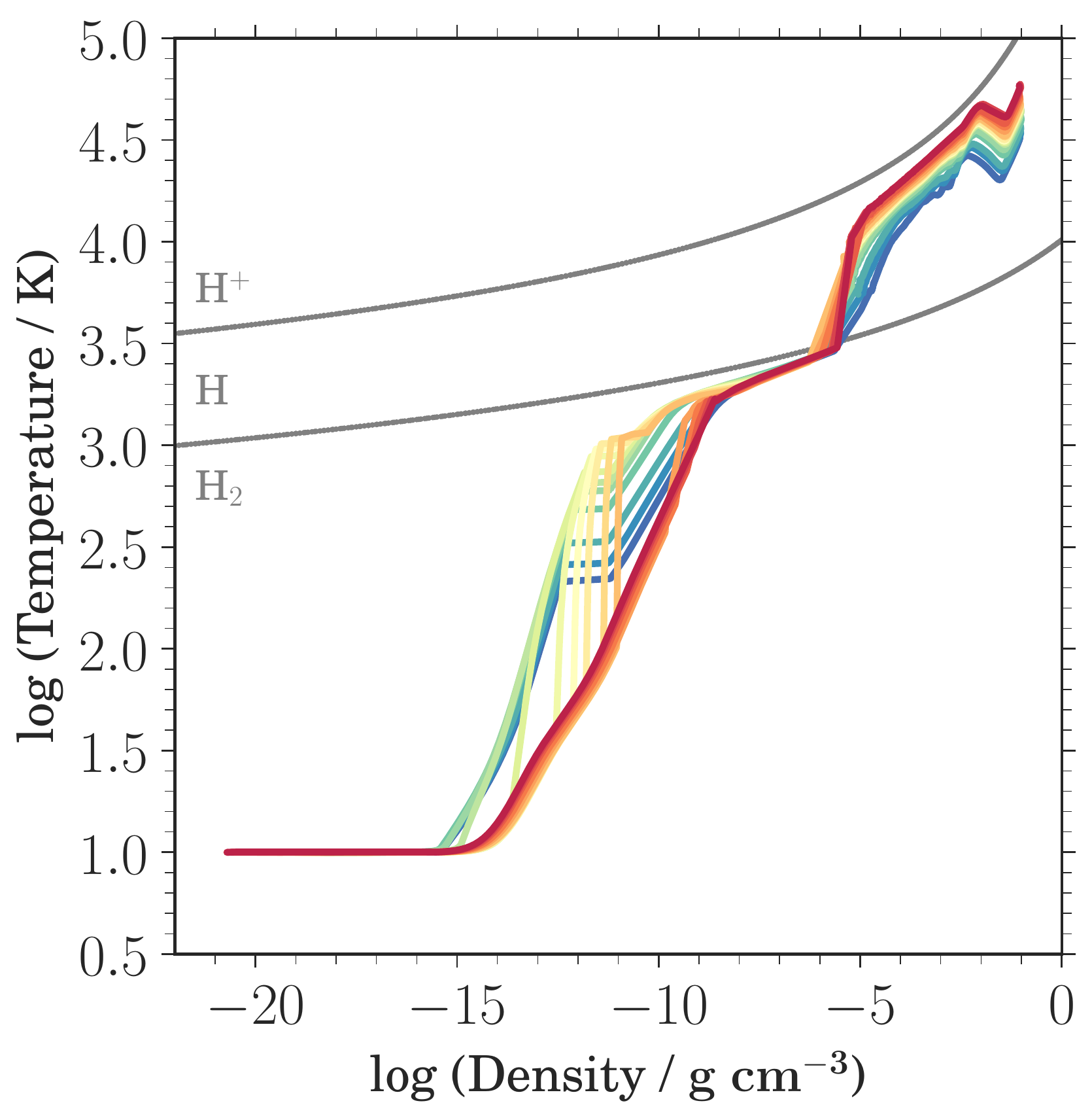}
\end{subfigure}
\begin{subfigure}{0.243\textwidth}
\includegraphics[width= 1.2\textwidth]{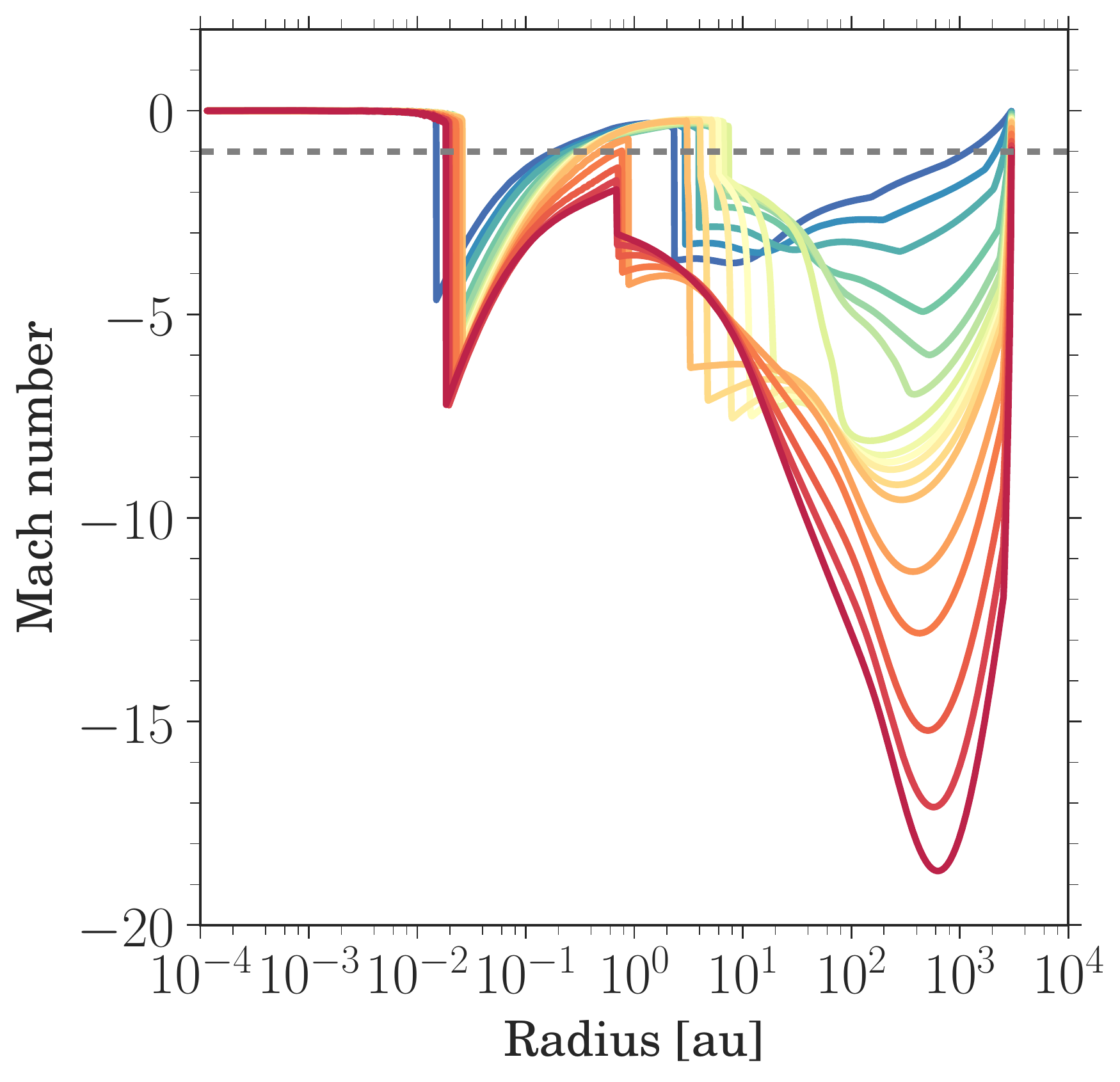}
\end{subfigure}
\hspace{0.5in}
\begin{subfigure}{0.243\textwidth}
\includegraphics[width= 1.2\textwidth]{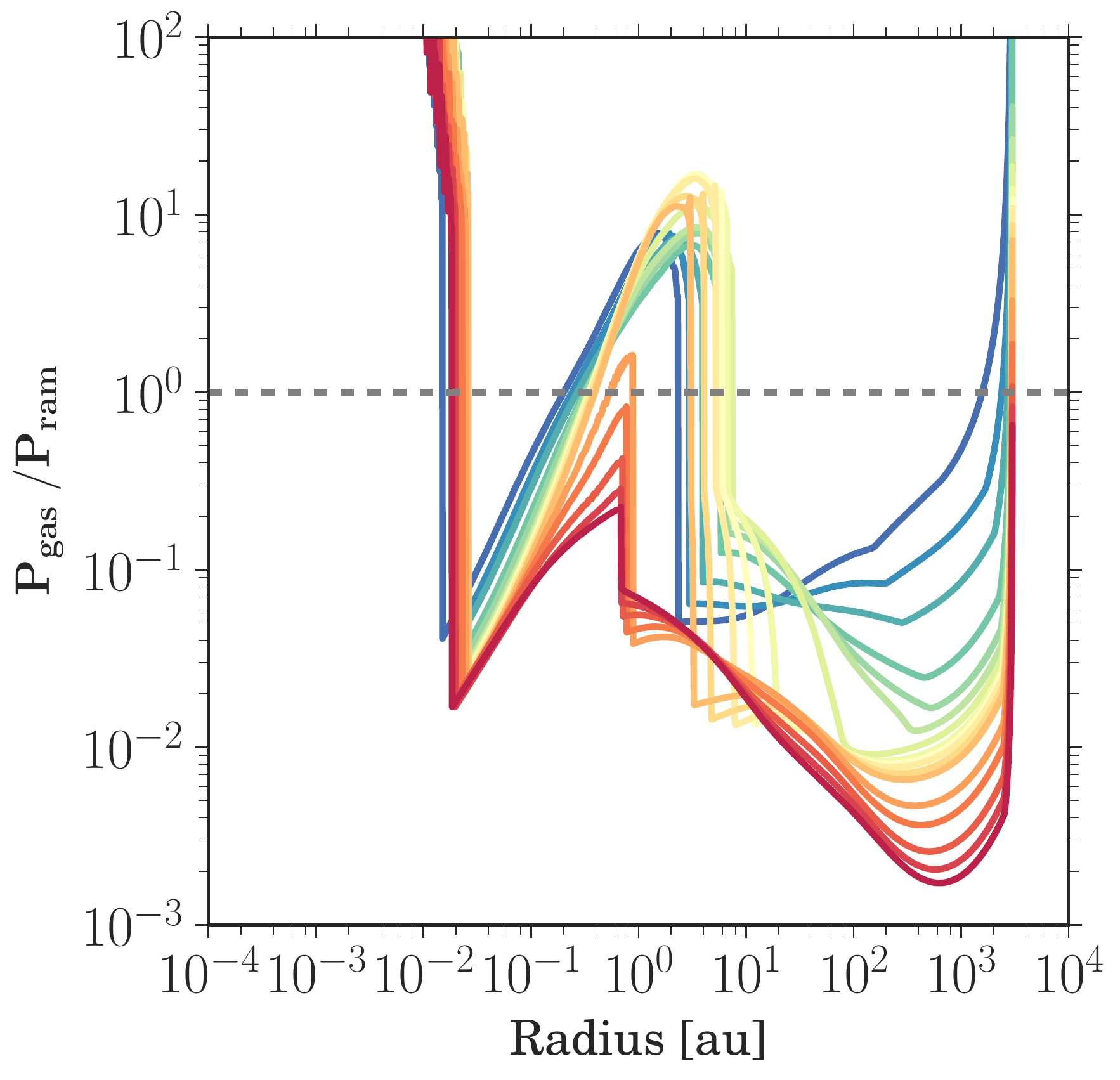}
\end{subfigure}
\hspace{0.5in}
\begin{subfigure}{0.243\textwidth}
\includegraphics[width= 1.2\textwidth]{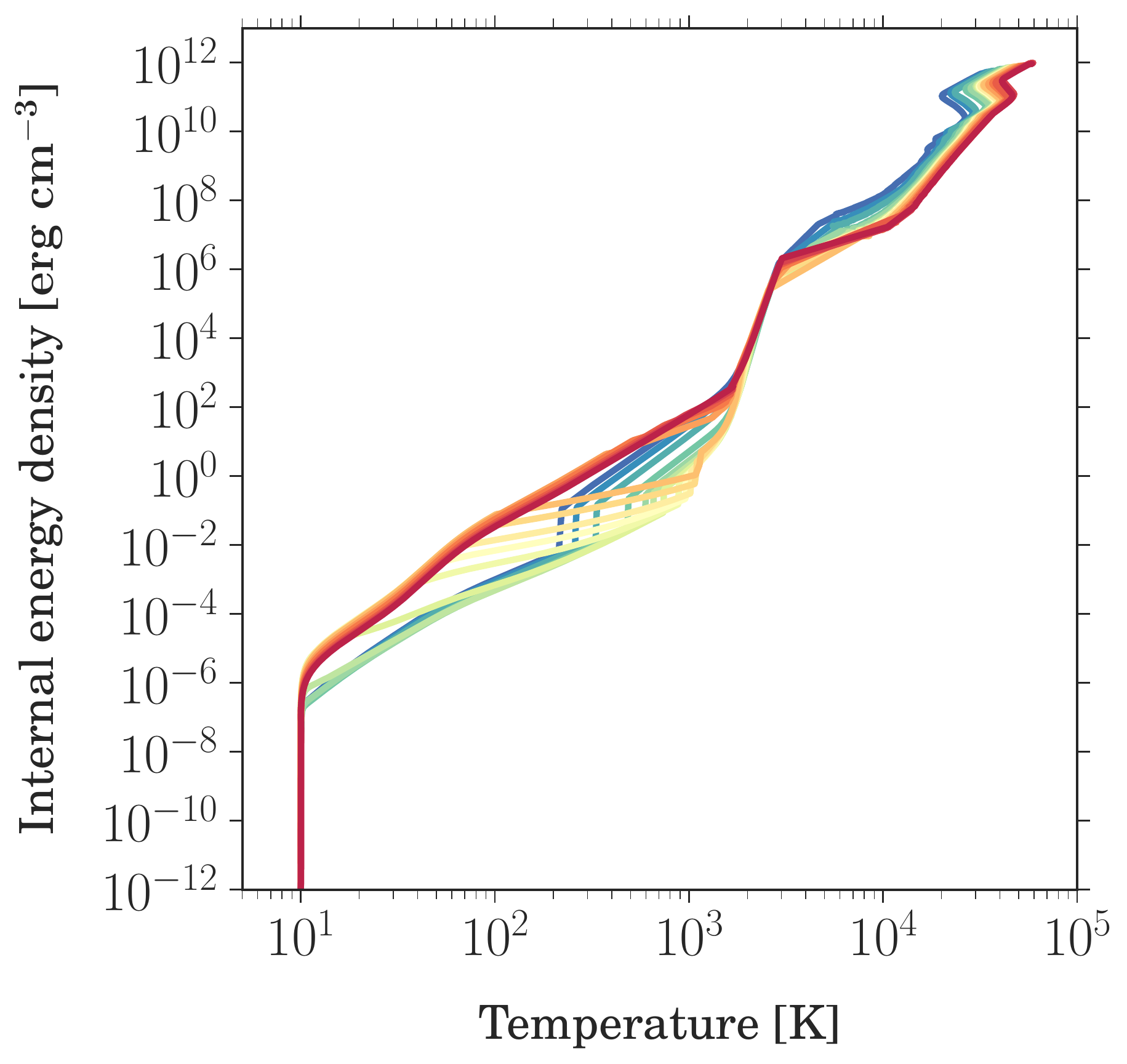}
\end{subfigure}
\begin{subfigure}{0.243\textwidth}
\includegraphics[width= 1.2\textwidth]{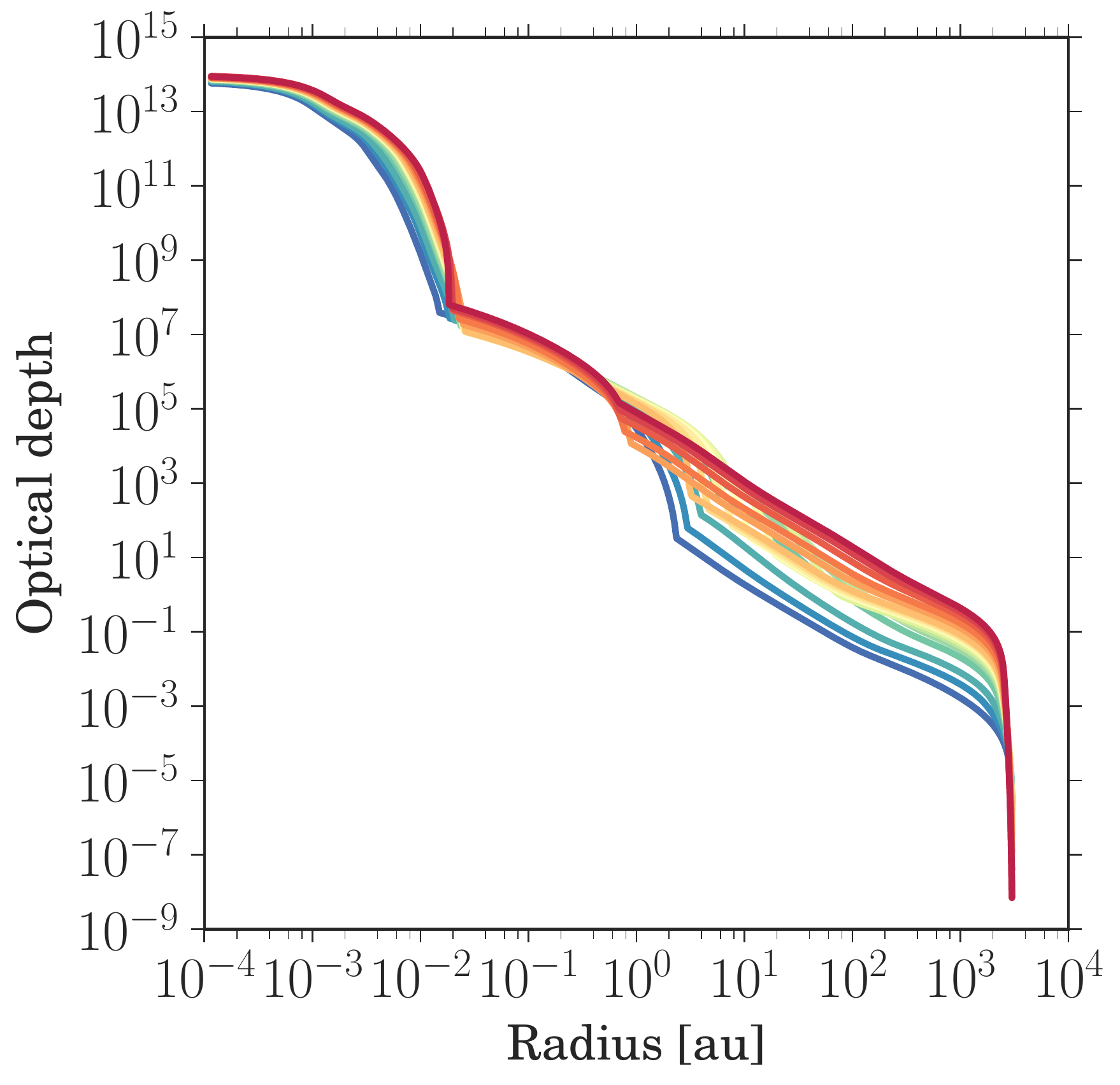}
\end{subfigure}
\hspace{0.5in}
\begin{subfigure}{0.243\textwidth}
\includegraphics[width= 1.2\textwidth]{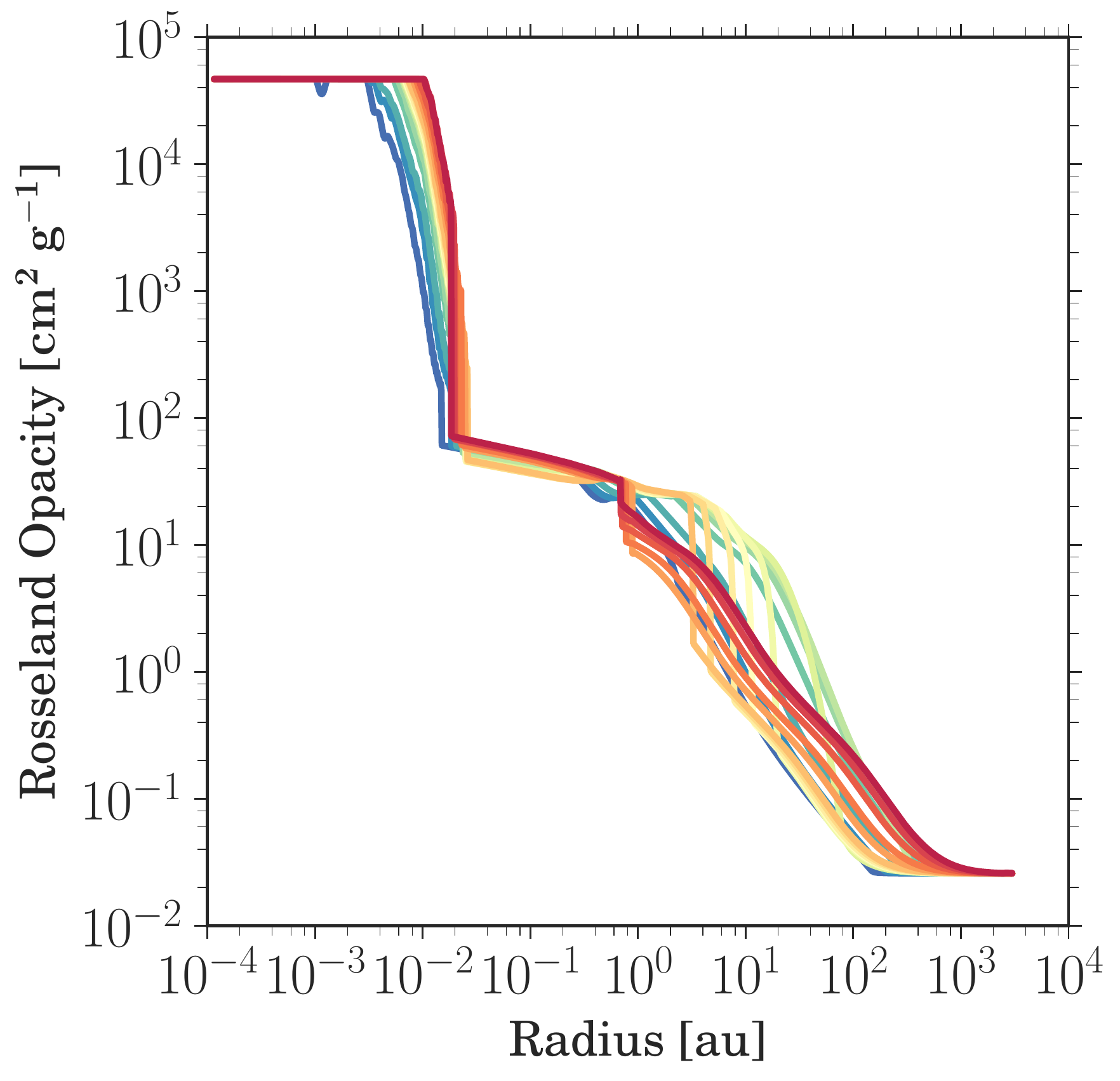}
\end{subfigure}
\hspace{0.5in}
\begin{subfigure}{0.243\textwidth}
\includegraphics[width= 1.2\textwidth]{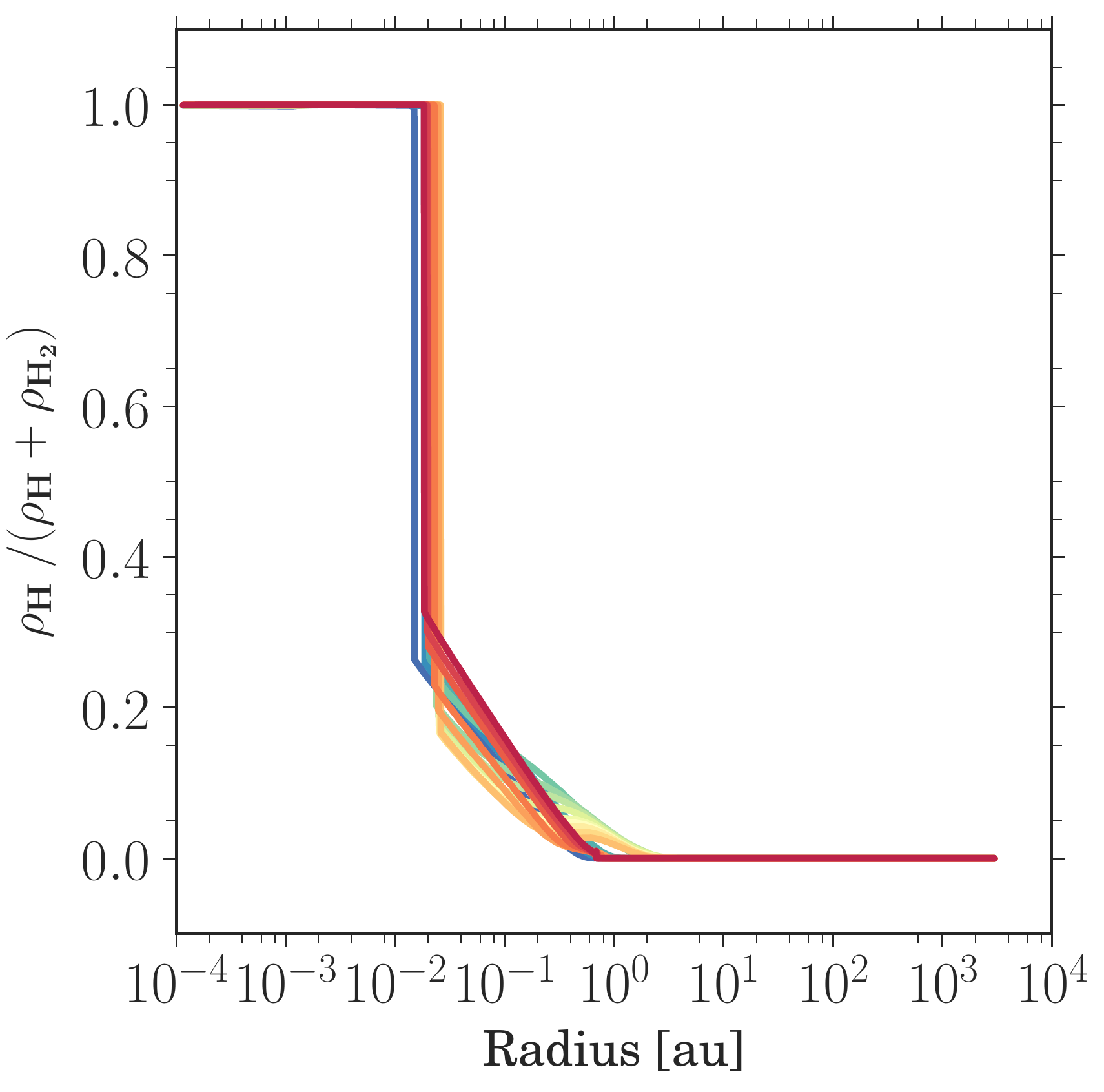}
\end{subfigure}
\hspace{0.5in}
\begin{subfigure}{0.7\textwidth}
\includegraphics[width=1.1\textwidth]{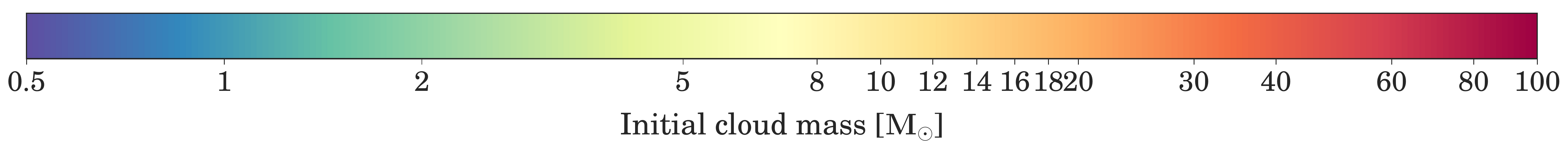}
\end{subfigure}
\caption{Shown above are the radial profiles (across and down) of \mbox{\bf a)} density, \mbox{\bf b)} pressure, \mbox{\bf c)} gas temperature, \mbox{\bf d)} velocity, and \mbox{\bf e)} enclosed mass as well as the \mbox{\bf f)} thermal structure, \mbox{\bf g)} Mach number, \mbox{\bf h)} ratio of gas to ram pressure, \mbox{\bf i)} internal energy density as a function of temperature, \mbox{\bf j)}~optical depth, \mbox{\bf k)} Rosseland mean opacity and \mbox{\bf l)} dissociation fraction at the snapshot after second core formation. Different colors indicate clouds with different initial masses as seen in the color bar. The gray lines in the thermal structure plot show the 50\,\%\ dissociation and ionization curves according to Eq. \eqref{eq:H} and Eq. \eqref{eq:H2}. }
\label{fig:masscomparison}
\end{figure*}

\begin{table*}[!htp]
\centering
\caption{Properties of the first and second cores estimated at the snapshot after second core formation for different initial cloud masses $M_{0}$ with a fixed outer radius $R_\mathrm{out} = 3000$ au and initial temperature $T_{\mathrm{0}}$ = 10 K. }
\resizebox{\textwidth}{!}{
\begin{tabular}[t]{ccccccccccc} 
\hline	 
$M_{0} ~\mathrm{[M_{\odot}]}$ & $R_{\mathrm{fc}}$ [au]  & $M_{\mathrm{fc}} ~\mathrm{[M_{\odot}]}$ & $T_{\mathrm{fc}}$ [K]  & $\dot{M}_\mathrm{fc} ~\mathrm{[M_{\odot}/yr]}$  & $R_{\mathrm{sc}}$ [au]  & $M_{\mathrm{sc}} ~\mathrm{[M_{\odot}]}$ & $T_{\mathrm{sc}}$ [K]  & $\dot{M}_\mathrm{sc} ~\mathrm{[M_{\odot}/yr]}$  \TBstrut\\ \hline \hline 
0.5  & 2.35 & 2.53e-02 & 2.13e+02 & 3.47e-05 & 1.46e-02 & 3.78e-03 & 6.12e+03 & ~2.61e-02 \Tstrut \\
1.0  & 2.96 & 3.18e-02 & 2.58e+02 & 6.32e-05 & 1.84e-02 & 4.68e-03 & 6.16e+03 & 3.02e-02 \\
2.0  & 3.96 & 4.40e-02 & 3.30e+02 & 1.38e-04 & 2.00e-02 & 5.18e-03 & 6.14e+03 & 2.37e-02 \\
5.0  & 5.76 & 7.20e-02 & 4.91e+02 & 6.63e-04 & 2.34e-02 & 6.32e-03 & 6.82e+03 & 2.70e-02 \\
8.0  & 6.76 & 8.69e-02 & 6.03e+02 & 9.52e-04 & 2.33e-02 & 6.74e-03 & 6.87e+03 & 1.96e-02 \\
10.0 & 7.24 & 9.24e-02 & 6.62e+02 & 1.34e-03 & 2.43e-02 & 7.29e-03 & 6.82e+03 & 2.31e-02 \\
12.0 & 7.22 & 9.08e-02 & 7.59e+02 & 1.32e-03 & 2.42e-02 & 7.47e-03 & 7.51e+03 & 2.56e-02 \\
14.0 & 6.54 & 8.10e-02 & 8.97e+02 & 1.19e-03 & 2.32e-02 & 7.55e-03 & 8.63e+03 & 2.98e-02 \\
15.0 & 5.91 & 7.37e-02 & 9.71e+02 & 1.18e-03 & 2.48e-02 & 7.73e-03 & 7.90e+03 & 3.03e-02 \\
16.0 & 5.16 & 6.58e-02 & 1.04e+03 & 1.16e-03 & 2.53e-02 & 7.91e-03 & 7.33e+03 & 2.25e-02 \\
18.0 & 3.99 & 5.35e-02 & 1.11e+03 & 1.33e-03 & 2.55e-02 & 8.18e-03 & 8.01e+03 & 2.74e-02 \\
20.0 & 3.04 & 4.27e-02 & 1.17e+03 & 1.49e-03 & 2.48e-02 & 8.26e-03 & 8.33e+03 & 2.63e-02 \\ 
30.0 & 0.89 & 2.17e-02 & 1.46e+03 & 3.09e-03 & 2.45e-02 & 8.83e-03 & 9.28e+03 &
4.13e-02 \\ 
40.0 & 0.74 & 2.23e-02 & 1.56e+03 & 7.79e-03 & 2.23e-02 & 9.20e-03 & 1.15e+04 & 4.83e-02 \\ 
60.0 & 0.72 & 2.27e-02 & 1.65e+03 & 8.32e-03 & 2.00e-02 & 9.46e-03 & 1.32e+04 & 8.68e-02 \\
80.0 & 0.70 & 3.01e-02 & 1.68e+03 & 1.98e-02 & 1.83e-02 & 1.03e-02 & 1.48e+04 & 8.73e-02 \\
100.0 & 0.69 & 3.16e-02 & 1.67e+03 & 1.79e-02 & 1.79e-02 & 1.01e-02 & 1.50e+04 & ~9.53e-02 \Bstrut \\ \hline
\end{tabular}
} 
\vspace{0.1cm}
\caption*{Note: The properties listed are the first core radius $R_{\mathrm{fc}}$, mass $M_{\mathrm{fc}}$, temperature $T_{\mathrm{fc}}$, accretion rate $\dot{M}_\mathrm{fc}$, and second core radius $R_{\mathrm{sc}}$, mass $M_{\mathrm{sc}}$, temperature $T_{\mathrm{sc}}$, accretion rate $\dot{M}_\mathrm{sc}$.}
\label{tab:properties}
\end{table*}

\subsection{Effect of different initial cloud masses}
\label{sec:initcloudmass}

Here, we discuss the core collapse scenario for different initial cloud masses. We span a wide range of initial cloud masses from 0.5 to 100 $\mathrm{M_{\odot}}$. The clouds with different initial masses~$M_{0}$ and central densities $\rho_\mathrm{c}$ at the same initial temperature of 10~K and an outer radius of 3000 au follow a similar evolution as seen in Fig.~\ref{fig:masscomparison}. Most significant differences are seen outwards from the first shock as a horizontal spread. 

The thermal structure for cases with different initial cloud masses (Fig. \ref{fig:masscomparison}f) shows that the clouds begin with the same isothermal phase but eventually heat up at different densities. This difference in thermal evolution can have a significant effect on the properties of the first and second core since for the intermediate- and high-mass clouds ($M_0 \geq$ 8 $\mathrm{M_{\odot}}$) the dissociation temperature is reached earlier when the cloud is at a comparatively lower density which in turn affects the lifetime of the first and second core. The change in optical depth shown in Fig.~\ref{fig:masscomparison}j as the cloud evolves is mainly governed by the balance between radiative cooling and compressional heating. The sharp dissociation front seen in Fig.~\ref{fig:masscomparison}l indicates that most of the $\mathrm{H_2}$ is dissociated at the second core accretion shock. 

The first core radius $R_{\mathrm{fc}}$, defined by the position of the outer discontinuity in the density profile or shock position in velocity profile, is seen to increase with an increase in the initial cloud mass up to around 8 -- 10 $\mathrm{M_{\odot}}$ after which there is a decrease in the first core radius with an increase in the cloud mass (see inset in the radial density profile). 

For the initial clouds of mass 40 $\mathrm{M_{\odot}}$, 60 $\mathrm{M_{\odot}}$, 80 $\mathrm{M_{\odot}}$, and 100 $\mathrm{M_{\odot}}$, the first core barely forms and the evolution proceeds rapidly to the second collapse phase. For these cases, since the ram pressure $P_\mathrm{ram} = \rho u^2$ is always higher than the thermal pressure $P_\mathrm{gas}$ both above and below the first core radius (see Fig. \ref{fig:masscomparison}h), gravity acts as a dominant force which prevents a strong first accretion shock. These high-mass clouds also have the highest accretion rate and are the most unstable which is why they evolve faster. In summary, in the high-mass regime, first cores do not exist. 

For comparison with the low- and intermediate-mass regimes, the shock-like velocity structure is still referred to as an ``accretion shock'' and the first core-like structure is referred to as a ``first core'' even in the high-mass regime.
 
The second core radius $R_{\mathrm{sc}}$ is defined by the position of the inner discontinuity in the density profile or inner shock position in the velocity profile. The main properties of the first and second cores for each of the different cases are listed in Table~\ref{tab:properties}. These are the first core mass $M_{\mathrm{fc}}$, temperature $T_{\mathrm{fc}}$ and accretion rate $\dot{M}_\mathrm{fc}$ calculated at the first core radius $R_{\mathrm{fc}}$ as well as the second core mass $M_{\mathrm{sc}}$, temperature $T_{\mathrm{sc}}$ and accretion rate $\dot{M}_\mathrm{sc}$ calculated at the second core radius $R_{\mathrm{sc}}$. 

\subsection{First core properties}
\label{sec:firstcore}

In this section we focus on the dependence of the core properties on the initial cloud mass. Our results indicate slight differences (within an order of magnitude) in the first core radius and mass for the collapse simulations with different initial cloud masses. In our studies, since we span a wide range from 0.5 to 100 $\mathrm{M_{\odot}}$, we are able to see a transition region around 8 -- 10 $\mathrm{M_{\odot}}$. Hence although the differences in the first core properties are within an order of magnitude, we would like to draw more attention to the diminishing first core lifetimes for higher initial cloud masses. This in turn affects the size and mass of the first core.   
\begin{figure}[!htp]
\centering
\begin{subfigure}{0.37\textwidth}
\includegraphics[width= \textwidth]{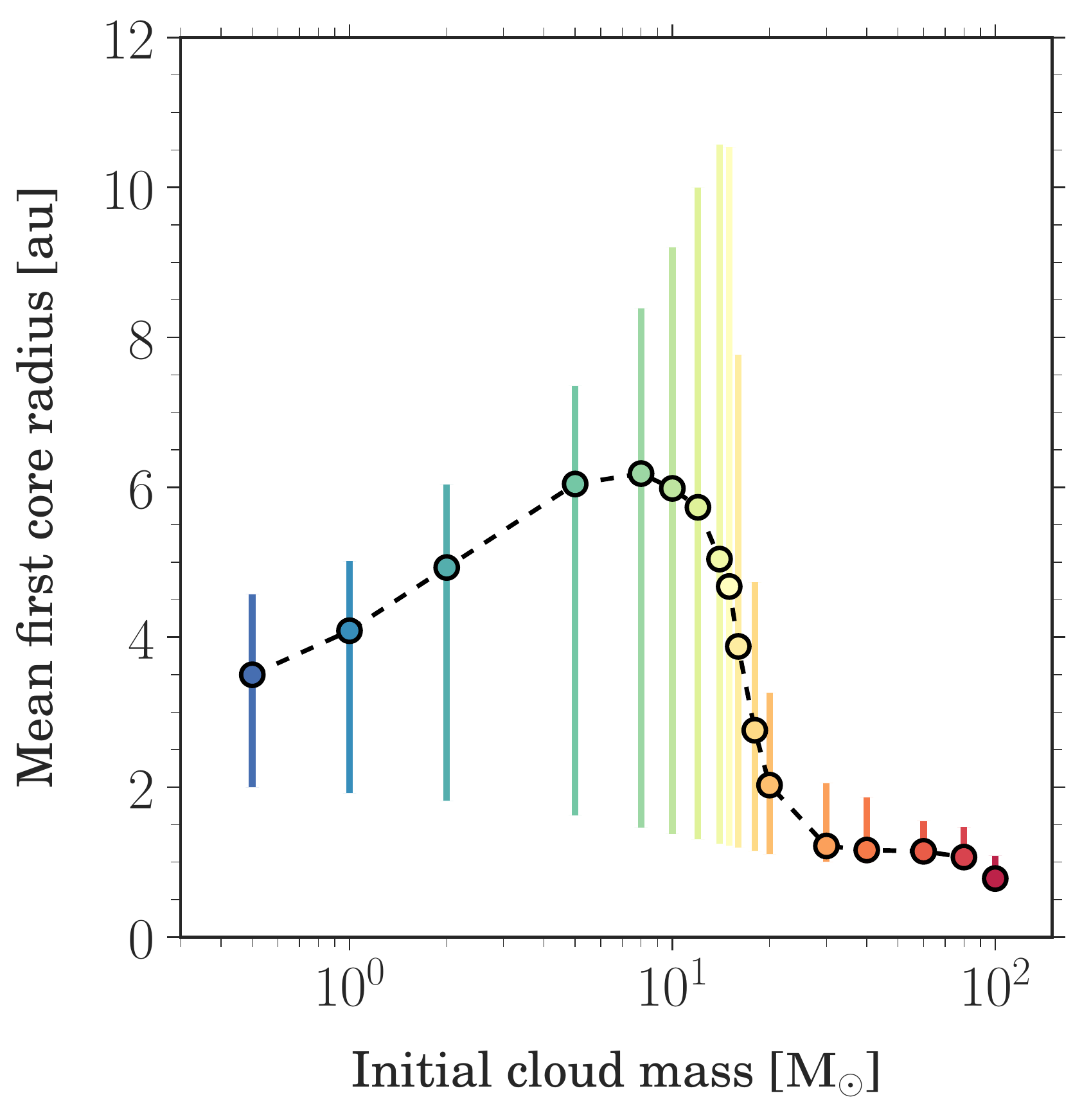}
\end{subfigure}
\begin{subfigure}{0.37\textwidth}
\includegraphics[width= \textwidth]{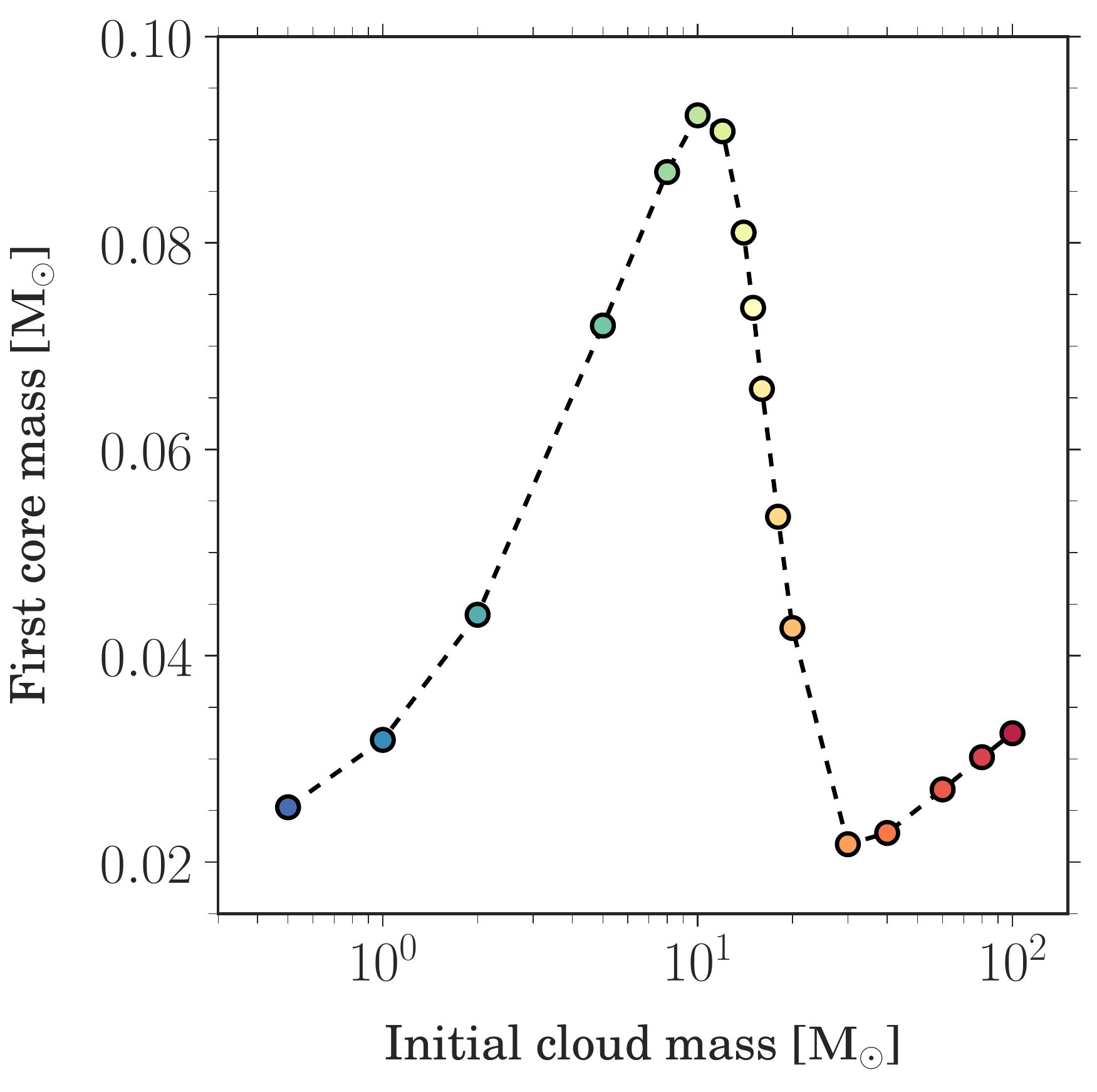}
\end{subfigure}
\begin{subfigure}{0.37\textwidth}
\includegraphics[width= \textwidth]{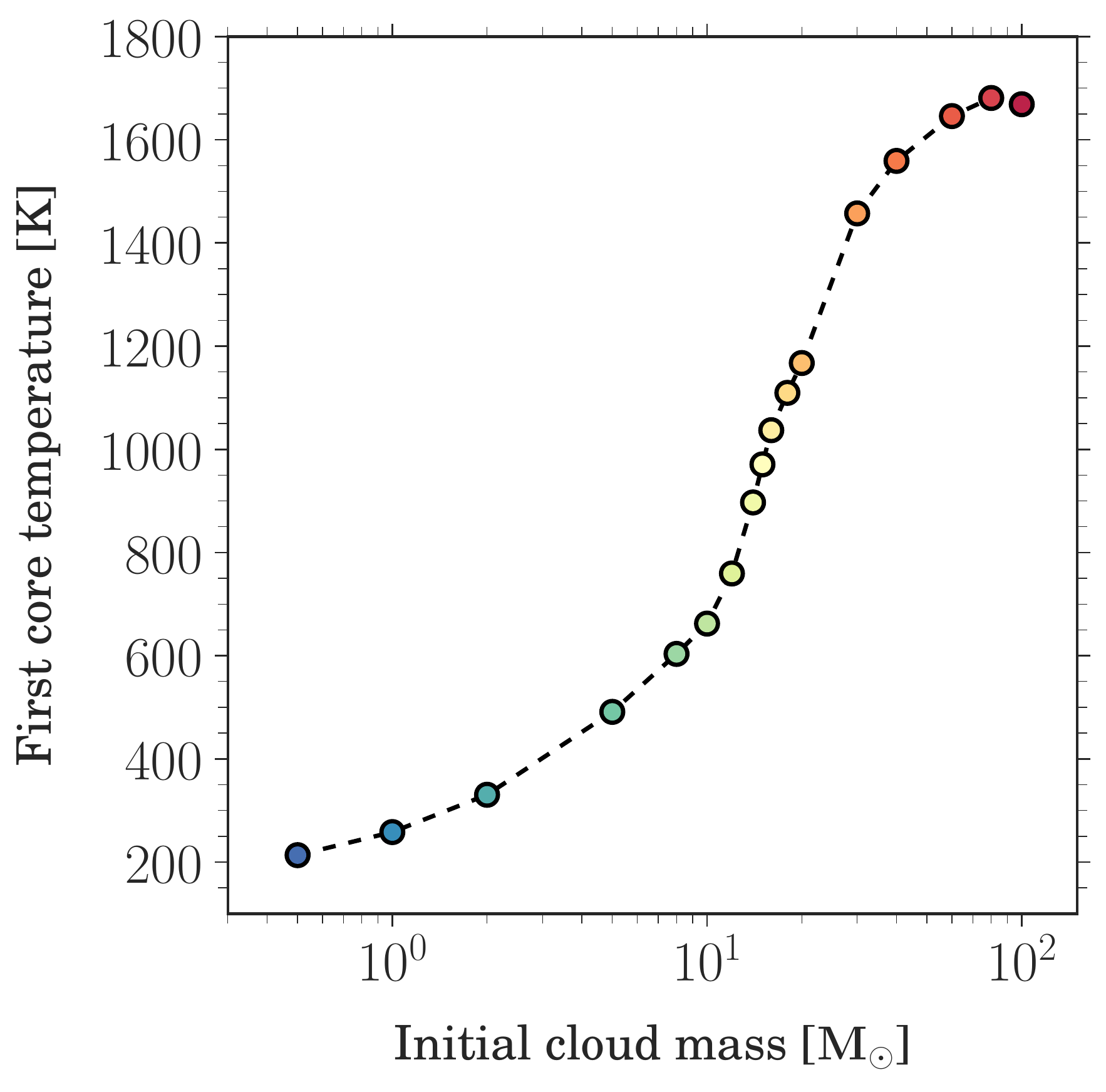}
\end{subfigure}
\caption{Dependence of the first core properties on initial cloud mass. Shown is \mbox{\bf a)} the mean first core radius (mean radius is calculated over the time from the onset of the first core formation until the second core formation). The vertical lines span the minimum to maximum first core radius as the core evolves. Also shown are, \mbox{\bf b)} the first core mass and \mbox{\bf c)} outer shock temperature as a function of initial cloud mass as estimated at a time after the second core formation when the first core is stable and no longer evolves. A transition region is seen around 8 -- 10 $\mathrm{M_{\odot}}$ indicating the diminishing first core radius and mass towards the high-mass regime.}
\label{fig:coreproperties}
\end{figure}
\begin{figure}[!htp]
\centering
\begin{subfigure}{0.47\textwidth}
\includegraphics[width=\textwidth]{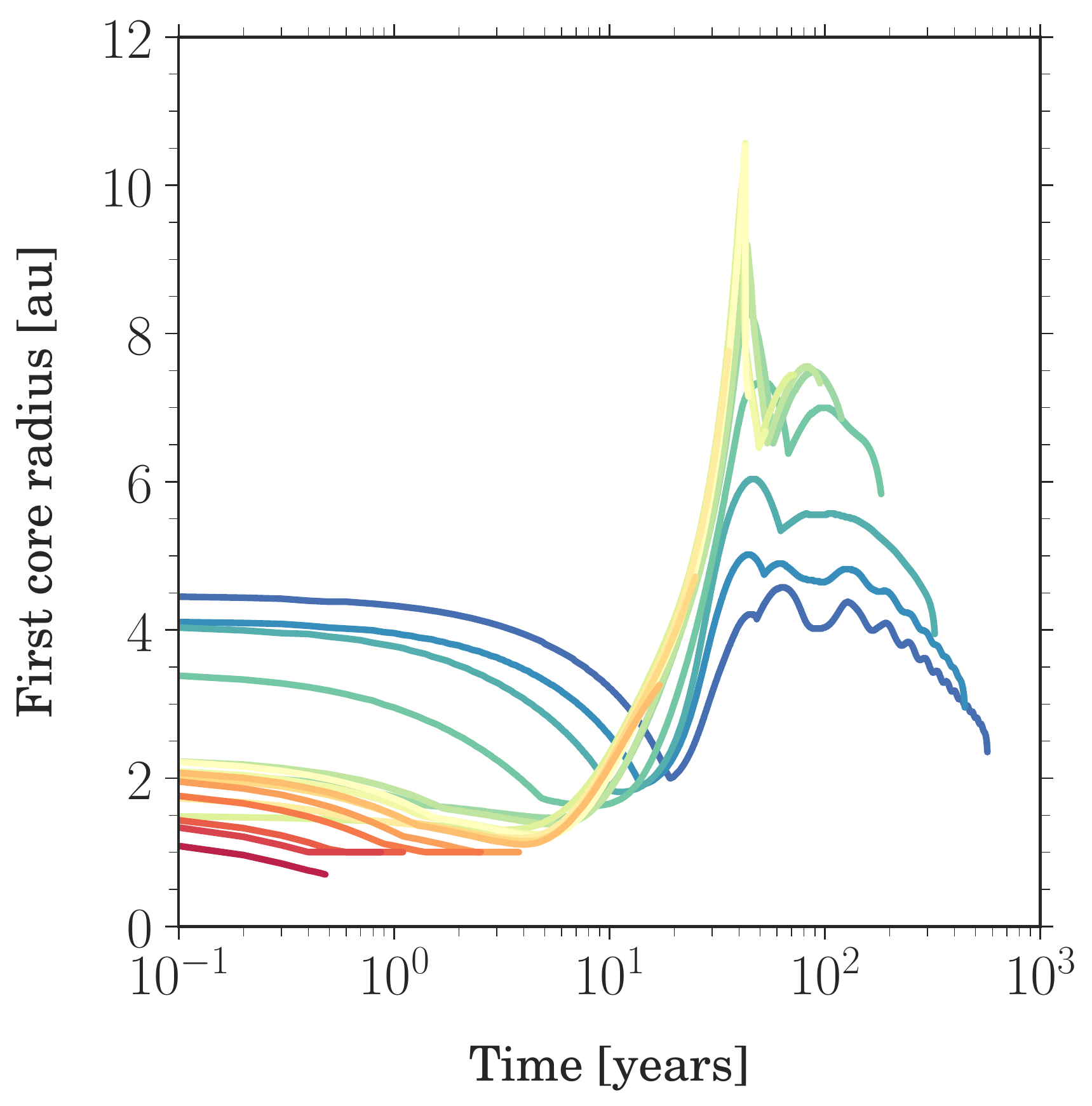}
\end{subfigure}
\begin{subfigure}{0.4\textwidth}
\hspace{0.4cm}
\includegraphics[width=\textwidth]{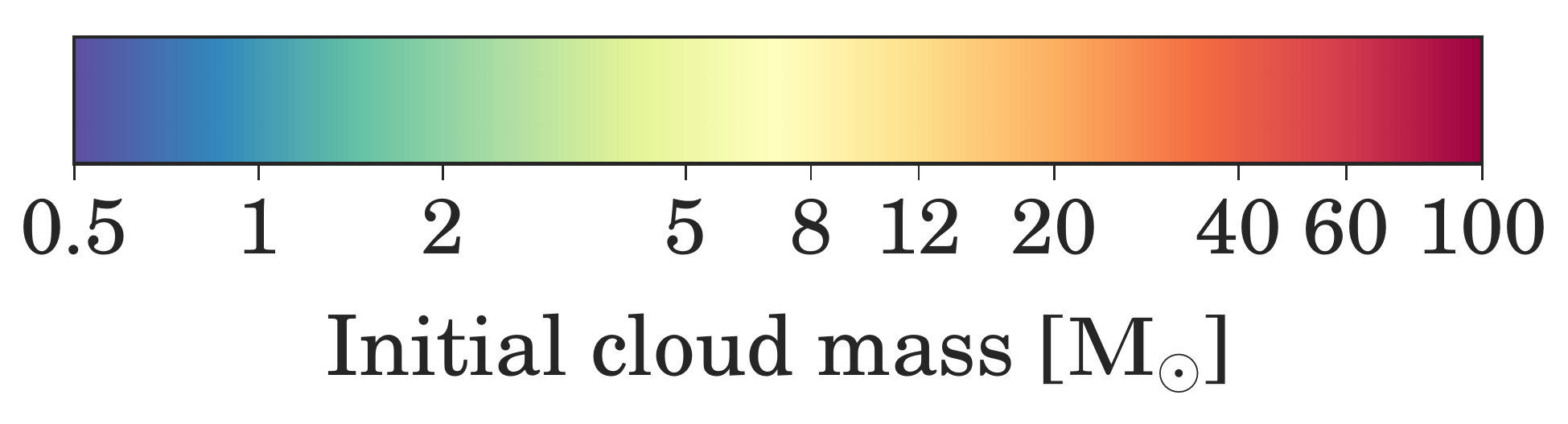}
\end{subfigure}   
\caption{Time evolution of the first core radius showing an initial contraction phase followed by a rapid expansion and a second contraction phase. The colors indicate the different initial cloud masses ranging from 0.5 to 100 $\mathrm{M_{\odot}}$ as shown in the color bar.}
\label{fig:evolution}
\end{figure}

As already noted, the first core radius increases with an increase in the cloud mass until around 8 -- 10 $\mathrm{M_{\odot}}$ after which it decreases. Figure \ref{fig:coreproperties}a shows the mean first core radius as a function of the initial cloud mass where the mean radius is calculated over time from the onset of the first core formation until the second core formation. The vertical lines span the minimum to maximum first core radius as the core evolves. The transition around 8 -- 10 $\mathrm{M_{\odot}}$ is also seen for the first core mass (see Fig.~\ref{fig:coreproperties}b), whereas the first core temperature always increases with an increase in the initial cloud mass (see Fig. \ref{fig:coreproperties}c). 

The evolution of the first core radius from the onset of the first core formation until the second core formation is shown in Fig.~\ref{fig:evolution}. The first core undergoes an initial contraction phase followed by a rapid expansion and a second contraction phase. 

\begin{figure}[t]
\centering
\includegraphics[width=0.465\textwidth]{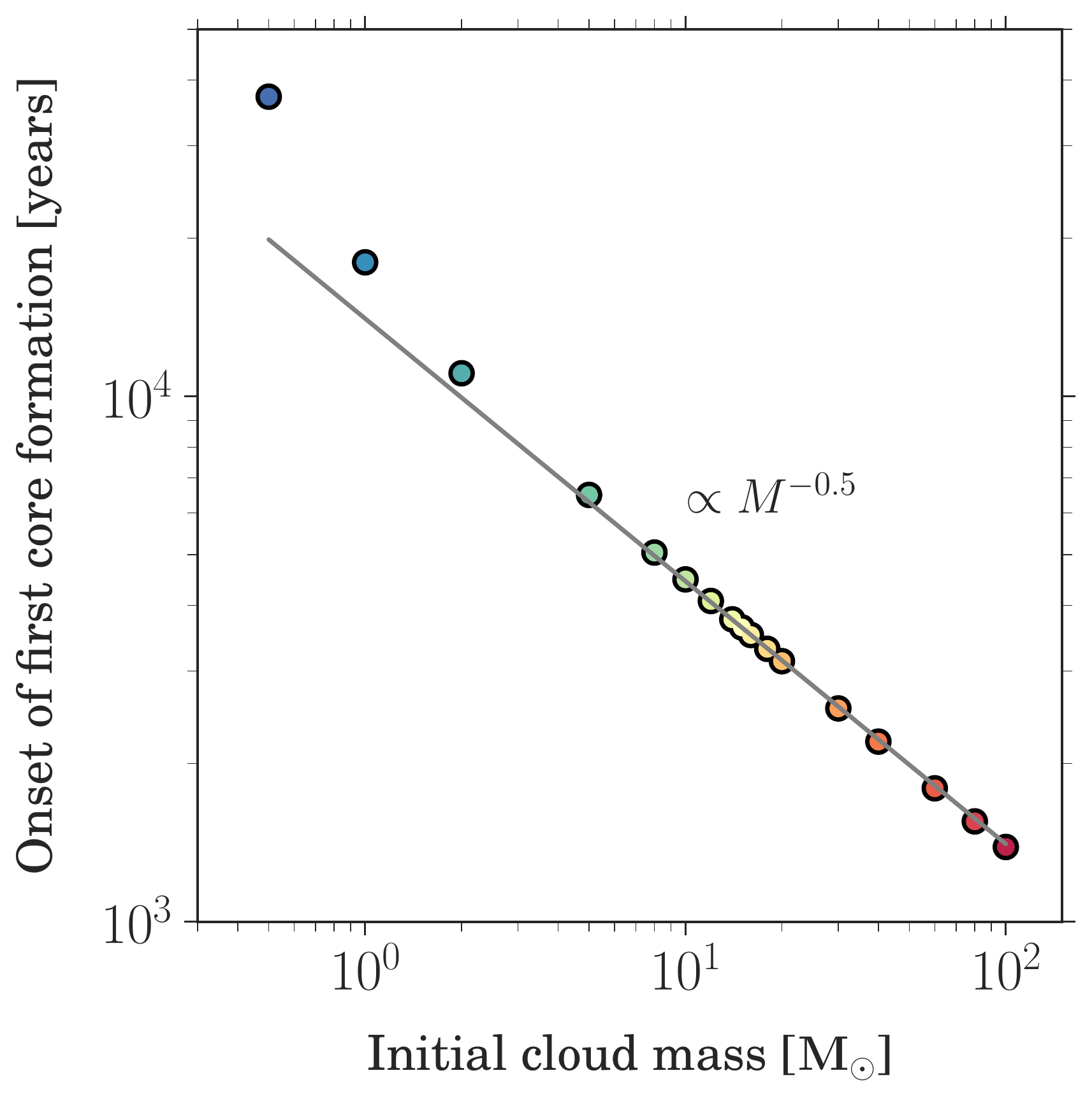}
\caption{Onset of formation of the first core for different initial cloud masses. Initially higher mass clouds tend to collapse faster in comparison to the low-mass regime. }
\label{fig:onset}
\end{figure}

Figure~\ref{fig:onset} shows the onset of the first core formation as a function of initial cloud mass. In the low-mass range ($M_0 \leq 8 ~\mathrm{M_{\odot}}$), the cloud is seen to undergo a comparatively slower collapse hence initiating the first core formation after $\approx$ 5000 - 18000 years. On the other hand, in the intermediate- and high-mass regime, the collapse is seen to be much faster with the first core forming after a few thousand years ($\leq$ 5000 years) followed by an instantaneous second collapse phase which prevents the first core from growing. 

\begin{figure}[!t]
\centering
\includegraphics[width=0.47\textwidth]{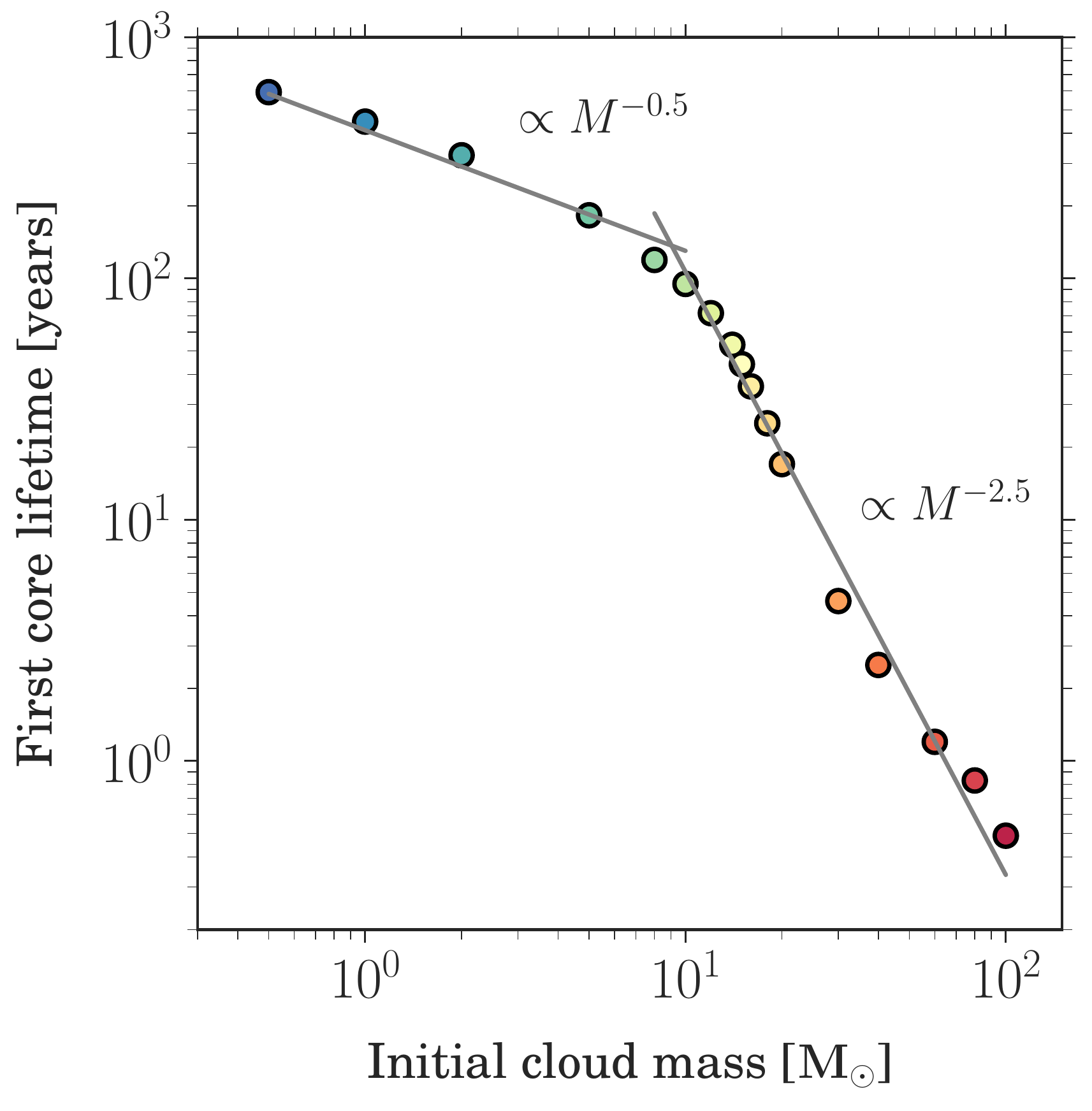}
\caption{First core lifetime i.e. time between the onset of formation of the first and second cores for different initial cloud masses. }
\label{fig:lifetime}
\end{figure}

Figure \ref{fig:lifetime} shows the first core lifetime as a function of initial cloud mass. The first core lifetime is defined as the time between the onset of formation of the first core until the onset of the second core formation. Since currently we stop our simulations a few years after the second core formation, the total simulation time minus the first core formation time is almost equivalent to that of the first core lifetime. 

As seen in all the previous studies, we also note that in the low-mass regime ($\leq$~8 $\mathrm{M_{\odot}}$) the first core lifetime scales as $M^{-0.5}$ as seen in Fig. \ref{fig:lifetime}. 
 
In the intermediate- and high-mass regime due to the vanishing thermal pressure support, this dependence changes to $M^{-2.5}$ (see Fig. \ref{fig:lifetime}), which can be analytically derived as follows. The accretion energy $\dot{E}$ is given as
\begin{align}
\mbox{$\dot{E} = \dfrac{E_\mathrm{fc}}{\tau_\mathrm{fc}} \propto {\dfrac{G M_\mathrm{fc}}{R_\mathrm{fc}}} \dot{M}_\mathrm{fc}$},
\label{eq:accretion_lifetime}
\end{align}
where $M_\mathrm{fc}$ is the mass enclosed within the first core, $R_\mathrm{fc}$ is the first core radius, $\tau_\mathrm{fc}$ is the first core lifetime and $\dot{M}_\mathrm{fc}$ is the accretion rate. The internal energy profiles seen in Fig.~\ref{fig:masscomparison}i look strikingly similar at the onset of the second collapse phase (i.e. at T $\approx$ 2000 K) for all the different initial cloud masses. This indicates that indeed the internal energy of the first core $E_\mathrm{fc}$, at the onset of the second collapse phase, is independent of the initial cloud mass. 

Now, consider the ratio $M_\mathrm{fc}/R_\mathrm{fc}$ and multiply and divide by the velocity $u_\mathrm{fc}$: 
\begin{align}
\mbox{$\dfrac{M_\mathrm{fc}}{R_\mathrm{fc}} = \dfrac{4 \pi}{3} ~\rho_\mathrm{fc} ~R_\mathrm{fc}^{2} = \dfrac{4 \pi ~\rho_\mathrm{fc} ~R_\mathrm{fc}^{2} ~u_\mathrm{fc}}{3 ~u_\mathrm{fc}} = \dfrac{\dot{M}_\mathrm{fc}}{3 ~u_\mathrm{fc}}$}.
\label{eq:massfc_radfc_ratio}
\end{align} 
Inserting this into the expression for the accretion energy $\dot{E}$ yields
\begin{align}
\mbox{$\dot{E} \propto {\dfrac{G M_\mathrm{fc}}{R_\mathrm{fc}}} \dot{M}_\mathrm{fc} \propto \dfrac{G \dot{M}_\mathrm{fc}^{2}}{3 ~u_\mathrm{fc}}$}.
\label{eq:accretion_energy}
\end{align}
In the intermediate- and high-mass regime, we assume the whole cloud to be in free-fall and hence we can relate the local properties to the large scale properties. The accretion rate $\dot{M}_\mathrm{fc}$ is then defined as
\begin{align}
\mbox{$\dot{M}_\mathrm{fc} = \dfrac{M_{0}}{t_\mathrm{ff}} $},
\end{align}
where $t_\mathrm{ff}$ is the free-fall time. We then assume that the accretion is constant in space (and time), which is valid only for a $\rho~\propto~R^{-2}$ profile, seen in the outer parts of a Bonnor--Ebert sphere like density profile. In this case, the mean velocity $u_\mathrm{fc}$ can be estimated as 
\begin{align}
\mbox{$ u_\mathrm{fc} = \dfrac{R_\mathrm{cloud}}{t_\mathrm{ff}} $}. 
\end{align}
Now, the free-fall time of a collapsing cloud is given by
\begin{align}
\mbox{$t_\mathrm{ff} = \sqrt{\dfrac{3 \pi}{32 G \rho_\mathrm{c}}} \propto \sqrt{\dfrac{R_\mathrm{cloud}^3}{M_{0}}} $}.
\end{align}
Using these relations in the expression for accretion energy $\dot{E}$, Eq. \eqref{eq:accretion_energy} yields
\begin{align}
\mbox{$\dot{E} \propto \dfrac{\dot{M}_\mathrm{fc}^{2}}{u_\mathrm{fc}} \propto \Bigg({\dfrac{M_{0}}{R_\mathrm{cloud}}}\Bigg)^{5/2}$}.
\end{align}
Furthermore, from Eq. \eqref{eq:accretion_lifetime}
\begin{align}
\mbox{$\tau_\mathrm{fc} \propto \dfrac{1}{\dot{E}} \propto \Bigg({\dfrac{R_\mathrm{cloud}}{M_{0}}}\Bigg)^{5/2}$}.
\end{align}	
Thus in the intermediate- and high-mass regime, the first core lifetime scales as $M^{-2.5}$ as seen in Fig. \ref{fig:lifetime}. The dependence on the cloud radius is seen in Fig. \ref{fig:lifetimeR5} and discussed in \cref{sec:Rout}. 

\subsection{Dependence on initial conditions}
\label{sec:initialsetup}

\begin{figure}[t]
\centering
\includegraphics[width=0.45\textwidth]{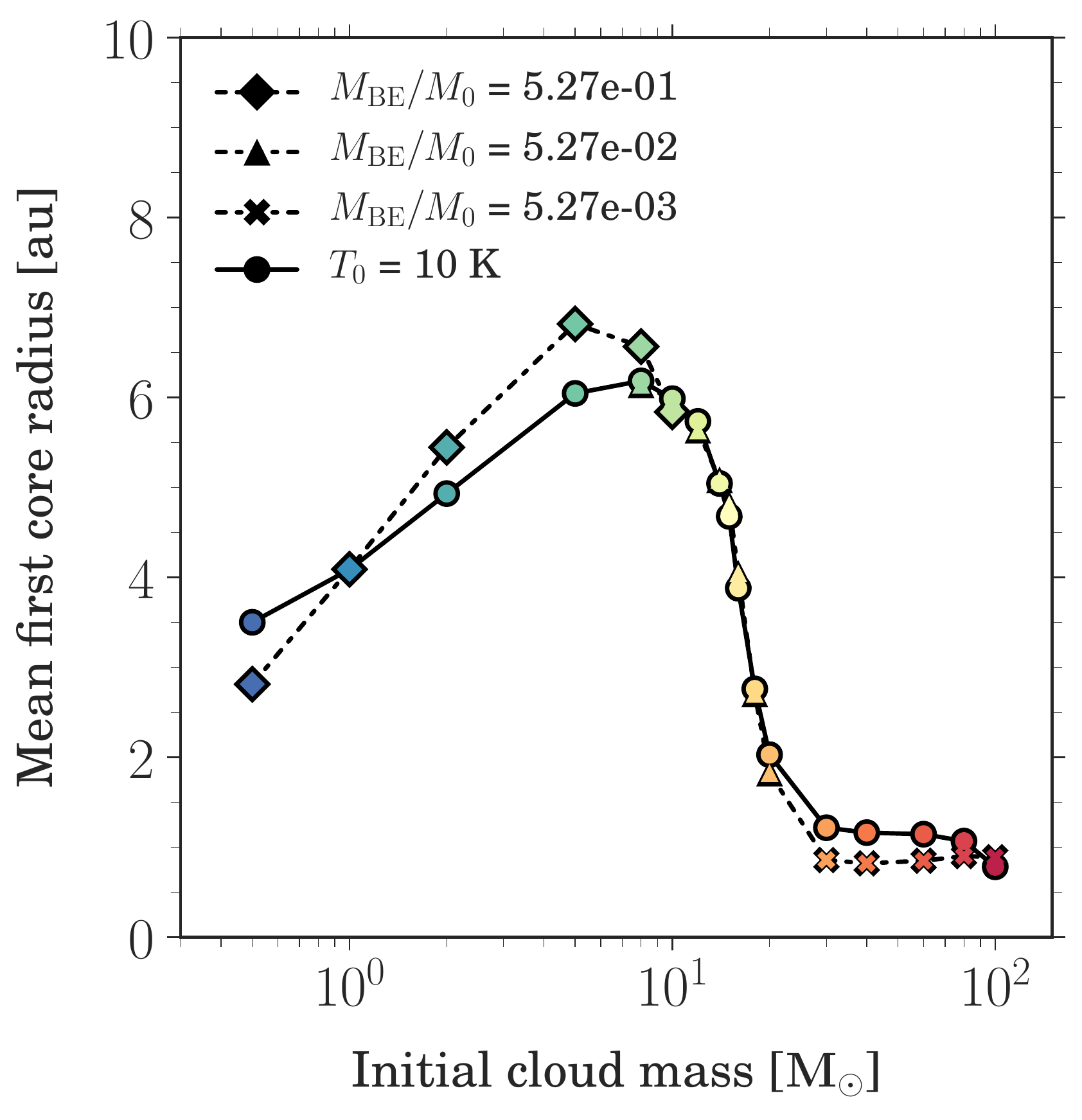}
\caption{Mean first core radius as a function of initial cloud mass where the mean radius is calculated over the time from the onset of the first core formation until the second core formation. The circles indicate results from the simulation runs with a constant initial temperature of 10 K, whereas the diamonds, triangles and crosses indicate results from the simulation runs with constant stability parameters $M_{\mathrm{BE}}/M_{\mathrm{0}}$ of 5.27e-01, 5.27e-02 and \mbox{5.27e-03} for the low-, intermediate- and high-mass regimes, respectively.}
\label{fig:radiusstability}
\end{figure}

\begin{figure}[t]
\centering
\includegraphics[width=0.46\textwidth]{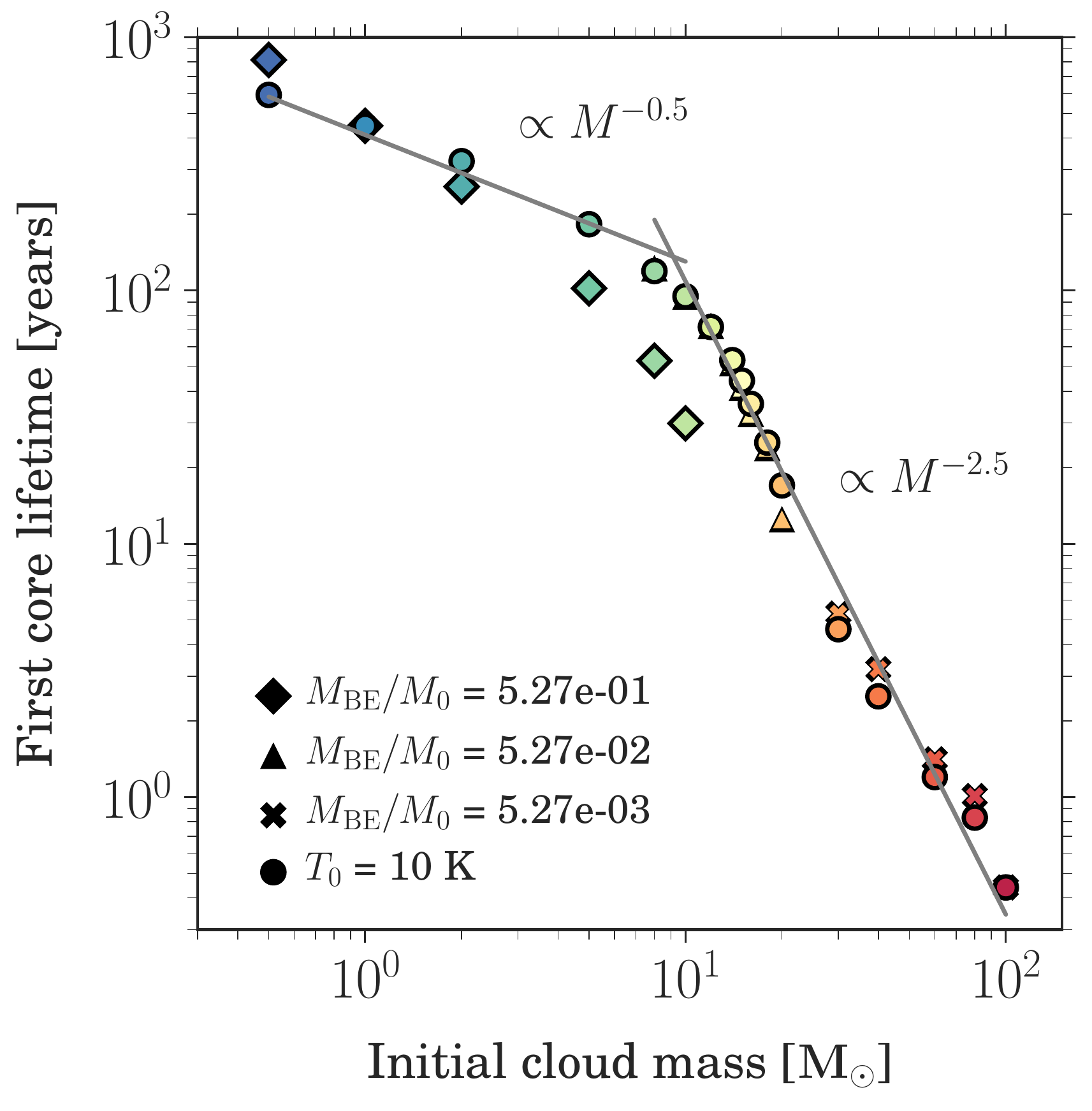}
\caption{First core lifetime, i.e. time between the onset of formation of the first and second cores for different initial cloud masses. The circles indicate results from the simulation runs with a constant initial temperature of 10 K, whereas the diamonds, triangles and crosses indicate results from the simulation runs with constant stability parameters $M_{\mathrm{BE}}/M_{\mathrm{0}}$ of \mbox{5.27e-01}, 5.27e-02 and 5.27e-03 for the low-, intermediate- and high-mass regimes, respectively. }
\label{fig:lifetimestability}
\end{figure}

\begin{table}[!]
\centering
\caption{Initial cloud properties}
\begin{tabular}[t]{ccccc}
\hline	
$M_{0} ~\mathrm{[M_{\odot}]}$ & $R_\mathrm{out}$ [au] & $T_{\mathrm{0}}$ [K] & $M_{\mathrm{BE}}/M_{\mathrm{0}}$ & $\rho_\mathrm{c} ~[\mathrm{g ~cm^{-3}}]$  
\TBstrut\\ \hline \hline 
0.5   & 3000   & 5.0        & 5.27e-01    & ~1.16e-17  \Tstrut \\
1.0   & 3000   & 10.0       & 5.27e-01    & 2.33e-17  \\
2.0   & 3000   & 20.0       & 5.27e-01    & 4.66e-17  \\
5.0   & 3000   & 50.0       & 5.27e-01    & 1.16e-16  \\
8.0   & 3000   & 80.0       & 5.27e-01    & 1.86e-16  \\
10.0  & 3000   & 100.0      & 5.27e-01    & 2.33e-16  \\ \hline
8.0   & 3000   & 8.0        & 5.27e-02    & ~1.86e-16  \Tstrut \\
10.0  & 3000   & 10.0       & 5.27e-02    & 2.33e-16  \\
12.0  & 3000   & 12.0       & 5.27e-02    & 2.80e-16  \\
14.0  & 3000   & 14.0       & 5.27e-02    & 3.26e-16  \\
15.0  & 3000   & 15.0       & 5.27e-02    & 3.50e-16  \\
16.0  & 3000   & 16.0       & 5.27e-02    & 3.73e-16  \\
18.0  & 3000   & 18.0       & 5.27e-02    & 4.20e-16  \\
20.0  & 3000   & 20.0       & 5.27e-02    & 4.66e-16  \\ \hline
30.0  & 3000   & 3.0        & 5.27e-03    & ~6.99e-16   \Tstrut \\
40.0  & 3000   & 4.0        & 5.27e-03    & 9.33e-16  \\
60.0  & 3000   & 6.0        & 5.27e-03    & 1.40e-15  \\
80.0  & 3000   & 8.0        & 5.27e-03    & 1.86e-15  \\
100.0 & 3000   & 10.0       & 5.27e-03    & ~2.33e-15   \Bstrut \\ \hline
\end{tabular}
\vspace{0.2cm}
\caption*{Note: Listed above are the cloud properties for runs with different initial cloud mass $M_{0} ~\mathrm{[M_{\odot}]}$, outer radius $R_\mathrm{out}$ [au], temperature $T_{\mathrm{0}}$ [K], stability parameter $M_{\mathrm{BE}}/M_{\mathrm{0}}$ and central density $\rho_\mathrm{c} ~[\mathrm{g ~cm^{-3}}]$. }
\label{tab:newruns}
\end{table}

In order to assess the robustness of the transition region seen in properties of the first core, we performed three additional set of simulations using a different constant stability parameter $M_{\mathrm{BE}}/M_{\mathrm{0}}$ for the low-mass (0.5 to 10 $\mathrm{M_{\odot}}$), intermediate-mass (8 to 20 $\mathrm{M_{\odot}}$) and high-mass regime (30 to 100 $\mathrm{M_{\odot}}$), respectively, with some overlap between the low- and intermediate-masses. This implies a change in the initial temperature for the different cases, which now varies from 5 -- 100 K. We use three different stability parameters in order to avoid extremely high initial cloud temperatures in the intermediate- and high-mass regime. The outer radius for all the different simulations is fixed to 3000 au. The different runs with constant stability parameters are listed in Table \ref{tab:newruns}. 

We find a transition region in the intermediate-mass regime similar to the one described in \cref{sec:firstcore}. The mean first core radius increases with an increase in the initial cloud mass until around 5 to 8 $\mathrm{M_{\odot}}$ and then decreases towards the intermediate- and high-mass regime. We compare this to the previously described runs with a fixed initial cloud temperature of 10 K in Fig. \ref{fig:radiusstability}. 

Figure~\ref{fig:lifetimestability} shows the dependence of the first core lifetime on the initial cloud mass. It is very similar to the one previously seen in Fig. \ref{fig:lifetime}, thereby confirming that the first cores are non-existent in the high-mass regime. These results also indicate that the first core properties do not have a very strong dependence on the initial cloud properties.

We also investigated the dependence of the first core properties on the outer cloud radius by performing a set of simulations for the different cases from 1 to 100 $\mathrm{M_{\odot}}$ using an outer radius of 5000 au. The initial temperature for these runs was kept constant at 10 K. We see a similar transition region in the intermediate-mass regime. The initial cloud properties and first core properties for these runs are described in appendix \ref{sec:Rout}. 

\subsection{Comparison to previous results}
\label{sec:comparisons}

\begin{figure*}[!htp]
\centering
\begin{subfigure}{0.25\textwidth}
\includegraphics[width= 1.2\textwidth]{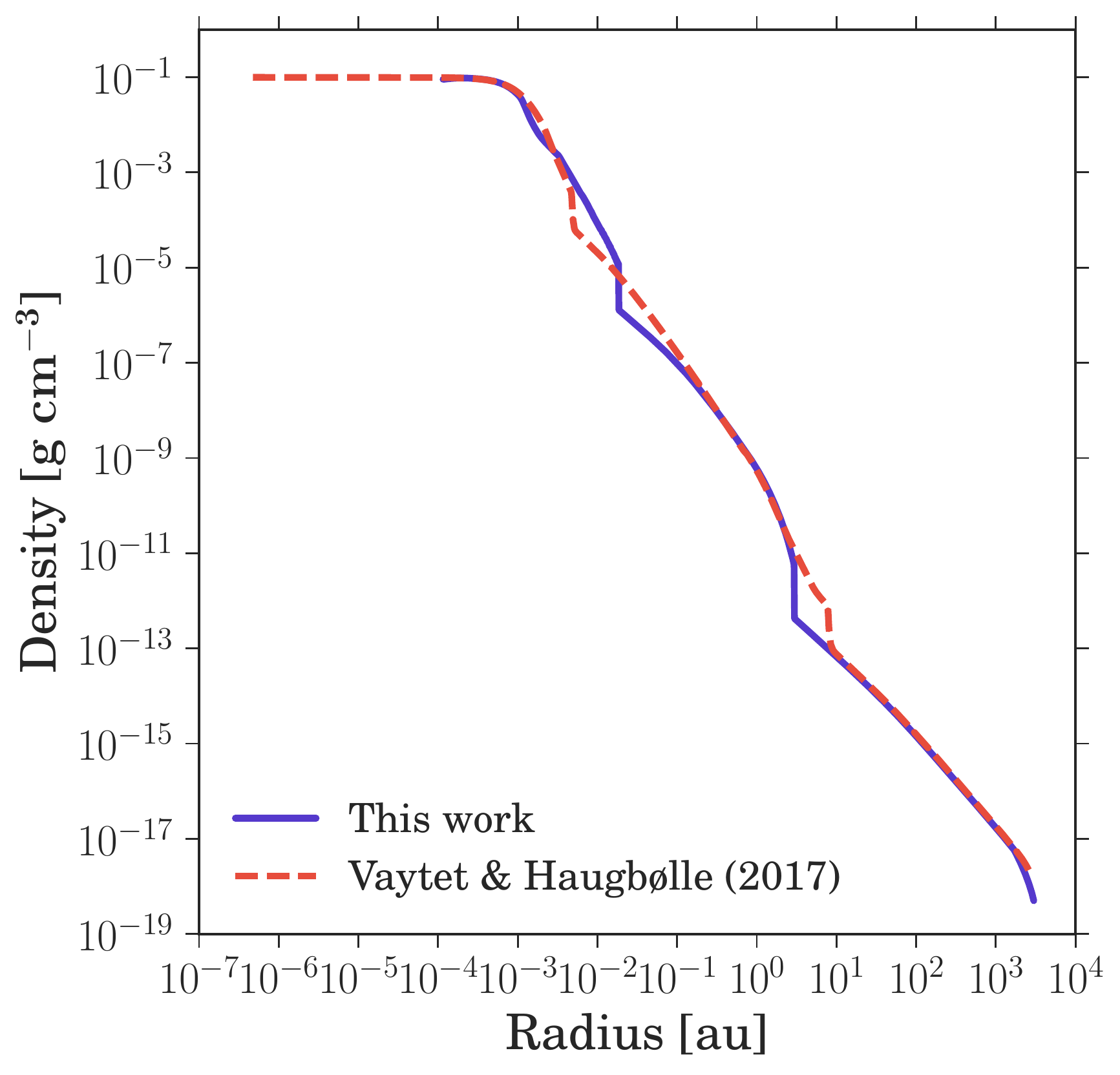}
\end{subfigure}
\hspace{0.5in}
\begin{subfigure}{0.25\textwidth}
\includegraphics[width= 1.2\textwidth]{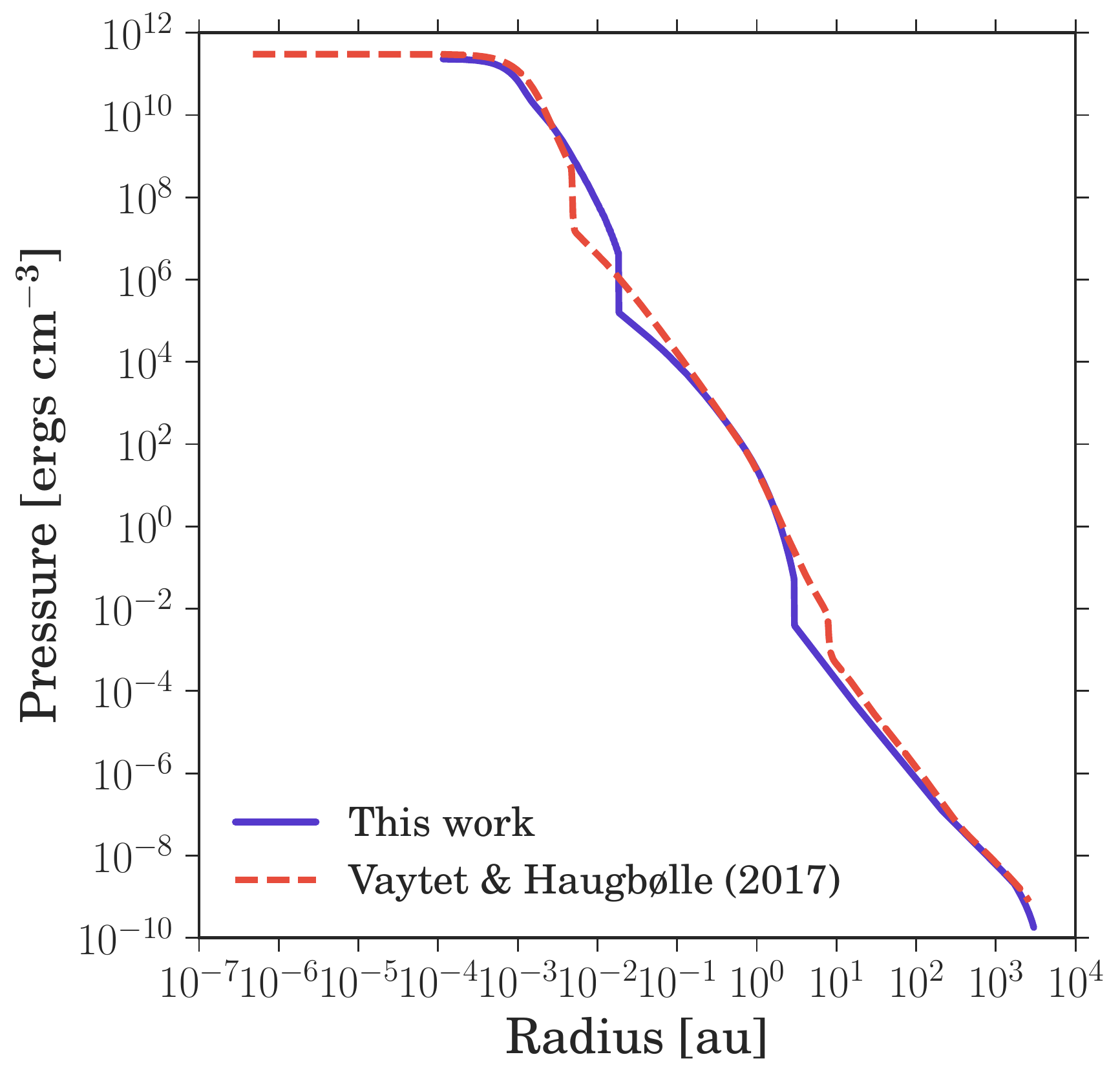}
\end{subfigure}
\hspace{0.5in}
\begin{subfigure}{0.25\textwidth}
\includegraphics[width= 1.2\textwidth]{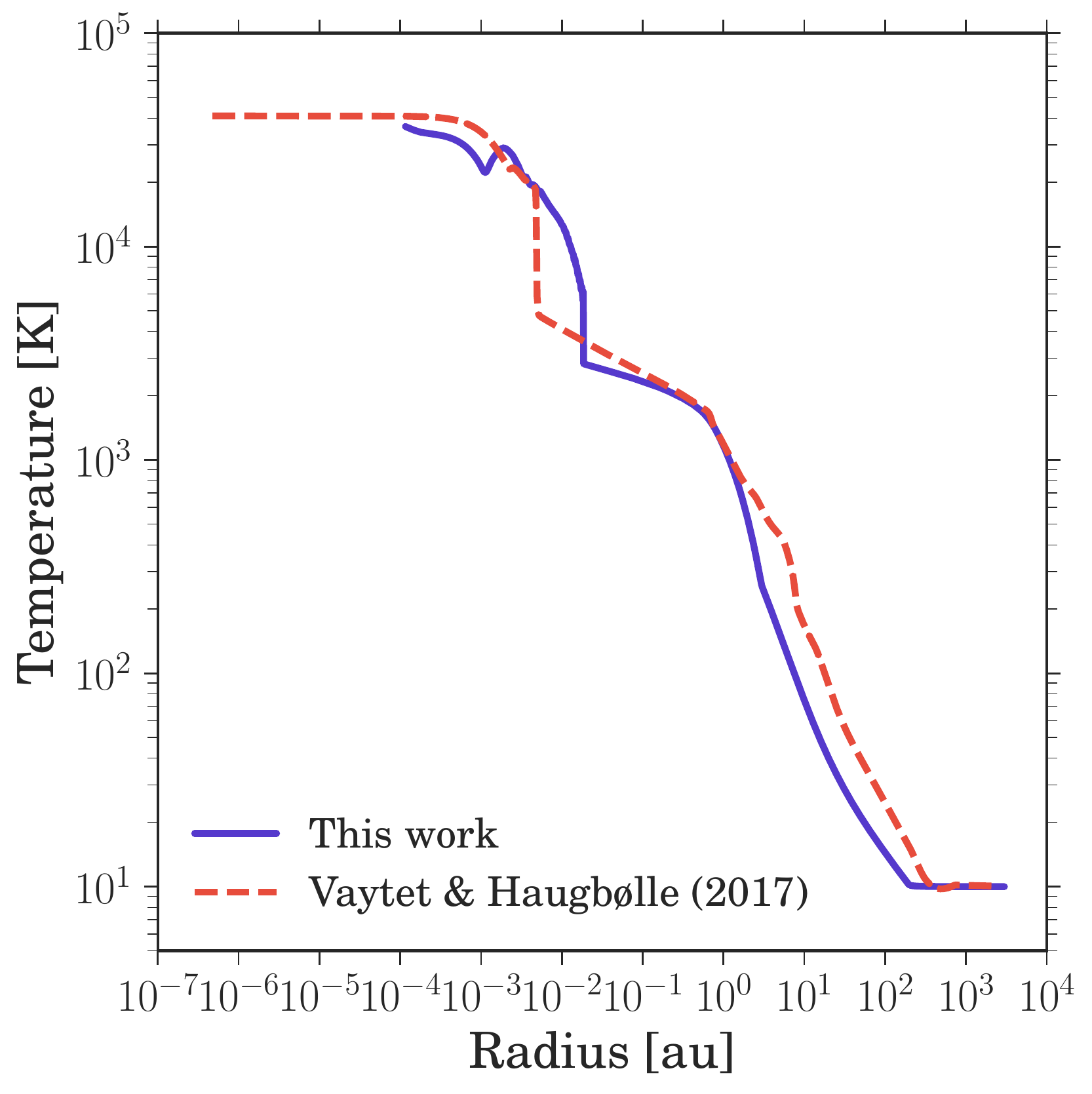}
\end{subfigure}
\begin{subfigure}{0.25\textwidth}
\includegraphics[width= 1.2\textwidth]{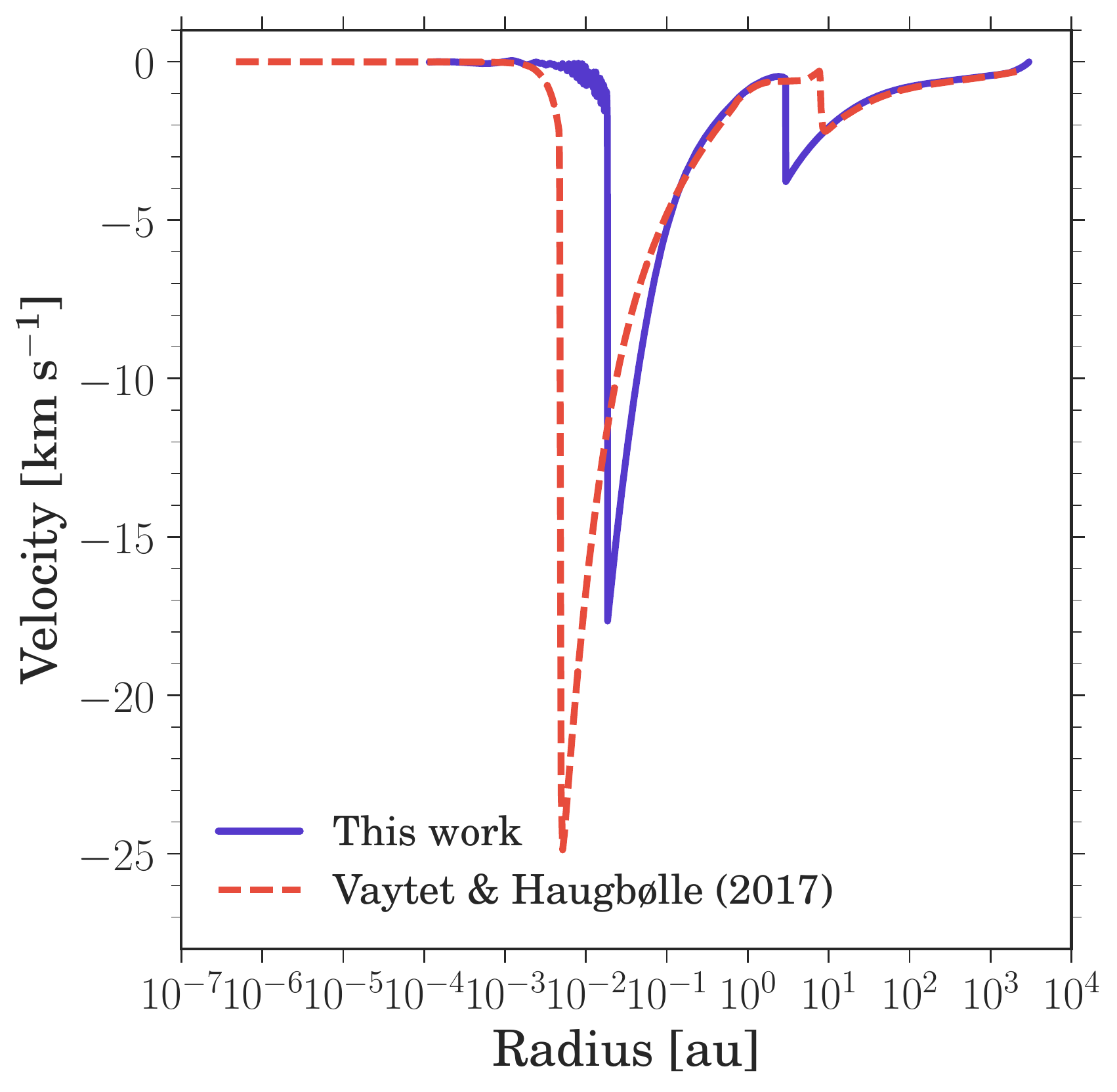}
\end{subfigure}
\hspace{0.5in}
\begin{subfigure}{0.25\textwidth}
\includegraphics[width= 1.2\textwidth]{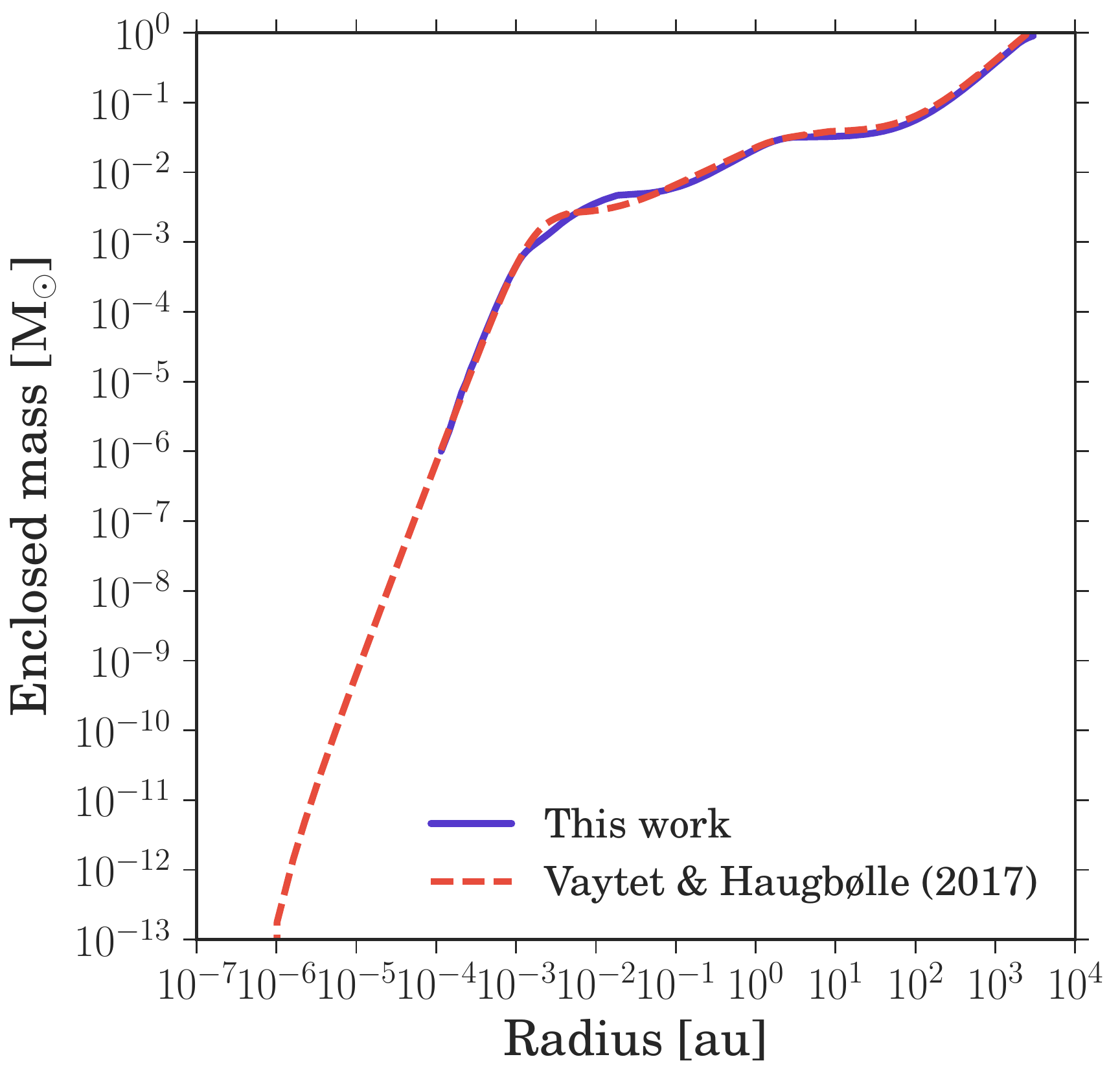}
\end{subfigure}
\hspace{0.5in}
\begin{subfigure}{0.25\textwidth}
\includegraphics[width= 1.2\textwidth]{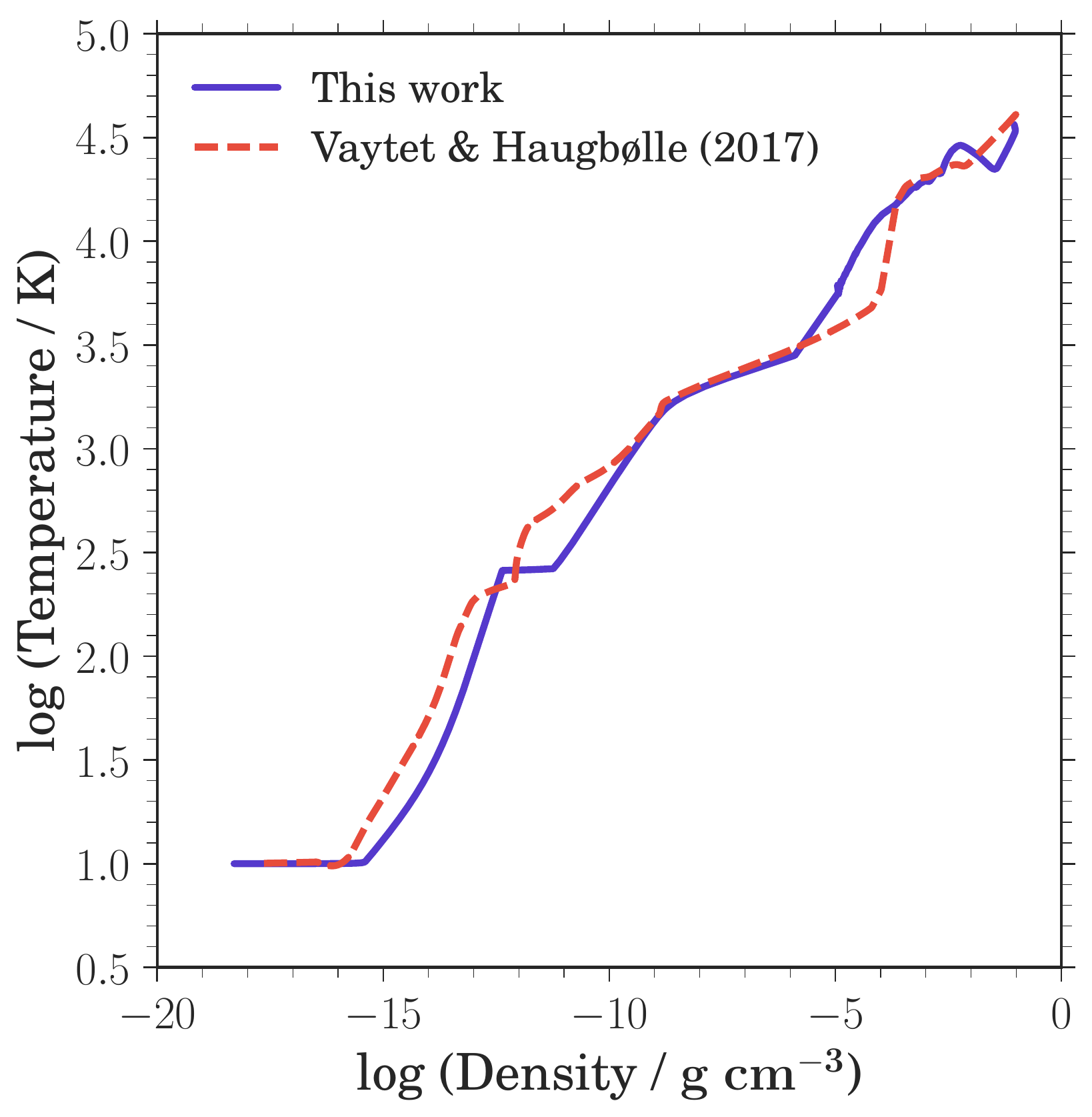}
\end{subfigure}
\caption{Comparisons of our results for an initial 1 $\mathrm{M_{\odot}}$ cloud indicated in bluish purple to those by \citet{Vaytet2017} shown using dashed red line. Radial profiles (across and down) of the \mbox{\bf a)} density, \mbox{\bf b)} pressure, \mbox{\bf c)} gas temperature, \mbox{\bf d)} velocity, \mbox{\bf e)} enclosed mass, and \mbox{\bf f)} thermal structure are shown at the time when central density $\rho_\mathrm{c}$ in both simulations reach roughly $10^{-1} ~\mathrm{g ~cm^{-3}}$. }
\label{fig:vaytet}
\end{figure*}

In this paper, we expand the collapse simulations for the first time, to cover a wide range of initial cloud masses from 0.5 up to 100~$\mathrm{M_{\odot}}$. Figure~\ref{fig:vaytet} shows comparisons from our run for an initial 1 $\mathrm{M_{\odot}}$ cloud (bluish purple line) to those by \citet{Vaytet2017} (dashed red lines). Both the simulations use an initial Bonnor--Ebert density profile with an outer boundary $R_\mathrm{out} \approx$~3000 au, \mbox{$\rho_\mathrm{c} \approx 10^{-17} \mathrm{~g ~cm^{-3}}$} and initial temperature of 10~K. 

Note that since the temporal evolution is slightly different in both our studies owing to the differences in the gas equation of state (\citet{Saumon1995} used by \citet{Vaytet2017} vs \citet{Dangelo2013} used in this work), opacities and griding scheme (Lagrangian vs Eulerian), the comparisons are not made at the exact same time but when the central density $\rho_\mathrm{c}$ in both simulations reaches $\sim 10^{-1}$ g cm$^{-3}$. 

\citet{Vaytet2017} report a first core radius of roughly 2~au at the time of formation which then expands to about 5 au and stays roughly constant for a few hundred years and undergoes a second expansion phase which increases the core radius to $\sim$ 8 au. In our simulations, the radius is also roughly 2 au at the time of formation which then grows to about 5 au and gradually contracts back to $\sim$ 3 au. Some earlier studies also estimate a first core radius of roughly 3 au, however the core is seen only to be contracting with time \citep{Masunaga1998,Tomida2013}. The first core lifetime is $\sim$ 450 years in comparison to the $\sim$ 415 years obtained by \citet{Vaytet2017} and $\sim$~650~years by \citet{Masunaga2000} and \citet{Tomida2013}. 

For an initial 1 $\mathrm{M_{\odot}}$ cloud, at the end of our simulation, the second core radius is $\sim$ 3.95 $\mathrm{R_{\odot}}$ in agreement with \mbox{\citet[][$\sim$ 4 $\mathrm{R_{\odot}}$]{Masunaga2000}} and still expanding as seen by \citet[][$\sim$ 10 $\mathrm{R_{\odot}}$]{Tomida2013}. We note an initial contraction phase followed by expansion due to heating or mass accumulation as also seen by earlier studies \citep{Larson1969, Masunaga2000, Tomida2013}. In comparison, \citet{Vaytet2017} obtain a much smaller second core of roughly 1 $\mathrm{R_{\odot}}$ but they expect the core to expand to larger radii. 

Note that \citet{Masunaga2000} supposed an initial uniform density profile with \mbox{$\rho$ = 1.415 $\times 10^{-19} \mathrm{~g ~cm^{-3}}$}, an outer boundary $R_\mathrm{out} = 10^4$ au and initial temperature $T_0$ of 10 K whereas \citet{Tomida2013} adopted an initial Bonnor--Ebert density profile with $\rho_\mathrm{c}$ = 1.2 $\times 10^{-18} \mathrm{~g ~cm^{-3}}$, an outer boundary \mbox{$R_\mathrm{out} \approx$ 8800 au} and $T_0$ = 10 K. 

Since the studies by \citet{Vaytet2017} are closest to our approach, we further investigated the differences between our results for the collapse of an initial 1 $\mathrm{M_{\odot}}$ cloud by using the same temperature-dependent opacities instead of opacity tables (see \cref{sec:opacities}). 

All of the previous one-dimensional spherically symmetric radiation hydrodynamic (RHD) studies using frequency-dependent \citep{Masunaga1998,Masunaga2000} and gray FLD approximation \citep{Vaytet2017} as well as 3D radiation magnetohydrodynamic (RMHD) simulations without rotation and magnetic fields \citep{Tomida2013} were limited to the low-mass regime ($M_0 \leq 10 ~\mathrm{M_{\odot}}$). The thermal evolution and properties of the first and second cores from our low-mass runs are in good agreement with these previous works. 

In their collapse calculations for the low-mass regime, \citet{Vaytet2017} show comparisons for different initial cloud masses ($M_0 \leq$ 8 $\mathrm{M_{\odot}}$) at a time after the formation of the second core which indicate that most significant differences in the radial profiles of different core properties are seen outwards from the first shock as a horizontal spread (see their Fig.~4) which is similar to our results presented in \cref{sec:initcloudmass}. 

\citet{Baraffe2012}, \citet{Vaytet2013} and \citet{Vaytet2017} find the first core radius and mass to be similar within an order of magnitude for their collapse simulations with different initial cloud masses similar to the results presented herein for the low-mass regime (see \cref{sec:firstcore}). \citet{Masunaga1998} note that the first core radius and mass are independent of the initial cloud mass and density profile, but are weakly dependent on initial cloud temperature and opacity. We also find this weak dependence on initial cloud temperature as discussed in \cref{sec:initialsetup}. 

\citet{Tomida2010a} suggest that the thermal evolution may depend on the initial conditions such as cloud mass, opacities, temperature etc. In our studies, since we span a wide range of initial cloud masses beyond 10 $\mathrm{M_{\odot}}$, we find a transition region in the intermediate-mass regime which indicates a dependence on the initial cloud mass as discussed in the previous \cref{sec:firstcore}. We also find a linear dependence of this transition region on the initial cloud radius (see \cref{sec:Rout}).


\subsection{Limitations}
\label{sec:limitations}

In our studies, we use spherically symmetric models which neglect the effects of rotation and turbulence. It is however important to take into account effects due to non-negligible internal motions in molecular clouds. Rotation and magnetic fields are expected to have a significant effect on the evolution of the cloud and properties of the hydrostatic cores. These effects will be investigated in our future work. 

In comparison to pure radiation hydrodynamic simulations without rotation, depending on an ideal or resistive magnetohydrodynamics model and how slow or fast the rotation is, \citet{Tomida2013} find significant differences mostly in the first core lifetime and second core radius (see their Table 2). \citet{Tomida2013} and \citet{Vaytet2013} suggest that the first core lifetime increases slightly in the presence of rotation since it would slow down the collapse. 

The lifetimes estimated in our studies can thus be considered as lower limits. Despite the absence of rotation and magnetic fields, our results can still be used as initial conditions in stellar evolution simulations. 

\section{Summary}
\label{sec:Summary}

We performed 1D radiation hydrodynamic simulations to model the gravitational collapse of a molecular cloud through the formation of the first and second hydrostatic cores. As done by some previous studies, we emphasize on the importance of using a realistic gas equation of state which takes into account effects such as dissociation, ionization, rotational and vibrational degrees of freedom for the molecules and which also plays a significant role to account for the phase transitions from the monatomic to diatomic gas. 

Using an initial constant cloud temperature ranging from 5~to 100 K and an outer radius of 3000 au and 5000 au, we model clouds with different initial masses spanning a range from 0.5 to 100 $\mathrm{M_{\odot}}$. For each of these cases, we trace the evolution through an initial isothermal collapse phase, first core formation, adiabatic contraction, $\mathrm{H_2}$ dissociation, second collapse phase and the second core formation. The thermal evolution of the cloud for the 1 $\mathrm{M_{\odot}}$ cloud is summarized in Fig. \ref{fig:thermalevolution}. 

First, we varied the initial cloud mass, keeping a constant initial temperature of 10~K and an outer radius of 3000~au. We note the differences (within an order of magnitude) in the first core properties (listed in Table~\ref{tab:properties}), although the clouds with different initial masses follow a similar evolution. We examine the dependence of the first core properties on the initial cloud mass and find a transition region in the intermediate-mass regime. Our results indicate an increase in the first core radius with an increase in the initial cloud mass until around 8~--~10~$\mathrm{M_{\odot}}$, after which the first core radius decreases towards the higher initial cloud masses. This trend is also observed when comparing the first core mass for different initial cloud masses. 

We would like to draw more attention to the diminishing first core lifetimes for higher initial cloud masses which in turn affects the size and mass of the first core. It is also highly unlikely to observe first cores with such small lifetimes. Massive clouds have the highest accretion rate and are the most unstable which is why they evolve faster. For these cases, since the ram pressure is higher than the gas pressure, gravity acts as a dominant force which prevents a strong accretion shock. Hence, we predict that the first cores are non-existent in the high-mass regime. 

We confirmed the presence of the transition region in the intermediate-mass regime by performing an additional set of simulations. We use a different constant stability parameter $M_{\mathrm{BE}}/M_{\mathrm{0}}$ for the low-mass (0.5 to 10 $\mathrm{M_{\odot}}$), intermediate-mass (8 to 20 $\mathrm{M_{\odot}}$) and high-mass regime (30 to 100 $\mathrm{M_{\odot}}$), respectively. This implies that the initial cloud temperatures range from 5~to~100~K. The outer cloud radius for these runs is always fixed to 3000~au. We also investigated the influence of the outer cloud radius on the first core properties by performing simulations with a cloud radius of 5000 au for a constant initial temperature of 10~K. We found a similar transition region in the intermediate-mass regime. These results also indicate the weak dependence of the first core properties on the initial cloud temperature and outer radius. 

We note that the results for the first core lifetimes presented here should be treated as lower bounds on the core properties since we neglect the effects of rotation and magnetic fields which could slow down the collapse and in turn affect the core properties. These effects will be taken into account in future studies.   

\begin{acknowledgements} 
We thank the referee for the constructive comments which helped improve this work. A.B. would like to thank Neil Vaytet for the useful discussions during this work and for performing dedicated comparison simulations discussed in Appendix D. We would also like to thank Mykola Malygin for providing us with the gas opacity tables. Simulations shown here were run on the Isaac cluster at the Rechenzentrum Garching (RZG) of the Max Planck Society. We further acknowledge computing time on the BinAC cluster from the bwHPC-C5 initiative, funded by the Ministry of Science, Research and the Arts of the State of Baden-W\"urttemberg, Germany. R.K. and A.K. acknowledge financial support via the Emmy Noether Research Group on Accretion Flows and Feedback in Realistic Models of Massive Star Formation funded by the German Research Foundation (DFG) under grant no. KU 2849/3-1. G.D.M. acknowledges support from the Swiss National Science Foundation under grant BSSGI0\_155816 ``PlanetsInTime'' and from the DFG priority program SPP 1992 ``Exploring the Diversity of Extrasolar Planets'' (KU 2849/7-1). Parts of this work have been carried out within the frame of the National Centre for Competence in Research PlanetS supported by the SNSF. The authors acknowledge support by the High Performance and Cloud Computing Group at the Zentrum für Datenverarbeitung of the University of T\"ubingen, the State of Baden-W\"urttemberg through bwHPC and the German Research Foundation (DFG) through grant no INST 37/935-1 FUGG.
		
\end{acknowledgements} 

\bibliographystyle{yahapj}

\bibliography{Bibliography}

\begin{appendix}
	
\section{Comparisons to a uniform density cloud}
\label{sec:density}

\begin{figure}[!htp]
\centering
\begin{subfigure}{0.4\textwidth}
\includegraphics[width=\textwidth]{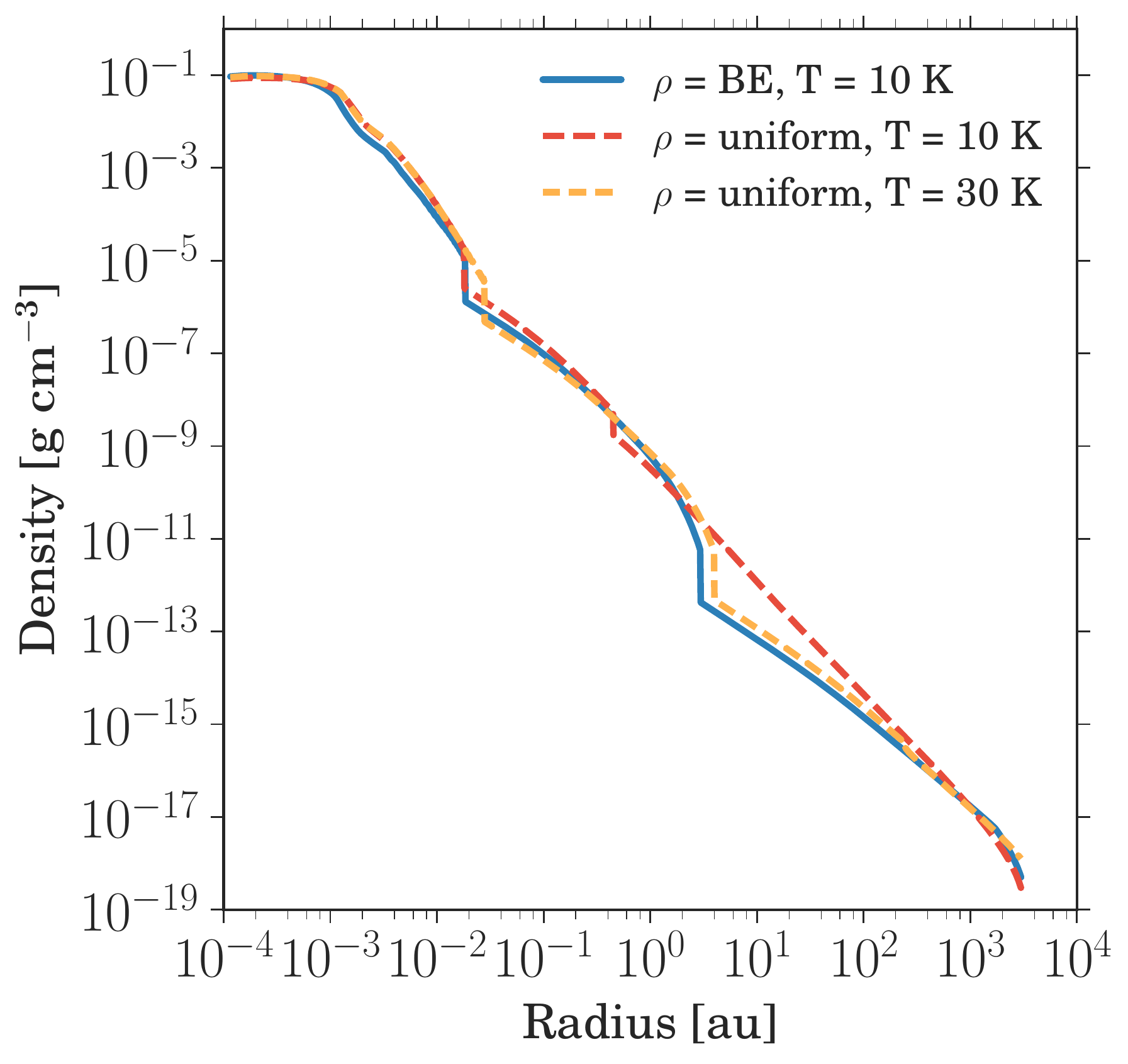}
\end{subfigure}
\begin{subfigure}{0.4\textwidth}
\includegraphics[width=\textwidth]{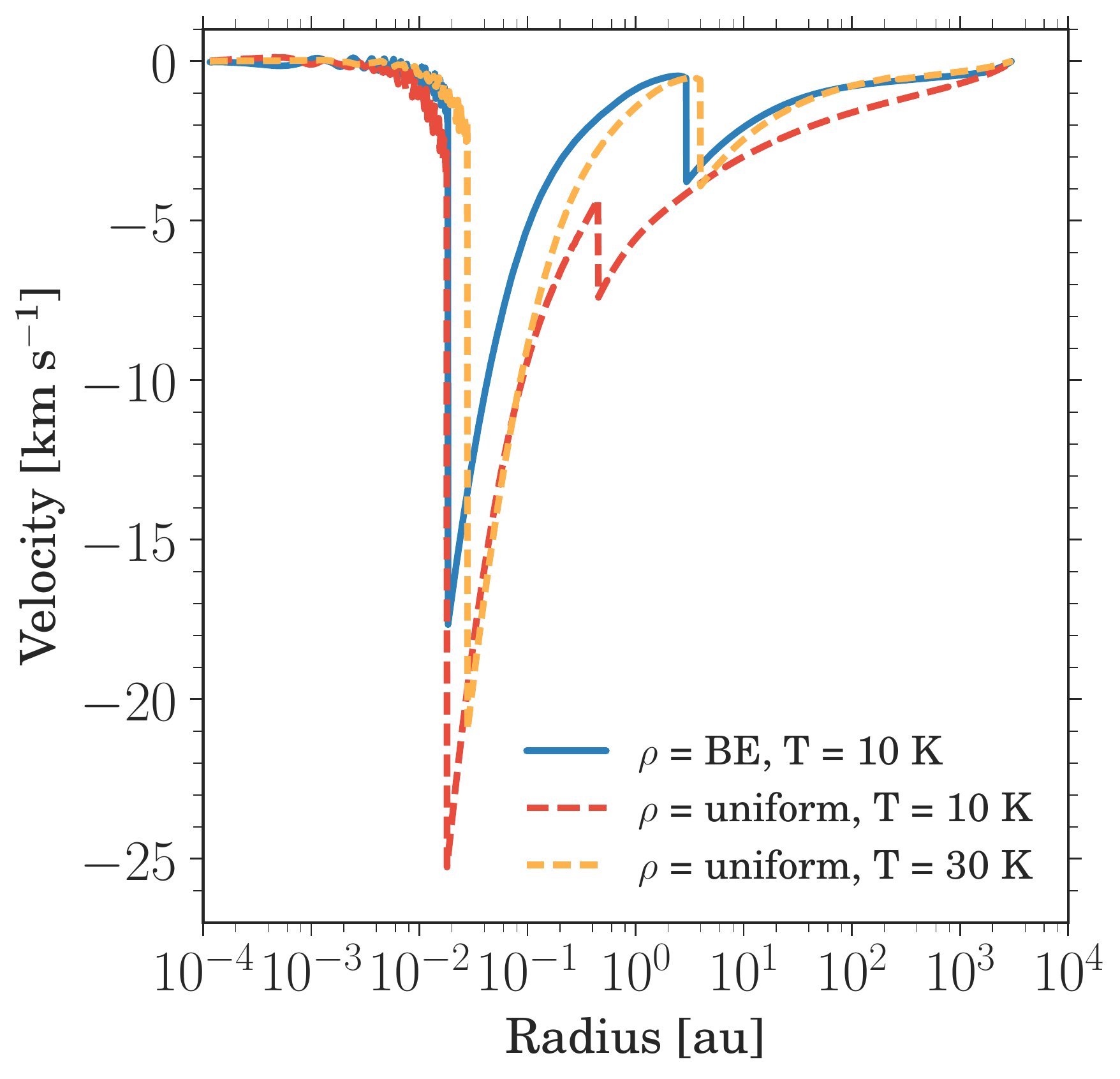}
\end{subfigure} 
\begin{subfigure}{0.4\textwidth}
\includegraphics[width=\textwidth]{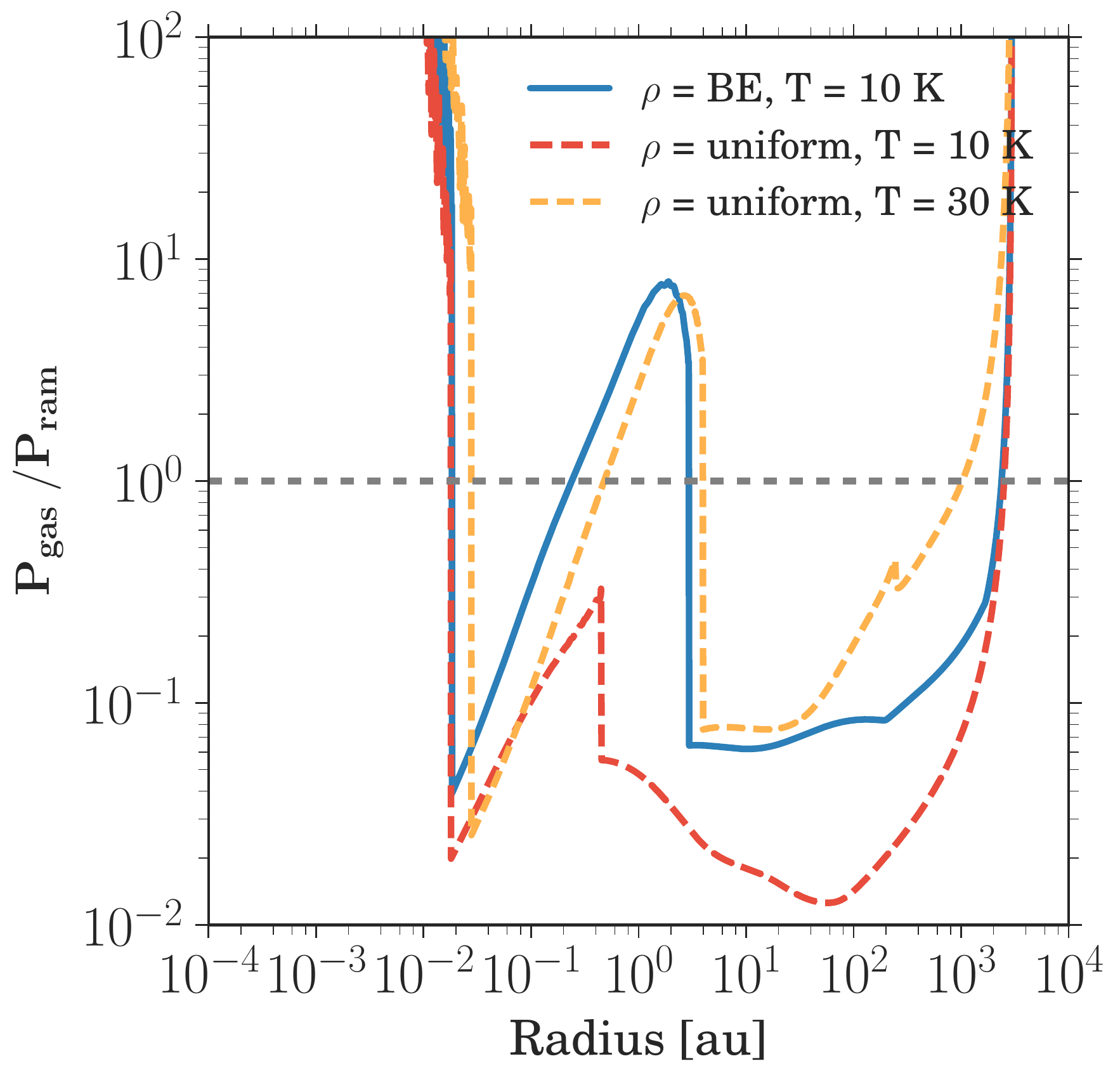}
\end{subfigure} 
\caption{Radial profiles of the density, velocity and the ratio of gas to ram pressure for collapse of a 1 $\mathrm{M_{\odot}}$ cloud for three different cases, using an initial Bonnor--Ebert sphere (blue) at 10~K, uniform cloud at 10 K (dashed red) and uniform cloud at 30 K (dashed yellow) are shown at a time step after the second core formation.}
\label{fig:rho}
\end{figure}

There have been previous collapse studies using a uniform density cloud as an initial setup instead of the Bonnor--Ebert sphere as considered in this work. However, the uniform density cloud eventually evolves into a Bonnor--Ebert like profile \citep{Masunaga1998,Larson1969}. Here, we compare the effect of a uniform density and Bonnor--Ebert like density profile on the core properties. Figure~\ref{fig:rho} shows the radial profiles of the density and velocity and the ratio of gas to ram pressure for collapse of a 1 $\mathrm{M_{\odot}}$ cloud for three different cases, using an initial Bonnor--Ebert sphere at 10 K (blue) and a uniform cloud at 10 K (dashed red) and 30 K (dashed yellow). 

For the 30 K uniform density cloud and the 10 K Bonnor--Ebert sphere clouds, we note that the initial density profile does not have a significant effect on evolution of the cloud as also seen by \citet{Vaytet2017}. However, in comparison to the Bonnor--Ebert sphere setup, the evolution of a uniform density cloud is much slower ($\sim 3 \times 10^{4}$ years). In contrast, \citet{Masunaga2000} argue that the initial density profile does affect the protostellar evolution due to different dynamics. Since there are no significant differences between the two density profiles, in our studies we use a Bonnor--Ebert sphere as an initial density profile. \cite{Vaytet2017} also suggest that a Bonnor--Ebert sphere is a better representation of the collapsing cloud. 

In case of a 10 K uniform density cloud, we note a different behavior. In this case, the strong ram pressure due to the high infall velocities is always higher than the gas pressure as seen in Fig.~\ref{fig:rho}. This may be because the clouds are highly unstable and gravity acts as the dominant force, with little effect due to pressure forces, which prevents the formation of the first hydrostatic core. A similar case devoid of the first accretion shock is seen by \cite{Vaytet2017}. They used an initial uniform density setup for a 4 $\mathrm{M_{\odot}}$ cloud collapse at an initial temperature of 5 K. This behavior of the 10 K uniform density cloud does not invalidate the previous studies that used initial uniform density, since the clouds were not unstable to skip the first core formation. 

\section{Numerical convergence}

\subsection{Resolution tests}
\label{sec:resolution}

\begin{figure}[!htp]
\centering
\begin{subfigure}{0.4\textwidth}
\includegraphics[width=\textwidth]{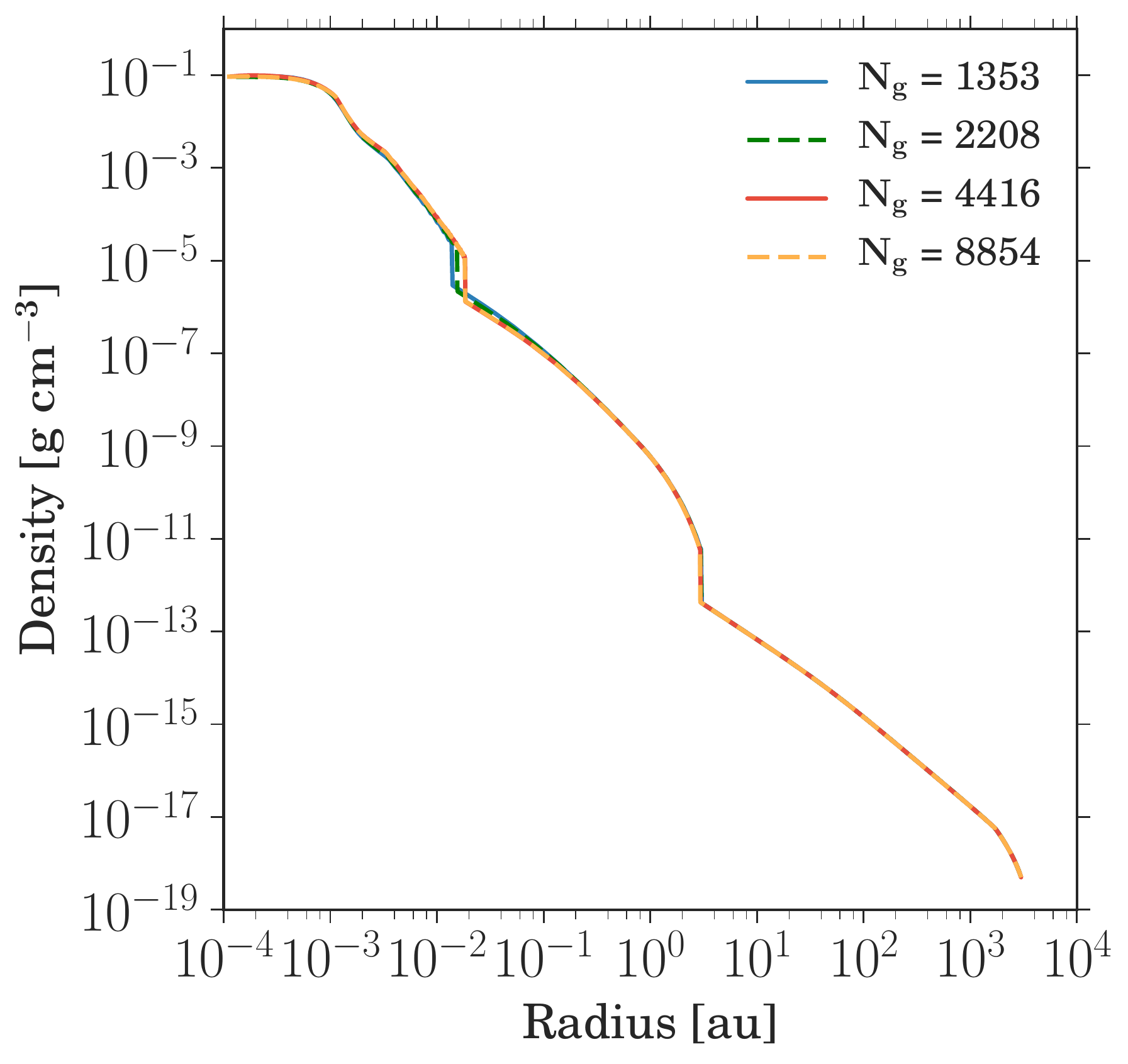}
\end{subfigure}
\begin{subfigure}{0.4\textwidth}
\includegraphics[width=\textwidth]{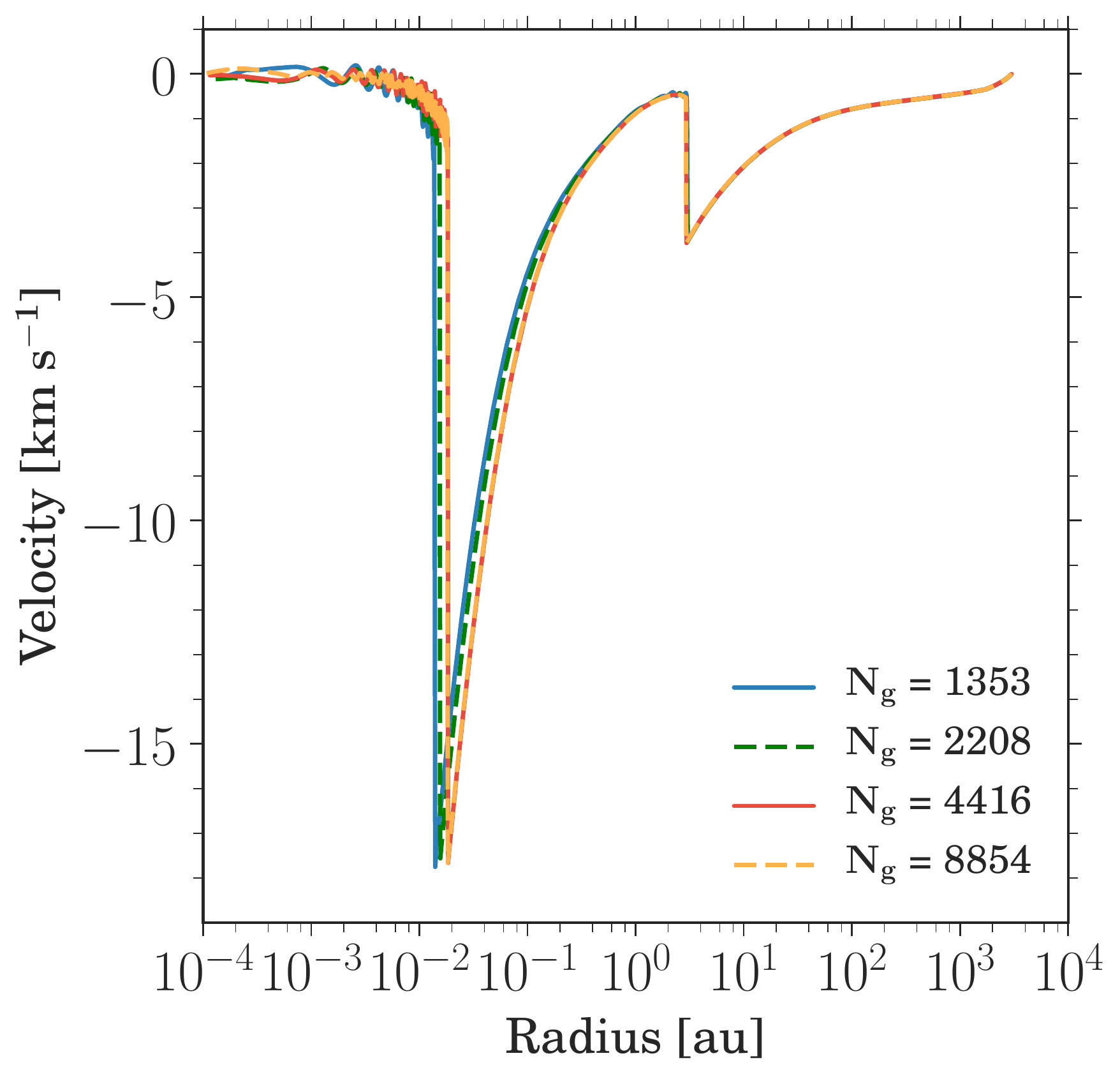}
\end{subfigure} 
\begin{subfigure}{0.4\textwidth}
\includegraphics[width=\textwidth]{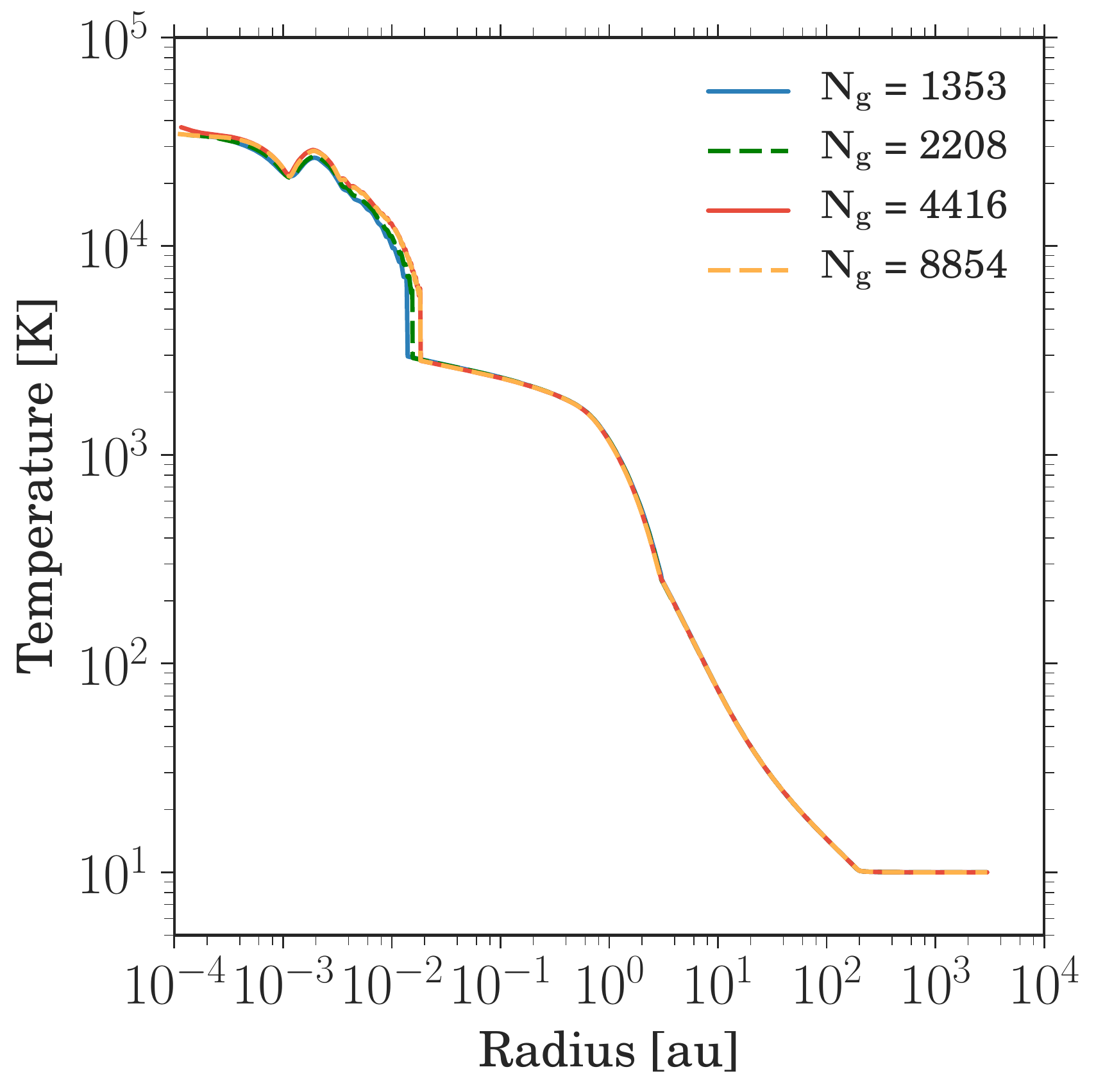}
\end{subfigure} 
\caption{Radial profiles of the density, velocity and gas temperature for an initial 1 $\mathrm{M_{\odot}}$ cloud at an initial temperature $T_\mathrm{0}$ of 10~K are shown at a time step after the second core formation. The different lines indicate the results using various grid resolutions. }
\label{fig:resolution}
\end{figure}

Resolution plays an important role especially when treating regions near accretion shocks. For an initial 1 $\mathrm{M_{\odot}}$ cloud, we performed core collapse simulations with the exact same initial conditions but using different resolutions.
The simulations using different resolutions have no significant effects on the evolution seen in Fig. \ref{fig:resolution} which indicates the numerically convergence for our studies. As expected for the lowest resolution, owing to fewer grid cells in the inner region, we see slight differences at the second shock position. These differences will probably increase for even lower resolutions. There seems to be a convergence around 4400 cells and above. This indicates a minimum resolution of around 4400 cells required for our simulations. 

\subsection{Comparisons for different inner radii}
\label{sec:rin}

\begin{figure}[!htp]
\centering
\begin{subfigure}{0.4\textwidth}
\includegraphics[width=\textwidth]{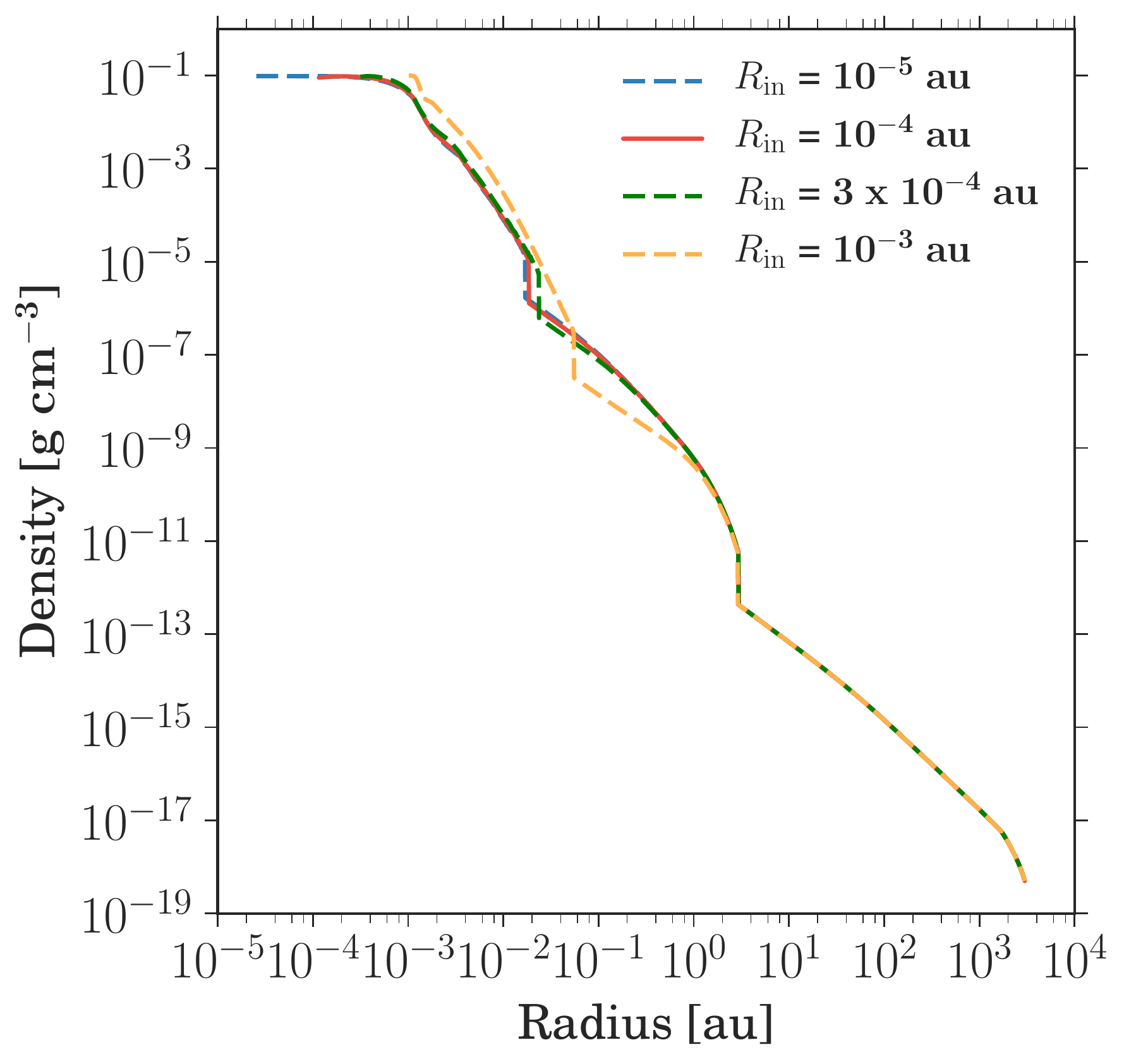}
\end{subfigure}
\begin{subfigure}{0.4\textwidth}
\includegraphics[width=\textwidth]{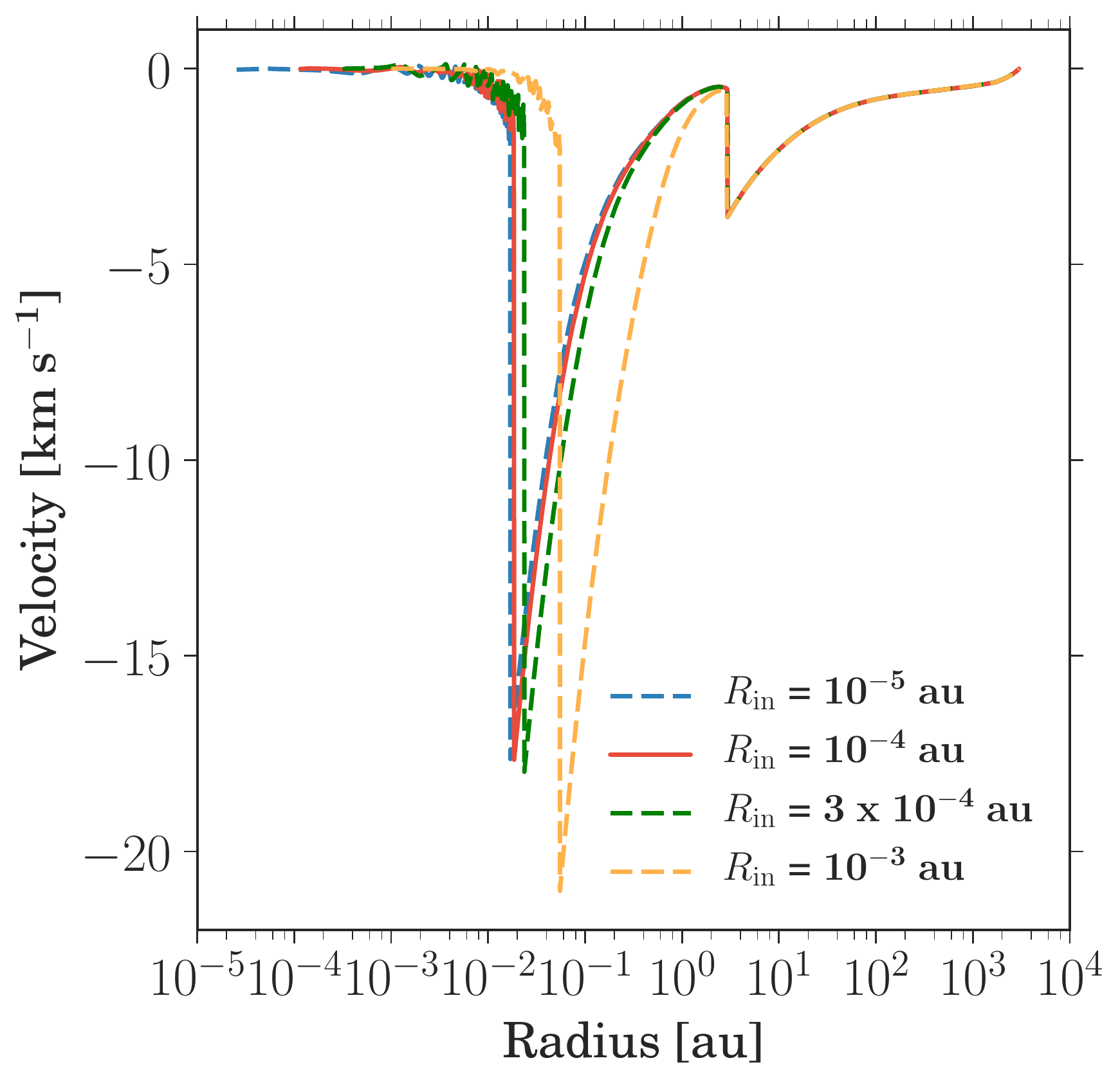}
\end{subfigure} 
\begin{subfigure}{0.4\textwidth}
\includegraphics[width=\textwidth]{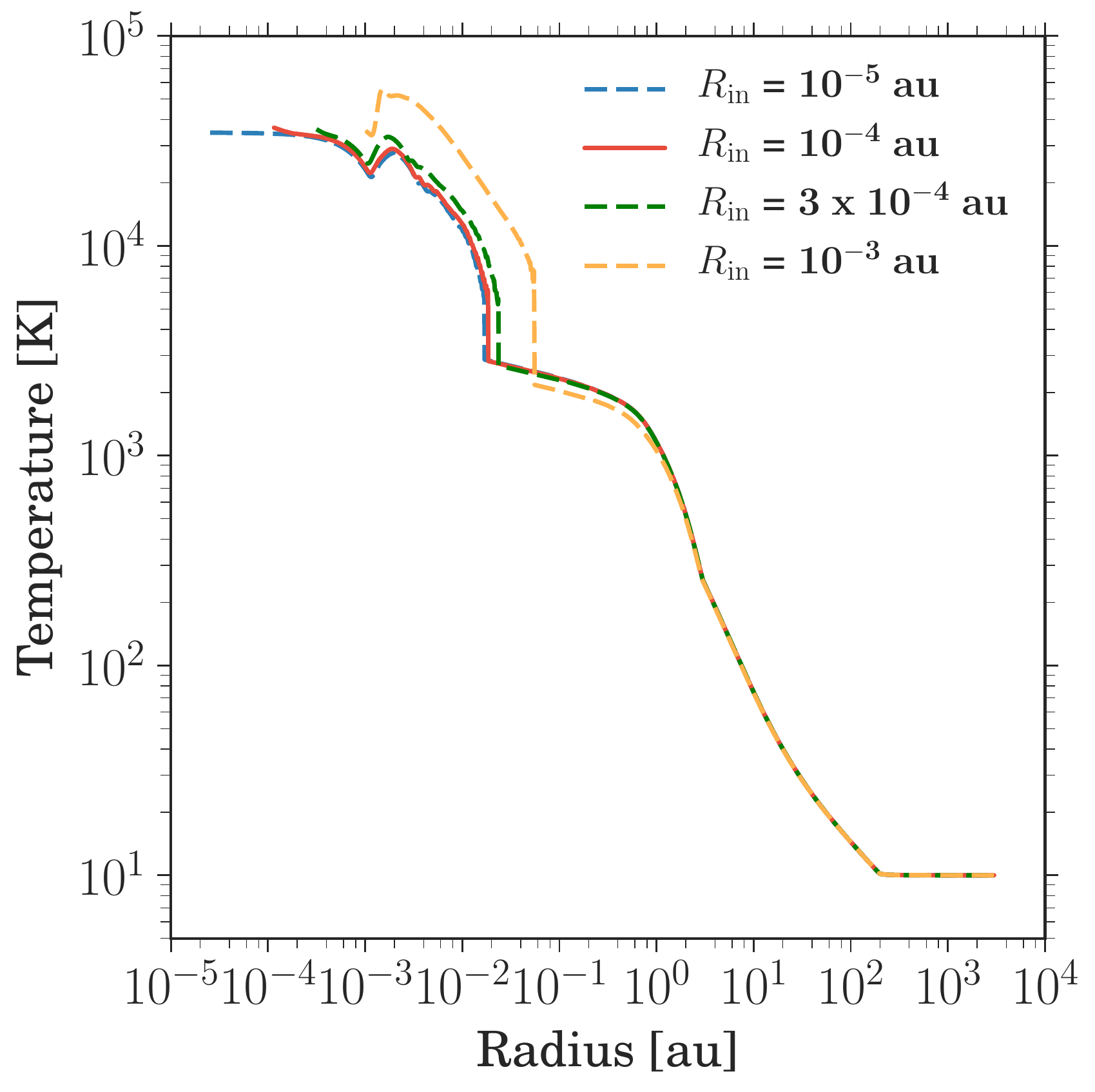}
\end{subfigure} 
\caption{Radial profiles of the density, velocity and gas temperature for an initial 1 $\mathrm{M_{\odot}}$ cloud are shown at a time step after the second core formation. The different lines indicate the results using different inner radius of the cloud. Dashed blue line for $10^{-5}$~au, red line for $10^{-4}$ au, dashed green line for $3 \times 10^{-4}$ au, and dashed yellow line for $10^{-3}$ au. }
\label{fig:rin}
\end{figure}

In order to ensure that the inner radius does not affect the second shock position, we ran some tests with different inner radii. As seen in Fig. \ref{fig:rin}, all of the runs evolve in a similar manner. We note the differences for the simulations with $R_\mathrm{in} = 3 \times 10^{-4}$ au and $R_\mathrm{in} = 10^{-3}$ au. However, there seems to be a convergence for an inner radius around $10^{-4}$ au. For our studies, we choose an inner radius of $10^{-4}$ au to avoid the boundary being too close to the second shock.

\section{Dependence on initial cloud radius}
\label{sec:Rout}

In order to test the robustness of the transition region seen in the first core properties as described in our results, we performed an additional set of simulations spanning initial cloud masses from 1 -- 100~$\mathrm{M_{\odot}}$ at a constant initial temperature of 10 K, but with a larger outer radius $R_\mathrm{out}$ of 5000~au. The computational grid for these simulations comprises of 4568 cells. We use 320 uniformly spaced cells from $10^{-4}$ to $10^{-2}$ au and 4248 logarithmically spaced cells from $10^{-2}$ to 5000 au. We again make sure that the last uniform cell and the first logarithmic cell are identical in size as described previously in \cref{sec:Method}. 

\begin{figure}[!tp]
\centering
\includegraphics[width=0.46\textwidth]{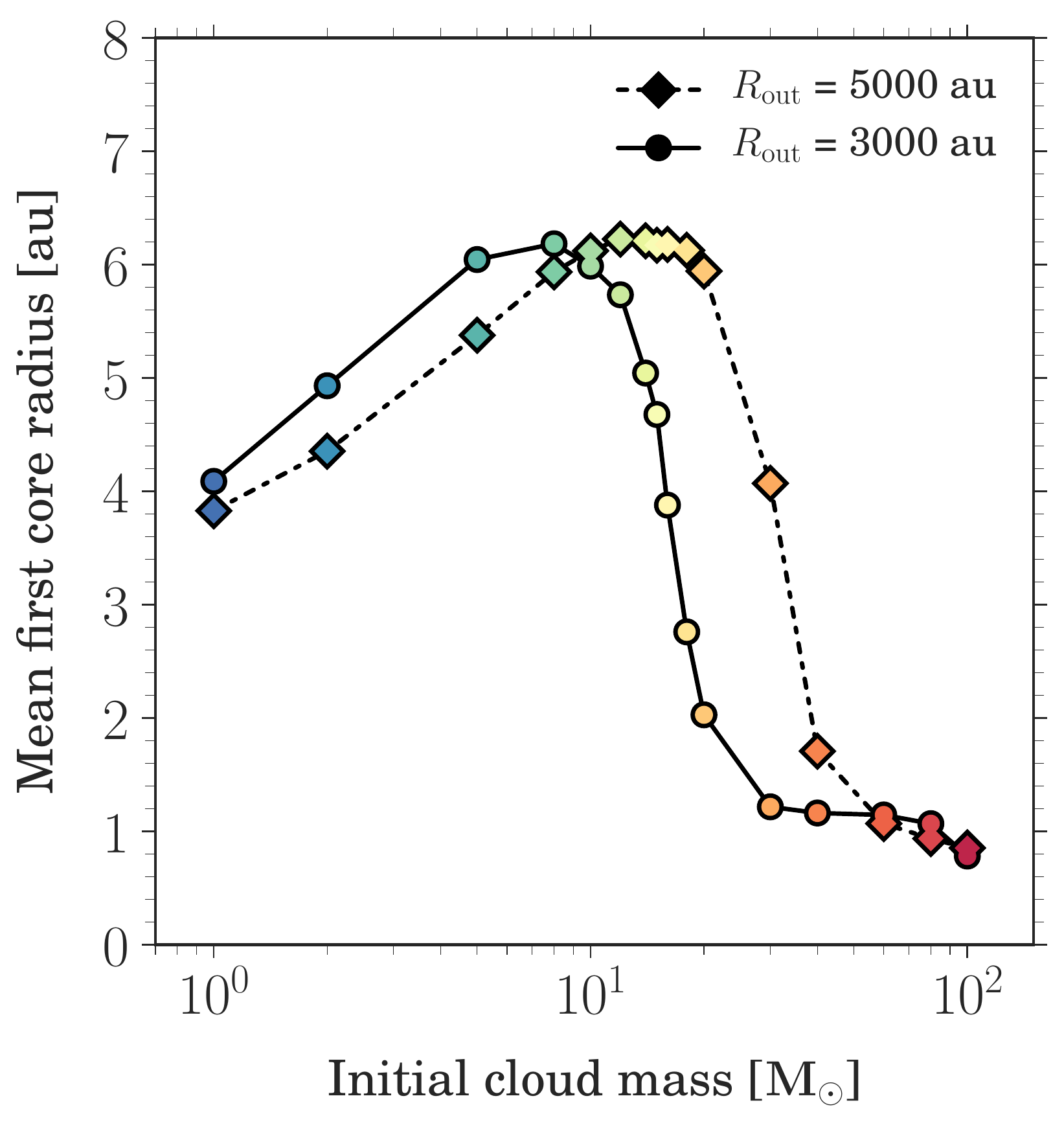}
\caption{Mean first core radius as a function of initial cloud mass where the mean radius is calculated over the time from the onset of the first core formation until the second core formation. Comparisons from two different set of simulations with an outer cloud radius of 3000 au (circles) and 5000 au (diamonds) are shown. }
\label{fig:radiusR5}
\end{figure}

\begin{figure}[!tp]
\centering
\includegraphics[width=0.47\textwidth]{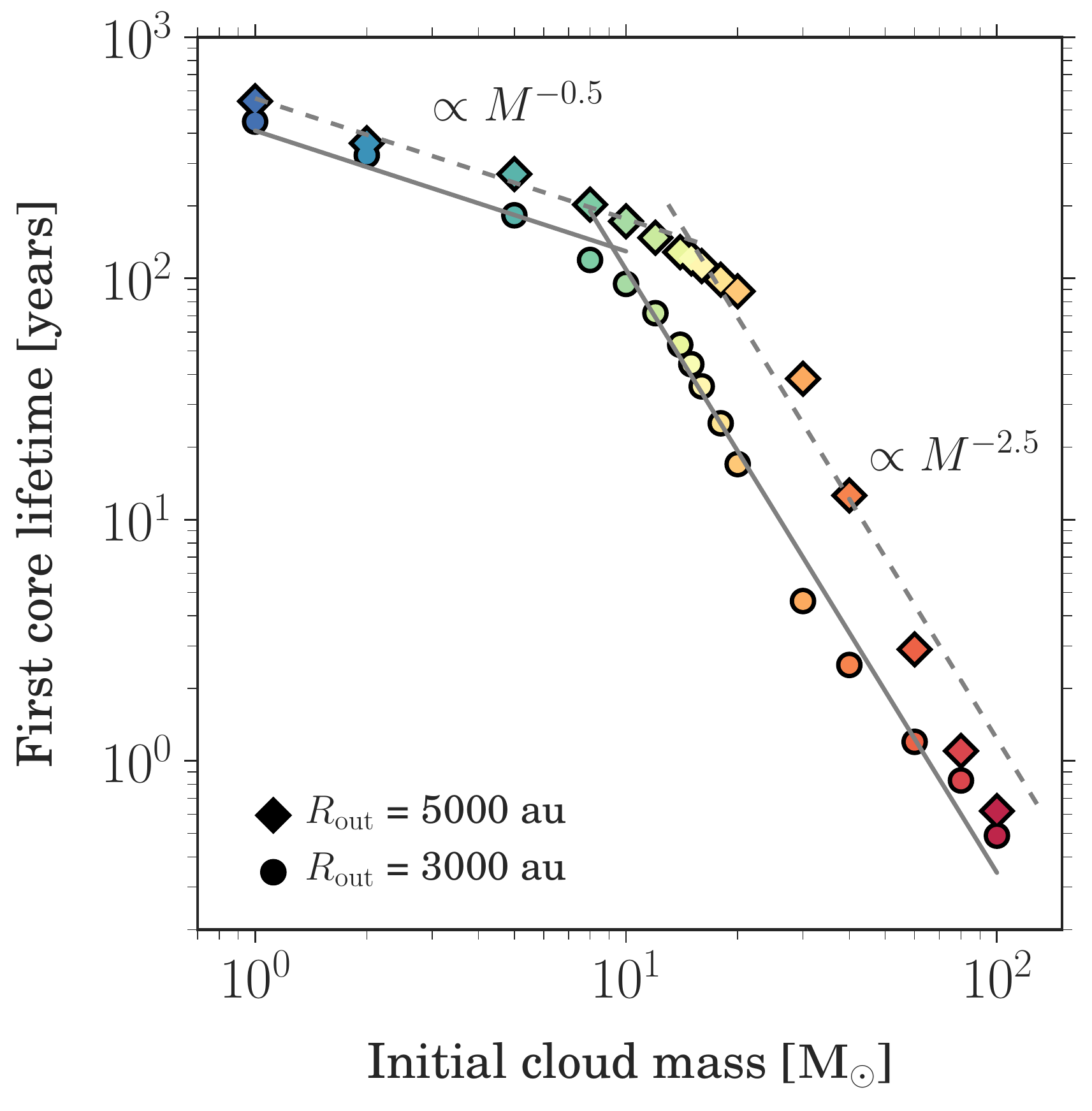}
\caption{First core lifetime i.e. time between the onset of formation of the first and second cores for different initial cloud masses. Comparisons from two different set of simulations with an outer cloud radius of 3000 au (circles) and 5000 au (diamonds) are shown. }
\label{fig:lifetimeR5}
\end{figure}

Figure \ref{fig:radiusR5} shows an increase in the mean first core radius until around 12 to 14 $\mathrm{M_{\odot}}$ beyond which it decreases towards the high-mass regime. Figure \ref{fig:lifetimeR5} indicates a shorter first core lifetime towards intermediate- and high-mass regime. Here, we compare the mean first core radius and first core lifetime from the runs with an outer radius of 3000 au (shown as circles) to those with an outer radius of 5000 au (shown as diamonds). We see that the lifetime scales as $M^{-2.5}$ in the intermediate- and high-mass regime. In this case, the fit (dashed line) also incorporates the radial dependence of $R^{-2.5}$ as derived in \cref{sec:firstcore}. 

We thus confirm the presence of a transition region in the intermediate-mass regime seen in the first core radius and lifetime which indicates that first cores are non-existent in the high-mass regime. We find a linear dependence of the transition mass on the initial cloud radius. 

\begin{table}[h]
\centering
\caption{Initial cloud properties}
\begin{tabular}[t]{ccccc}
\hline	
$M_{0} ~\mathrm{[M_{\odot}]}$ & $R_\mathrm{out}$ [au] & $T_{\mathrm{0}}$ [K] & $M_{\mathrm{BE}}/M_{\mathrm{0}}$ & $\rho_\mathrm{c} ~[\mathrm{g ~cm^{-3}}]$  
\TBstrut\\ \hline \hline 
1.0   & 5000   & 10.0       & 8.78e-01    & ~5.04e-18  \Tstrut \\
2.0   & 5000   & 10.0       & 4.39e-01    & 1.01e-17   \\
5.0   & 5000   & 10.0       & 1.76e-01	  & 2.52e-17   \\
8.0	  & 5000   & 10.0       & 1.01e-01    & 4.03e-17   \\
10.0  & 5000   & 10.0       & 8.78e-02    & 5.04e-17   \\
12.0  & 5000   & 10.0       & 7.32e-02	  & 6.04e-17   \\
14.0  & 5000   & 10.0       & 6.27e-02	  & 7.05e-17   \\
15.0  & 5000   & 10.0       & 5.85e-02    & 7.55e-17   \\
16.0  & 5000   & 10.0       & 5.49e-02    & 8.06e-17   \\
18.0  & 5000   & 10.0       & 4.88e-02    & 9.06e-17   \\
20.0  & 5000   & 10.0       & 4.39e-02    & 1.01e-16   \\
30.0  & 5000   & 10.0       & 2.93e-02    & 1.51e-16  \\
40.0  & 5000   & 10.0       & 2.19e-02    & 2.01e-16   \\
60.0  & 5000   & 10.0       & 1.46e-02    & 3.02e-16   \\
80.0  & 5000   & 10.0       & 1.01e-02    & 4.03e-16   \\
100.0 & 5000   & 10.0       & 8.78e-03    & ~5.04e-16  \Bstrut \\ \hline
\end{tabular}
\vspace{0.2cm}
\caption*{Note: Listed above are the cloud properties for runs with different initial cloud mass $M_{0} ~\mathrm{[M_{\odot}]}$, outer radius $R_\mathrm{out}$ [au], temperature $T_{\mathrm{0}}$ [K], stability parameter $M_{\mathrm{BE}}/M_{\mathrm{0}}$ and central density $\rho_\mathrm{c} ~[\mathrm{g ~cm^{-3}}]$. }
\label{tab:R5000}
\end{table}

\section{Effect of opacities}
\label{sec:opacities}

Here, we present comparison studies between our simulations and those kindly provided by N. Vaytet (2017, priv. comm.) mainly focusing on the effect of opacities. 

As described in \cref{sec:comparisons}, since the studies by \citet{Vaytet2017} are closest to our approach we compared our results for the collapse of a 1 $\mathrm{M_{\odot}}$ cloud at an initial temperature of 10 K. We note the discrepancies owing to the differences in the gas equation of state (\citet{Saumon1995} used by \citet{Vaytet2017} vs \citet{Dangelo2013} used in this work), opacity tables and griding scheme (Lagrangian vs Eulerian) as seen in Fig. \ref{fig:vaytet}. 

\begin{figure}[!htp]
\centering
\begin{subfigure}{0.38\textwidth}
\includegraphics[width=\textwidth]{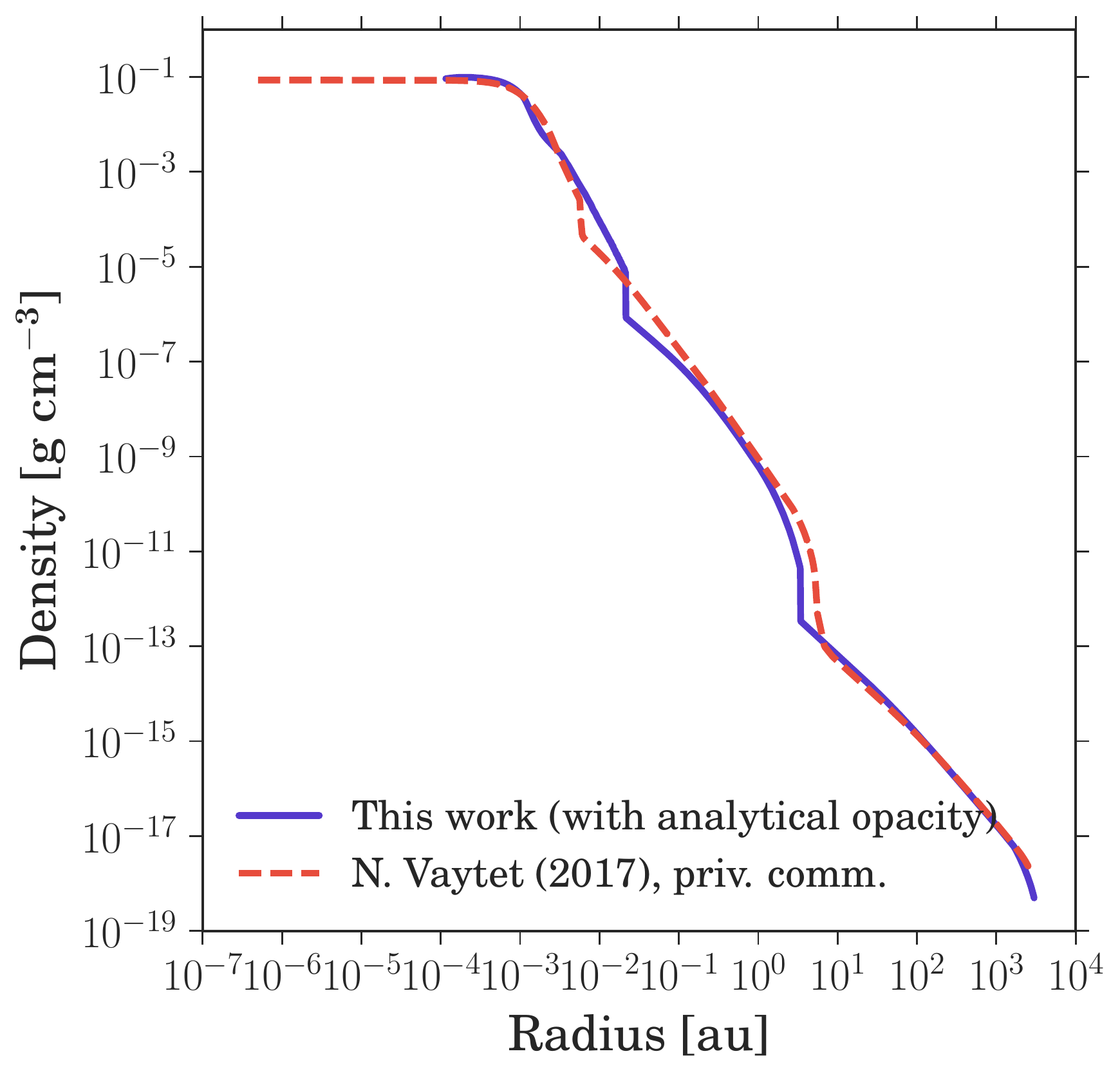}
\end{subfigure}
\begin{subfigure}{0.38\textwidth}
\includegraphics[width=\textwidth]{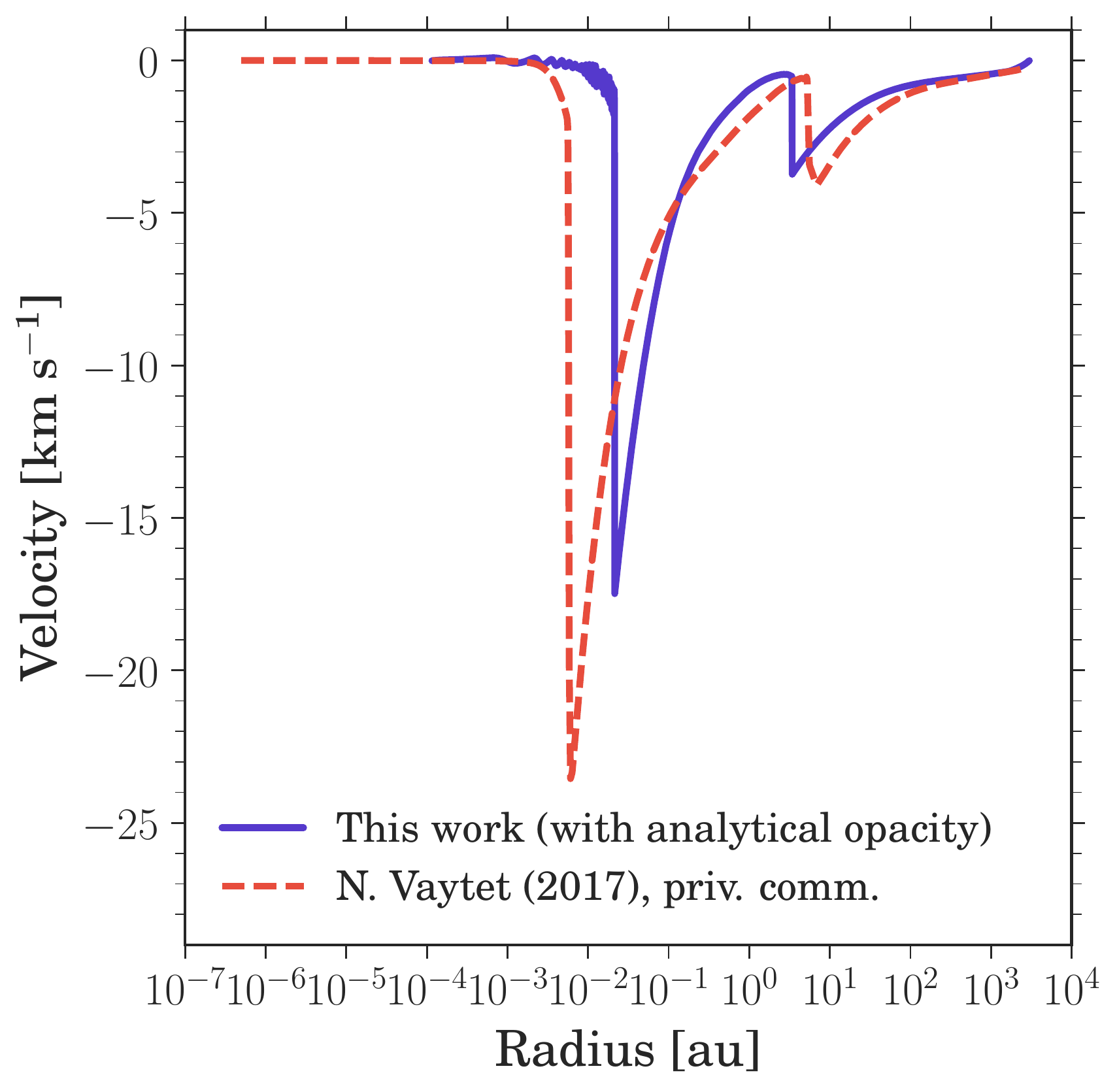}
\end{subfigure} 
\begin{subfigure}{0.38\textwidth}
\includegraphics[width=\textwidth]{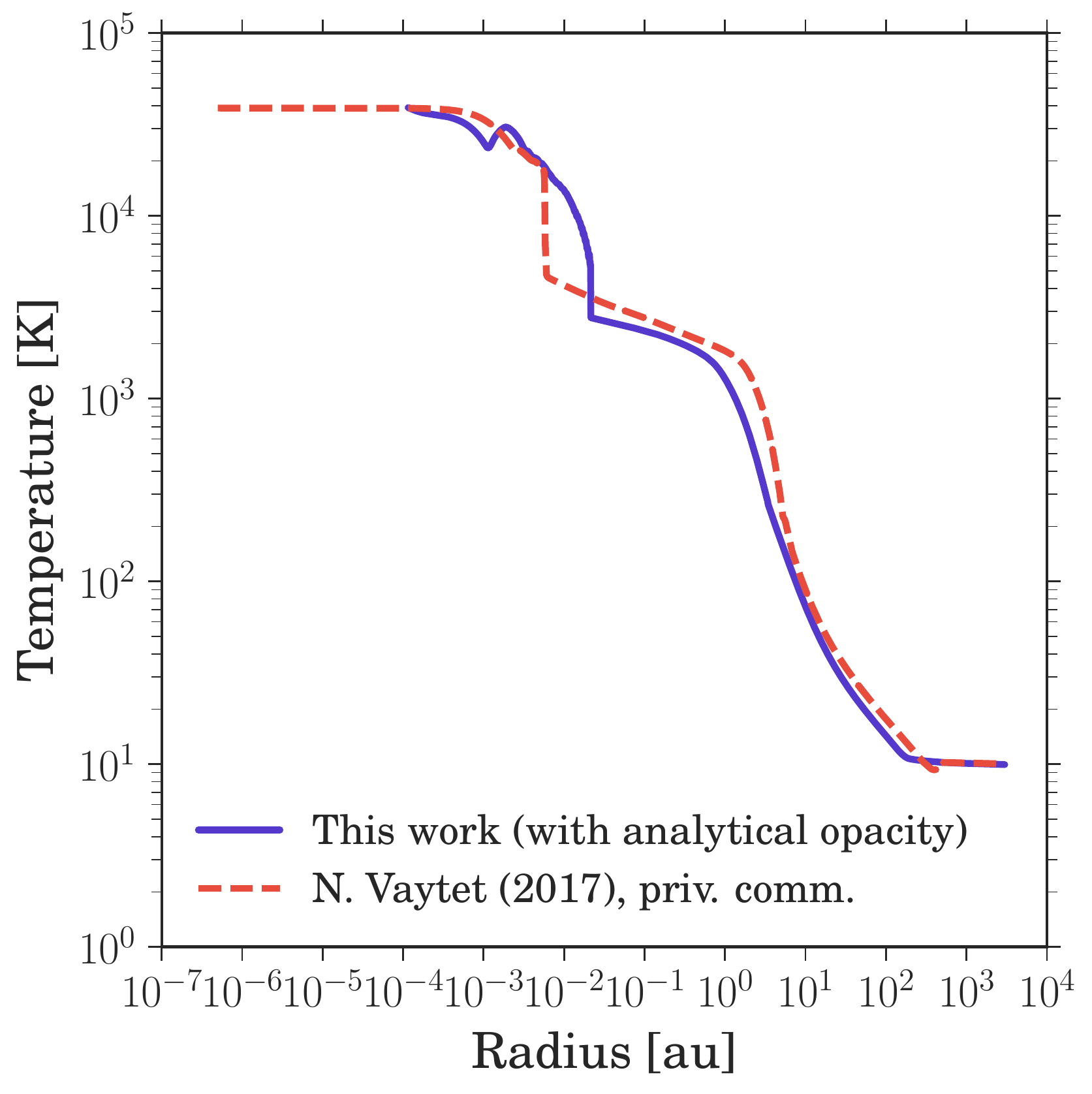}
\end{subfigure} 
\caption{Radial profiles of the density, velocity and gas temperature of an initial 1 $\mathrm{M_{\odot}}$ cloud at an initial temperature $T_\mathrm{0}$ of 10~K are shown at the time when central density $\rho_\mathrm{c}$ in both simulations reach roughly $10^{-1} \mathrm{~g ~cm^{-3}}$. The bluish purple solid lines show results from simulations described in \cref{sec:Method} while the dashed red line represents results from simulations provided by N. Vaytet (2017, priv. comm.). Note that for this comparison both codes use the same temperature-dependent opacity $\kappa = 0.02 ~(T/T_\mathrm{0})^2 ~\mathrm{cm^{2} ~g^{-1}}$.}
\label{fig:kappaT}
\end{figure}

In order to investigate the effect of opacities, we compared our simulation for the collapse of a 1 $\mathrm{M_{\odot}}$ cloud using a temperature-dependent opacity $\kappa = 0.02 ~(T/T_\mathrm{0})^2 ~\mathrm{cm^{2} ~g^{-1}}$ to the simulation provided by N. Vaytet (2017, priv. comm.) performed for an identical initial setup using the same temperature-dependent opacity. In both these runs, $T_\mathrm{0}$ = 10~K. As seen in Fig.~\ref{fig:kappaT}, although his simulations (dashed red line) still tend to produce a bigger first core radius, the difference is smaller compared to using different opacity tables (see Fig. \ref{fig:vaytet}). The second core does not contract as much in our simulations (bluish purple line), however as predicted the second core in his simulation may expand to obtain a value close to ours. These comparisons indicate that opacities play a role in determining the core properties but only provide some fine-tuning. Thus the main properties derived herein are still robust.  

In addition, the different treatment of the gas equation of state and griding scheme may also contribute to the differences. Note that since the temporal evolution is slightly different in both our studies owing to these differences the comparisons are not made at the exact same time but when the central density $\rho_\mathrm{c}$ in both simulations reaches $\sim$ $10^{-1} \mathrm{~g ~cm^{-3}}$.

\end{appendix}

\end{document}